\documentclass[11pt]{article}
\usepackage{amssymb}

\usepackage{graphicx}
\usepackage{amsmath}
\usepackage[all]{xy}

\newcounter{resultnum}[section]

\newcounter{conclusionnum}[section]

\newcounter{conditionnum}[section]

\newcounter{conjecturenum}[section]

\newcounter{examplenum}[section]

\newcounter{exercisenum}[section]

\newcounter{lemmanum}[section]

\newcounter{notationnum}[section]
\newtheorem{theorem}{Theorem}[section]

\newcounter{theoremnum}[section]
\newtheorem{definition}{Definition}[section]

\newcounter{definitionnum}[section]
\newtheorem{corollary}{Corollary}[section]

\newcounter{corollarynum}[section]

\newcounter{remarknum}[section]
\newtheorem{proposition}{Proposition}[section]

\newcounter{propositionnum}[section]

\newcounter{acknowledgementnum}[section]

\newcounter{algorithmnum}[section]

\newcounter{axiomnum}[section]

\newcounter{casenum}[section]

\newcounter{claimnum}[section]

\newcounter{summarynum}[section]

\newcounter{problemnum}[section]
\newenvironment{proof}[1][]{\textbf{Proof.} }{}
\newcommand{ \R} {\mbox{\rm I$\!$R}}

\font\fr=eufm10  scaled \magstep 1

\begin{document}

\title{Nonholonomic Algebroids, Finsler Geometry,\\
 and Lagrange--Hamilton Spaces }

\date{May 1, 2007}

\author{ Sergiu I. Vacaru\thanks{%
sergiu$_{-}$vacaru@yahoo.com, svacaru@fields.utoronto.ca } \\
%EndAName
{\quad} \\
\textsl{The Fields Institute for Research in Mathematical Science} \\
\textsl{222 College Street, 2d Floor, } \textsl{Toronto \ M5T 3J1, Canada} }

\maketitle

\begin{abstract}
We elaborate an unified geometric approach to classical mechanics,
Rie\-mann--Finsler spaces and gravity theories on Lie algebroids provided
with nonlinear connection (N--connection) structure. There are investigated
the conditions when the fundamental geometric objects like the anchor,
metric and linear connection, almost sympletic and related almost complex
structures may be canonically defined by a N--connection induced from a
regular Lagrangian (or Hamiltonian), in mechanical models, or by generic
off--diagonal metric terms and nonholonomic frames, in gravity theories.
Such geometric constructions are modelled on nonholonomic manifolds provided
with nonintegrable distributions and related chains of exact sequences of
submanifolds defining N--connections. We investigate the main properties of
the Lagrange, Hamilton, Finsler--Riemann and Einstein--Cartan algebroids and
construct and analyze exact solutions describing such objects.


\vskip0.2cm

2000 AMS Subject Classification:\

70G45, 70H03, 70H05, 53D17, 53B40, 83C99
\end{abstract}

\newpage

%\tableofcontents

%%%%%%%%%%%%%%%%%%%%%%%%%%%%%%%%%%%%%%%%%%%%%%%%%%%%%%%%%%%%%%%%%%%%

\section{Introduction}

The theory of Lie algebroids (see a mathematical background, discussion,
first applications and bibliography in \cite{acsw,mcz,cr,nistor}) has
received recently considerable attention in geometric mechanics \cite%
{weins1,lib,mart,dl1}, control theory \cite{cm1}, \ geometry of gauge
fields, string and gravity \cite{strobl1,strobl2,v501,v502,v503,v504}. In
the present paper, we study the canonical realization of the
Lagrange--Hamilton, Riemann--Finsler and Einstein--Cartan geometry (see
details, for instance, in Refs. \cite{ma1,bej,ma2,mhss,bcs}; for extensions
to superspaces, spinors and noncommutative spaces, see \cite{vnp,v0408121})
on Lie algebroids enabled with nonlinear connection structure and analyze
some important examples of such nonholonomic configurations and exact
solutions of the Einstein equations modelling algebroid structures.

There are two general approaches to geometrization of mechanics on the
tangent/cotangent bundle:

Roughly speaking, the first approach follows the idea to describe the
me\-cha\-nics in terms of sympletic geometry by developing certain
procedures of geometrization of the Euler--Lagrange equations (we cite a
recent a review \cite{ml} of results in mechanics and classical field theory
based on the (multi)sympletic formalism, differential forms, jets ....).

In a quite alternative form, i.e. in the second approach, the Lagrange and
Hamilton mechanics \footnote{%
in general, with spinor, supersymmetric, quantum group and another type of
noncommutative variables} was geometrized \cite{ma1,ma2,mhss,vnp,v0408121}
by using the methods of Finsler geometry, see also \cite{mats,bej,bcs} for
alternative researches and applications, on tangent and vector bundles and
generalizations to Clifford bundles, superbundles and projective modules in
noncommutative geometry..... Such spaces are enabled with nonholonomic
structures, i.e nonintegrable distributions, defined by nonlinear
connections (in brief, N--connection). Following this approach, the
N--connection and the bulk of fundamental geometric structures (metric,
canonical linear connection, almost sympletic and almost complex
structures...) are derived in general form starting from a regular (for
simplicity) Lagrangian and/or Hamiltonian. In such a case, the geometric
constructions are not related to the particular properties of corresponding
systems of partial differential equations, symmetries and constraints of
mechanical and field models, i. e. to the Euler--Lagrange equations which
are equivalently transformed into a (semi)spray configuration of
''nonlinear'' geodesics, but canonically defined by certain classes of
Sasaki type metrics, distinguished linear connections (adapted to the
N--connection) and corresponding torsions and curvatures.

The first mentioned approach was recently developed into some descriptions
of mechanics on Lie algebroids \cite{weins1,lib,mart,dl1} and the second one
has natural extensions to the noncommutative geometry of mechanics,
Clifford--Lagrange spaces and nonholonomic Dirac operators \cite{v0408121}.
Of course, both approaches are inter--related: For instance, certain (semi)
sprays and N--connecti\-on configurations were considered in the mentioned
first direction of researches and the almost sympletic/ complex/ tangent
structures were derived for the second ones. There are also works applying
both types of geometrizations of mechanics with explicit purposes to
elaborate a geometric quantization formalism for nonholomic mechanics \cite%
{v0407495,vaism1}.

There are a number of features motivating rigorous studies of new types of
nonholonomic algebroid structures and their applications. Here, we enumerate
\textbf{twelve} already distinguished directions:

The \textbf{first} one come from the geometry of nonholonomic manifolds and
bundles (in our case provided with N--connection structure). Such spaces
with nonholonomic distributions are characterized by generalized Lie type
nonholonomy relations for frames and admit quotients by the structure Lie
group which requests definition of a new class of nonholonomic Lie
algebroids (in brief, Lie N--algebroids). This is not a trivial academic
procedure of modelling physical theories and geometries on spaces provided
with algebroid structure because for general nonholonomic manifolds there is
a not completely solved problem of definition the curvature tensor\footnote{%
see detailed discussions and references, for instance, in \cite%
{v0407495,v0408121}}. In this work, we shall prove that it is possible to
construct curvature tensors for very general classes on nonholonomic
manifolds defined by nonlinear and linear connections on Lie N--algebroids.

The \textbf{second} direction arises from the modern gravity theories, in
general, with nontrivial torsion and nonmetricity. There are classes of such
spacetimes (considered also in this work) when the generic off--diagonal
metric, nonholonomic frames and nonlinear/linear connections mimic certain
Lie N--anholonomic algebroid structures constructed as exact solutions of
the generalized gravitational field equations (in particular, of the
Einstein equations). This can be related to new directions in constructing
exact solutions with Lie N--algebroid symmetries and investigation of their
symmetries, singular and nonholonomic configurations for which the
application of the methods of algebroid theory are crucial. We give in this
paper some explicit examples of such Einstein--Cartan algebroids.

There is a \textbf{third} motivation coming from the Riemann--Finsler
geometry, almost Kahler models and their generalizations. Such geometries
are naturally defined on tangent bundles and higher order extensions
provided with nonlinear connection structure, or on manifolds of even
dimensions (or containing embedding of such even dimension submanifolds)
provided with exact chains of submanifolds prescribing N--connection and
associated nonholonomic frame structures. It is not possible to define such
Finsler--Lagrange structures on general vector bundles with different
dimensions of the fibers and basic manifold. In another turn, the Lie
algebroid constructions related to the tangent bundles of the associated
vector bundles, allows to define new types of Finsler geometries (as well,
their Lagrange and/or Hamilton geometry extensions) because the even
dimensions arise naturally for the geometric objects transferred on
'tangents' to the fibers of a vector bundle. In this work, we define
explicitly and investigate such new type of Finsler/Lagrange/Hamilton
algebroids. As a matter of principle, such algebroid constructions are
defined by subclasses of structure functions of the Einstein--Cartan (and
more general metric--affine) algebroids provided with N--connection
structure.

The \textbf{forth} set of arguments for the theory of Lie of N--algebroids
results from the mentioned geometrization of mechanics with the Lagrange and
Hamilton functions defined on the Lie structure group quotients of the
tangent, respectively, cotangent bundles. This direction will be extended by
considering nonholonomic configurations (canonically defined by the
fundamental Lagrange or Hamilton functions and their respective homogeneous
variants for the Finsler and Cartan geometry). We study an application of
such Lie N--algebroid methods in order to elaborate a rigorous geometric
formalism for the optic--mechanical modelling of gravitational processes
like in analogous gravity.

There is the \textbf{fifth} approach related to investigation of the control
systems on the Lie algebroids \cite{mart1,cm1}. We shall not work in this
direction in this paper but we note here that the optimal control theory
having explicit relations to the Lagrange mechanics will obtain a number of
new features and possibilities by applying the formalism of Lie
N--algebroids provided with metric and distinguished connection (by
N--connection structures).

The \textbf{sixth} direction appears as the jet formalism elaborated on Lie
algebroids and related to models of classical field theory, time--dependent
and higher order mechanics, see the first results in \cite{cr,mart2,mart3}.
This concerns a further elaboration of algebroid multisympletic models for
noholonomic mechanics and classical field theories, sigma models,
gravitational and string actions and various type of topological theories.

We plan to elaborate a \textbf{seventh} direction devoted to N--connections
and field dynamics on Lie algebroid jets, revising the presympletic
formalism on the spaces of Cauchy data, classification of infinitesimal
symmetries, conservation lows in the geometric context of multisympletic
geometry and Ehresmann connections \cite{dlms}. The set of the invariants
and conservation lows on such algebroid spaces will be completed by the
corresponding fundamental system of nearly autoparallel maps and
conservation lows in the past investigated for the (pseudo) Riemannian and
generalized Finsler spaces \cite{vnam,dv}, see also Chapters 3 and 8 in \cite%
{vmon1}.

The \textbf{eights } direction may be related to the already stated
approaches to the ''Lie algebroid'' gauge theories and gravity models (of
string gravity and Einstein type) which in our opinion has certain
perspectives in gauge gravity modelling of the general relativity, brane
physics and string gravity. Such constructions may be derived from the gauge
locally anisotropic gravity and noncommutative gauge gravity \cite%
{vg,vncgg,vd}, see also Chapters 2 and 7 in \cite{vmon1}. The Lie algebroid
variants with singular maps, anchors and nonholonomic structure seem to
solve a number of problems concerning nonsemisimple realizations of
gravitational gauge theories and broken symmetries in such models.

The \textbf{ninths} new direction of the geometric and physical applications
of the algebroid mathematics is related to the Clifford algebroid structures
and various type of spinor--Finsler/Lagrange/Hamilton geometries and
redefinition of mechanics on spinor bundles provided with N--connection
structures \cite{vsp1,vshep,vstavr,vvicol} and Chapter 6 in \cite{vmon1}.
This is not a trivial rewriting of the Lagrange or Hamilton formalism in
spinor terms. An explicit geometrization of mechanics both in terms of
sympletic and N--connection structures on algebroids allows to work directly
with singular, nonholonomic and quotient symmetries which can be related to
the quantum ''world'' via spinor variables which in such cases are modelled
by nonholonomic Clifford structures, nonlinear connections, connections and
curvatures on such spaces.

It became already explicit the \textbf{tenths} direction which leads from
the Clifford--Lagrange and Clifford--Riemann--Finsler geometry to the
noncommutative geometry. Following the geometry of nonholonomic frames, the
Lagrange--Finsler algebroids can be treated as Riemann--Cartan manifolds
provided with corresponding prescribed types of N--connection and Lie
algebroid structures. The Riemann geometry can be 'extracted' from the
noncommutative geometry via the Dirac operator formalism \cite%
{connes1,bondia,rennie1}. A generalized Dirac operator approach was already
elaborated for the generalized Riemann--Finsler and Lagrange-Hamilton spaces %
\cite{v0408121} which emphasizes the possibility to define noncommutative
extensions of the Clifford--Lagrange--Hamilton algebroids.

The \textbf{eleventh} approach may be considered in connection to mechanical
integrators, numerical methods and applications in economics \cite%
{integrator}. In this case, it would be necessary the elaboration of a
discrete geometry and a corresponding calculus for nonholonomic algebroid
structures. The specific point would be that numeric and analytic methods of
the theory of differential equations will have to be elaborated in a form
preserving the prescribed algebroid configuration.

Finally, in the \textbf{twelfth}, less distinguished direction with possible
new subdirections, the mentioned classical commutative and noncommutative
geometry and algebroid methods seem to have a perspective to the geometric
quantization, spin networks and path quantum gravity, Fedosov spaces, Hopf
algebras and quantum group geometry, see references and algebroid related
discussions in \cite{acsw}. It also follows from the constructions with Lie
algebroids and nonholonomic geometries presented, for instance in \cite%
{v0408121}, and has a direct relation to the cohomology of Jacobi manifolds
and algebroids \cite{dlmp1,dlmp2}, but this can be considered as a long term
program of our further researches.

The purpose of this work is to elaborate the theory of Lie N--anholonomic
algebroids and to present a set of strong arguments and motivations for such
constructions, derived from the geometric mechanics and gravity theory. We
shall follow in the bulk the first mentioned four directions but also
formulate a Lie algebroid nonholonomic geometric background for a future
work related to the rest of eight directions.

This paper is organized into six sections. In Section 2, we recall some
necessary results on Lie algebroids and their prolongations. In Section 3,
we clarify the relevance of the geometry of nonlinear connections to
geometric models of mechanics on tangent bundles and Lie algebroids. We
elaborate the almost Hermitian model of Lagrange mechanics on Lie algebroids
and define the canonical nonlinear connection, metric and distinguished
connection, almost complex and sympletic structures all induced by regular
Lagrangians. The theory of Lie algebroids provided with nonlinear connection
structure is formulated in Section 4. We investigate the main classes of
nonlinear and linear connections and prove the main theorems on torsions and
curvatures of such nonholonomic manifolds provided with, in general,
nonintegrable distributions. In Section 5, we define the Finsler and
Hamilton algebroids and their generalizations and prove that there are
canonical Lie algebroid structures determined by the canonical sympletic,
metric and nonlinear connections induced by corresponding Lagrangians,
Hamitlonians and/or Finsler--Cartan fundamental functions. Section 6 is
devoted to a proof that a certain class of Lie algebroids can be associated
to exact solutions in gravity theories, parametrized by generic
off--diagonal metrics and nonholnomic frames. We formulate some criteria
when the gravitational processes may be modelled by optical and continuous
mechanics media and can geometrized in the Lie algebroid approach. Finally,
some examples of exact solutions defining such nonholonomic algebroid
configurations are constructed and analyzed.

\section{Preliminiaries: Lie Algebroids and Prolongati\-ons}

The section is an overview of the results and conventions on Lie algebroids
and vector/tangent bundles to be applied and developed in this work.

\subsection{Definition of Lie algebroids}

Let $\mathcal{E}=(E,\pi ,M)$ be a vector bundle defined by surjective
projection $\pi :E\longrightarrow M$ when the dimensions of the base and
toatal manifolds are respectively $\dim M=n$ and $\dim E=n+m.$ A Lie
algebroid $\mathcal{A}\doteqdot (E,\left[ \cdot ,\cdot \right] ,\rho )$ is
defined as the vector bundle $E$ provided with algebroid structure $(\left[
\cdot ,\cdot \right] ,\rho ),$ \footnote{%
Sometimes, we shall write for this vector bundle only the symbol $E$, if
there is not confusion.} where $\left[ \cdot ,\cdot \right] $ is a Lie
bracket on the $C^{\infty }(M)$--module of sections of $E,$ denoted $Sec(E),$
and the 'anchor' $\rho $ is defined as a bundle map $\rho :\ E\rightarrow TM$
($TM$ is the tangent bundle to $M$) such that
\begin{equation*}
\left[ X,fY\right] =f\left[ X,Y\right] +\rho (X)(f)Y
\end{equation*}%
for $X,Y\in Sec(E)$ and $f\in C^{\infty }(M).$ The anchor also induces an
homomorphism of $C^{\infty }(M)$-modules $\rho :Sec(A)\rightarrow \mathcal{X}%
^{1}(M)$ where $\wedge ^{r}(M)$ and $\mathcal{X}^{r}(M)$ denote,
respectively, the spaces of differential $r$--forms and $r$--multivector
fields on $M.$

Let us state the typical notations for abstract (coordinate) indices given
with respect to an arbitrary or coordinate basis. For a local basis on $%
\mathcal{E}$ we write $\mathbf{e}_{\alpha }=(e_{i},v_{a}).$ The small Greek
indices $\alpha ,\beta ,\gamma ,...$ are to be considered general ones when
the values $1,2,\ldots ,n+m$ and $i,j,k,...$ and $a,b,c,...$ label
respectively the geometrical objects on the base and typical fiber. The
local coordinates of a point $u\in \mathcal{E}$ are written $\mathbf{u=}%
(x,u),\ $or $u^{\alpha }=(x^{i},u^{a}),$ where $u^{a}(\mathbf{u})$ is the $a$%
-th coordinate with respect to the basis $v_{a}$ and $(x^{i})$ are local
coordinates with respect to the basis $e_{i}$ on $M$. For our purposes, it
is convenient to write the local coordinates on a tangent bundle, when $%
\mathcal{E}=TM,$ in the form $(x^{i},y^{k}),$ where $i,j,k,...=1,...,n.$ We
shall write also briefly $E$ or $\mathcal{E}$ instead of the set of sections
of the total space $Sec(E),$ if such a notation will not result in
ambiguities.

In local form, the Lie algebroid structure is defined by its structure
functions $\rho _{a}^{i}(x)$ and $C_{ab}^{f}(x)$ on $M,$ determined by the
relations
\begin{eqnarray}
\rho (v_{a}) &=&\rho _{a}^{i}(x)\ e_{i},  \label{anch} \\
\lbrack v_{a},v_{b}] &=&C_{ab}^{c}(x)\ v_{c}  \label{liea}
\end{eqnarray}%
and subjected to the structure equations
\begin{equation}
\rho _{a}^{j}\frac{\partial \rho _{b}^{i}}{\partial x^{j}}-\rho _{b}^{j}%
\frac{\partial \rho _{a}^{i}}{\partial x^{j}}=\rho _{c}^{j}C_{ab}^{c}~%
\mbox{\ and\ }\sum\limits_{cyclic(a,b,c)}\left( \rho _{a}^{j}\frac{\partial
C_{bc}^{d}}{\partial x^{j}}+C_{af}^{d}C_{bc}^{f}\right) =0.  \label{lasa}
\end{equation}

Roughly speaking, the concept of Lie algebroid $\mathcal{A}$ substitutes
that of the tangent bundle $TM,$ when an element $\sigma \in E$ is
considered as a generalized ``velocity'' and the actual velocity $V$ is
obtained via the anchor map, $V=\rho (\sigma ).$ Subjected to the conditions
(\ref{lasa}), the image $\rho (E)$ defines an integrable generalized
distribution; therefore, $M$ is foliated by the leaves of $\rho (E)$. We
will say that an algebroid is transitive if it has only one leaf which is
obviously equal to $M.$ This property holds if and only if the map $\rho $
is surjective (we shall construct mechanical models on such manifolds). For
certain gravity models, it is possible that $\mathcal{A}$ is not transitive
but the restriction of a Lie algebroid to a leaf $q\subset
M,E_{q}\rightarrow q$ is transitive (we can also consider a trivial
embedding of $E_{q}$ into a higher dimension which allows a surjective map).
One says that a Lie algebroid is locally transitive at a point $x\in M$ if $%
\rho _{x}:E_{x}\rightarrow T_{x}M$ \ is surjective. In this case, the point $%
x$ is contained in a leaf of maximal dimension.

If $\mathcal{A}$ is a Lie algebroid and $E^{\ast }$ is the dual of $E,$ we
can introduce the differential of $E,$ $d^{E}:Sec(\wedge ^{k}E^{\ast
})\rightarrow $ $Sec(\wedge ^{k+1}E^{\ast })$ as follows
\begin{eqnarray*}
&&d^{E}\lambda (X_{0},X_{1},...,X_{k})=\sum\limits_{r=1}^{k}(-1)^{r}\rho
(X_{r})\left( \lambda (X_{0},...,\widehat{X_{r}},...,X_{k})\right) \\
&&+\sum\limits_{r<r^{\prime }}^{k}(-1)^{r+r^{\prime }}\rho (X_{r})\left(
\lambda (\left[ X_{r},X_{r^{\prime }}\right] ,X_{0},...,\ \widehat{X_{r}}%
,...,\ \widehat{X_{r^{\prime }}},...,X_{k})\right)
\end{eqnarray*}%
for $\lambda $ being an element from the set of sections of the $E$--valued $%
k$--forms, $Sec(\wedge ^{k}E^{\ast }),$ and $X_{0},X_{1},...,X_{k}\in
Sec(E), $ where $\ \widehat{X_{r}}$ means that this term is omitted under
summation. It is obvious that $(d^{E})^{2}=0.$ The trivial examples of such
differentials are those for a function $f\in C^{\infty }(M)$ and $\theta
=\theta _{a}v^{a}\in Sec(E^{\ast })$ when, respectively,
\begin{equation}
d^{E}f=\frac{\partial f}{\partial x^{i}}\rho _{a}^{i}v^{a}\mbox{\
and\ }d^{E}\theta =\left( \rho _{a}^{i}\frac{\partial \theta _{b}}{\partial
x^{i}}-\frac{1}{2}\theta _{c}C_{ab}^{c}\right) v^{a}\wedge v^{b},
\label{form1a}
\end{equation}%
where $\left( d^{E}f\right) (X)=\rho (X)(f).$ Therefore,
\begin{equation*}
d^{E}x^{i}=\rho _{a}^{i}v^{a}\mbox{\ and\ }d^{E}v^{a}=-\frac{1}{2}%
C_{bc}^{a}v^{b}\wedge v^{c}.
\end{equation*}

We also define the Lie derivative with respect to $X$ as the operator $%
L_{X}^{E}:Sec(\Lambda ^{k}E^{\ast })\longrightarrow Sec(\Lambda ^{k}E^{\ast
})$ given by $L_{X}^{E}=i_{X}\circ d^{E}+d^{E}\circ i_{X}$.

On the other hand, for a function $f\in C^{\infty }(M),$ one introduces the
'complete' and 'vertical' lifts to $E$ of $f$ defined respectively by $%
~^{c}f(u)=\rho (u)(f)$ and $~^{v}f(u)=f(\pi (u))$ for all $u\in E.$ Let us
consider a section $X$ of $E.$ The vertical lift of $\ X$ is a vector field
on $E$ given by $~^{v}X(u)=\ ^{v}X(\pi (u))_{u},$ for all $u\in E,$ where \
a canonical isomorphism is defined by the map $_{\quad u}^{v}:E_{\pi
(u)}\rightarrow T_{u}(E_{\pi (u)}).$

Let us consider the notion of complete lift of a section. The complete lift $%
~^{c}X$ of a section $X$ of $E$ is the vector field on $E$ which satisfies
the following properties:

\begin{enumerate}
\item $~^{c}X$ is $\pi $-projectable on $\rho (X)$

\item $~^{c}X(\hat{\mu})=\widehat{L_{X}^{E}\mu }$, for all $\mu \in
Sec(E^{\ast })$.
\end{enumerate}

Here $\hat{\mu}$ denotes the linear function on $E$ defined by $\hat{\mu}%
(u)=\mu(\pi(u))(u)$, for all $u\in E$.

Now, we briefly consider the notion of prolongation of Lie algebroids [for
details, see \cite{dl1}], in order to generate a prolongation over a smooth
map; our notations are different, being adapted to those from Lagrange and
Finsler geometry \cite{ma1,ma2,v0408121}. The underlying motivation for
prolongations is that of formulating the ``second order dynamical models''
on $E$ and relating such constructions to similar models on $TM.$

\subsection{The prolongation of a Lie algebroid}

We consider a local basis $\{v_{a}\}$ of $Sec(E).\ $If $u\in E,\pi (u)=x\in
M,$ and $x^{i}$ are local coordinates around $x,$ we have $u=u^{a}v_{a}$ and
the bundle coordinates on $E$ are $(x^{i},u^{a}).$ In this case, the map $%
\rho $ acts in the form $\rho (x^{i},u^{a})=(x^{i},\rho _{a}^{i}(x))$ and $%
\rho (v_{a})=\rho _{a}^{i}(x)\frac{\partial }{\partial x^{i}}.$

Let $E\overset{\pi }{\rightarrow }M$ be a Lie algebroid with Lie bracket $%
[,] $ and anchor map $\rho :E\rightarrow TM.$ Consider the prolongation $%
\mathcal{L}^{\pi }E$ $\ $of $E$ as a subset $\mathcal{L}^{\pi }E\subset
E\times TE$ defined by
\begin{equation*}
\mathcal{L}^{\pi }E\doteqdot \{(u,z)\in E\times TE/\rho (u)=(T\pi )(z))\}
\end{equation*}%
where $T\pi :TE\rightarrow TM$ is the tangent map to $\pi .$ In particular,
we have the prolongation \ of $E$ over the vector bundle projection $\pi
:E\rightarrow M.$ \ In the case when $E$ is the standard algebroid on $TM,$
then $\ \mathcal{L}^{\pi }E=T(TM).$

The space $\mathcal{L}^{\pi }E$ is fibred over $E$ by the projection $\
^{\pi }\pi :\mathcal{L}^{\pi }E\longrightarrow E,$ given by $\ \ ^{\pi }\pi
(u,z)=\tau _{E}(z)$ where $\tau _{E}:TE\longrightarrow E$ is the tangent
projection. \ It is also interesting to define the projection into the
second factor: $\ ^{\pi }\rho :\mathcal{L}^{\pi }E\longrightarrow TE,$ \
given by $\ ^{\pi }\rho (u,z)=z.$

We denote respectively the section $s\in Sec(E)$ and the sections of the
modules of vector fields $\ ^{v}s\in {\hbox {\fr X}}(E),\ ^{c}s\in {%
\hbox
{\fr X}}(E),$ and $\ ^{\mathbf{v}}s\in {\hbox {\fr X}}(\mathcal{L}^{\pi
}E),\ ^{\mathbf{c}}s\in {Sec}(\mathcal{L}^{\pi }E),$ and define respectively
the vertical and complete lifts of sections of $E$ into sections of $%
\mathcal{L}^{\pi }E.$ \ In particular
\begin{equation}
\ ^{\mathbf{c}}s(u)=\left( s(\pi (u)),\ ^{c}s(u)\right) \mbox{ and }\ ^{%
\mathbf{v}}s(u)=\left( 0,\ ^{v}s(u)\right) .  \label{rul1a}
\end{equation}%
There is an unique Lie algebroid structure $(\ ^{\pi }[\cdot ,\cdot ],\
^{\pi }\rho )$ on $\ \mathcal{L}^{\pi }E$ \ which can be defined by
\begin{equation*}
~\ ^{\pi }[\ ^{\mathbf{v}}s,\ ^{\mathbf{v}}s]=0,~\ ^{\pi }[\ ^{\mathbf{c}%
}s,\ ^{\mathbf{v}}s]=\ ^{\mathbf{v}}[s,s],\ ^{\pi }[\ ^{\mathbf{c}}s,\ ^{%
\mathbf{c}}s]=\ ^{\mathbf{c}}[s,s].
\end{equation*}%
For the vertical and complete lifts of functions we have:
\begin{eqnarray*}
\ ^{\pi }\rho (\ ^{\mathbf{c}}s)(\ ^{c}f) &=&\ ^{c}\left( \rho (s)(f)\right)
,~\ ^{\pi }\rho (\ ^{\mathbf{c}}s)(\ ^{v}f)=\ ^{v}\left( \rho (s)(f)\right) ,
\\
\ ^{\pi }\rho (\ ^{\mathbf{v}}s)(\ ^{c}f) &=&\ ^{v}\left( \rho (s)(f)\right)
,~\ ^{\pi }\rho (s^{\mathbf{v}})(\ ^{v}f)=0.
\end{eqnarray*}

Other two interesting geometric objects on $\mathcal{L}^{\pi }E$ are the
Liouville section $\ ^{\pi }\Delta $ and the vertical endomorphism $\ ^{\pi
}S.$ The object $\ ^{\pi }\Delta $ is just the section of $\ ^{\pi }\pi $
defined by $\ ^{\pi }\Delta (u)=(0,\ ^{v}u_{u})$, for all $u\in E$ and the
object $\ ^{\pi }S$ is the section of the vector bundle $\mathcal{L}^{\pi
}E\oplus (\mathcal{L}^{\pi }E)^{\ast }\longrightarrow E$ defined by
\begin{equation*}
\ ^{\pi }S(\ ^{\mathbf{v}}s)=0,\quad \ ^{\pi }S(\ ^{\mathbf{c}}s)=\ ^{%
\mathbf{v}}s,
\end{equation*}%
for all sections $s$ of $E.$ For $E=TM,$ one has $\ \ ^{\pi }S=s$ and $\
^{\pi }\Delta =\Delta .$We will say that a section $Sec$ of $\ ^{\pi }\pi $
is a second order differential equation (SODE) or a semispray on $E$ if $\
^{\pi }S(Sec)=\ ^{\pi }\Delta $.

Let us consider local coordinates $x^{i}$ on $M$ and $(x^{i},u^{a})$ on $%
\mathcal{E},\ $a local basis $v_{b}$ of sections of $E$ and the Lie
algebroid structure functions $\rho _{a}^{i}(x)$ and $C_{be}^{a}(x).$ We can
define a local basis for the considered vertical and complete lifts,%
\begin{equation}
\ ^{c}e_{a}=\rho _{a}^{i}\frac{\partial }{\partial x^{i}}-C_{ae}^{b}u^{e}%
\frac{\partial }{\partial u^{b}}\mbox{ and }\ ^{v}e_{a}=\frac{\partial }{%
\partial u^{a}}  \label{bas2a}
\end{equation}%
which transform any section $s$ $=s^{a}v_{a}$ of $E,$ respectively, into the
vector fields $\ ^{v}s$ and $\ ^{c}s,$ when
\begin{equation*}
\ ^{c}s=s^{a}\rho _{a}^{i}\frac{\partial }{\partial x^{i}}+\left( \rho
_{a}^{i}\frac{\partial s^{b}}{\partial x^{i}}-s^{d}C_{da}^{b}\right) u^{a}%
\frac{\partial }{\partial u^{b}}\mbox{ and }\ ^{v}s=s^{a}\frac{\partial }{%
\partial u^{a}}.
\end{equation*}%
These are local expressions, for a complete definition see Ref. \cite{dl1}.

Following the rule (\ref{rul1a}) and putting $z_{a^{\prime }}=\
^{c}e_{a^{\prime }}$ and $v_{a}=\ ^{v}e_{a},$ we may transform the local
frame (\ref{bas2a}) into a local basis $(z_{a^{\prime }},v_{a})$ of $\
\mathcal{L}^{\pi }E$ when for $s=s^{a}v_{a}\in Sec(E)$ we have%
\begin{equation*}
\ ^{\mathbf{c}}s=(\rho _{b}^{i}\frac{\partial s^{a}}{\partial x^{i}}%
u^{b})v_{a}+s^{a}z_{a}\mbox{ and }\ ^{\mathbf{v}}s=s^{a}v_{a}.
\end{equation*}%
Hereafter we shall use primed indices $a^{\prime },b^{\prime },...$ running
the same values as $a,b,...$ if one would be necessary to distinguish the
objects decomposed with respect to the bases of type $z_{a^{\prime }}$ from
those decomposed with respect to the bases of type $v_{a}.$ It is convenient
to introduce a new local basis $\widetilde{c}_{A}=(\widetilde{z}_{a},%
\widetilde{v}_{a})$ on sections of $\ \mathcal{L}^{\pi }E$ when
\begin{equation}
\widetilde{c}_{A}\doteqdot (\widetilde{z}_{a}=C_{ae}^{b}u^{e}v_{b}+z_{a},%
\widetilde{v}_{a}=v_{a})  \label{basis1a}
\end{equation}%
with the components satisfying the typical Lie algebroid structure relations
(\ref{anch}) and (\ref{liea}), respectively,
\begin{equation*}
~\ ^{\pi }\rho (\widetilde{v}_{a})=\frac{\partial }{\partial u^{a}},\ ^{\pi
}\rho (\widetilde{z}_{a})=\rho _{a}^{i}\frac{\partial }{\partial x^{i}}
\end{equation*}%
and
\begin{equation*}
~\ ^{\pi }\left[ \widetilde{v}_{a},\widetilde{v}_{a}\right] =0,~\ ^{\pi }%
\left[ \widetilde{z}_{a},\widetilde{v}_{a}\right] =0,\ ^{\pi }\left[
\widetilde{z}_{a},\widetilde{z}_{a}\right] =C_{ab}^{e}\widetilde{z}_{a}.
\end{equation*}%
With respect to the $\widetilde{c}_{A}=(\widetilde{z}_{a},\widetilde{v}%
_{a}), $ for an element $\omega =\gamma ^{a}\widetilde{z}_{a}+\zeta ^{a}%
\widetilde{v}_{a}\in \ \mathcal{L}^{\pi }E,$ we can define the natural local
coordinates $(x^{i},u^{a},\gamma ^{a},\zeta ^{a})$ on $\ \mathcal{L}^{\pi
}E, $ when the point $\omega $ $\in \ ^{\pi }\pi (\pi ^{-1}(x))$ [for a
vector bundle projection $\ ^{\pi }\pi :\ \mathcal{L}^{\pi }E\rightarrow E$
and $x\in M,$ and $(x^{i},u^{a})$ considered also as the coordinates of the
point $\ ^{\pi }\pi (\omega )\in \pi ^{-1}(x)]$ may be expressed in
coordinate form
\begin{equation*}
\omega =\gamma ^{a}\widetilde{z}_{a}(\ ^{\pi }\pi (\omega ))+\zeta ^{a}%
\widetilde{v}_{a}(\ ^{\pi }\pi (\omega )).
\end{equation*}%
We note that $\ ^{\pi }\pi (\omega )=\ ^{\pi }\pi (u,z)=\tau _{E}(z)$ which
for
\begin{equation*}
u=\gamma ^{a}e_{a}\mbox{ and }z=\gamma ^{a}\rho _{a}^{i}\frac{\partial }{%
\partial x^{i}}+\zeta ^{a}\frac{\partial }{\partial u^{a}}
\end{equation*}%
results in\ the coordinate expression $\ ^{\pi }\pi (x^{i},u^{a},\gamma
^{a},\zeta ^{a})=\tau _{E}(z)=(x^{i},u^{a}).$ In coordinate form, the anchor
map is defined
\begin{equation*}
\ ^{\pi }\rho (x^{i},u^{a},\gamma ^{a},\zeta ^{a})=(x^{i},u^{a},\rho
_{a}^{i}\gamma ^{a},\zeta ^{a}).
\end{equation*}

We can elaborate a differential form calculus by stating an abstract
differential operator $d^{\mathcal{L}}\equiv d^{\mathcal{L}^{\pi }E}$ acting
in the form
\begin{eqnarray}
d^{\mathcal{L}}f &=&\rho _{a}^{i}\frac{\partial f}{\partial x^{i}}\widetilde{%
z}^{a}+\frac{\partial f}{\partial u^{a}}\widetilde{v}^{a},  \label{form1b} \\
d^{\mathcal{L}}\widetilde{z}^{a} &=&-\frac{1}{2}C_{be}^{a}\widetilde{z}%
^{b}\wedge \widetilde{z}^{e},~d^{\mathcal{L}}\widetilde{v}^{a}=0,  \notag
\end{eqnarray}%
where the local basis $\widetilde{c}^{A}=(\widetilde{z}^{a},\widetilde{v}%
^{a})$ is the dual to $\widetilde{c}_{A}=(\widetilde{z}_{a},\widetilde{v}%
_{a}).$

\section{N--anholonomic Lie Algebroids}

In this section, we formulate an approach to the theory of Lie algebroids
provided with a general N--connection structure. We define and investigate
the main properties of the metric and nonlinear and connection structures
and compute their torsions and curvatures and related almost Hermitian
models of N--anholonomic manifolds.

\subsection{Lie Algebroids with N--connection structure}

For convenience, we consider the main concepts and formulas for the
N--connection geometry both on vector bundles and related Lie algebroids.

\subsubsection{Lie N--anholonomic algebroids}

We start with the definition of nonlinear connection for vector bundles:

\begin{definition}
A nonlinear connection (in brief, N--connection) $\mathbf{N}$ on a vector
bundle $\mathcal{E}$ \ is defined by using the exact sequence%
\begin{equation*}
0\rightarrow v\mathcal{E}\overset{i}{\rightarrow }T\mathcal{E}\rightarrow T%
\mathcal{E}/vE\rightarrow 0,
\end{equation*}%
and giving a morphism $\mathbf{N}:TE\rightarrow vE$ such that $\mathbf{N}%
\circ i$ is the identity in the vertical subbundle $vE$ (the kernel $\ker
\pi ^{\intercal }\doteqdot vE,$ for $\pi ^{\intercal }:TE\rightarrow TM)$
where $i:vE\rightarrow TE$ is the inclusion mapping.
\end{definition}

We remit the reader to Refs. \cite{ma1,ma2,gr,v0408121} for historical
remarks and discussions on E. Cartan and A. Kawaguchi first definitions of \
N--connections, in Finsler geometry, and further generalizations (by
Ehresmann, Barthel, Miron, Grifone and others) to various type of spaces.%
\footnote{%
The term ''non-linear'' connection has already used in algebroid theory (for
instance, in \cite{cr}) for certain connections related to sections of
algebroids and associated bundles: such approaches do not follow the general
theory of N--connections in bundle spaces and/or on nonholonomic manifolds.}

We can say equivalently that a N--connection is defined by a Whitney sum
\begin{equation}
T\mathcal{E}=h\mathcal{E}\oplus v\mathcal{E}  \label{withns}
\end{equation}%
globally splitting $T\mathcal{E}$ into conventional horizontal (h--)
subspace, $h\mathcal{E},$ and vertical (v--) subspace, $v\mathcal{E},$
(subbundles). In general, a decomposition of type (\ref{withns}) defines a
nonintegrable (or nonholonomic; in literature, one uses also the equivalent
term 'anholonomic') distribution. Such spaces are called 'N--anho\-lonomic'
because their nonholonomy is related to the N--connection structure: we
follow the conventions from \cite{v0408121,v0407495}.

\begin{definition}
A Lie algebroid $\mathbf{A}\doteqdot (\mathbf{E},\left[ \cdot ,\cdot \right]
,\mathbf{\rho })$ is N--anholonomic (in brief, it is a Lie N--algebroid) if
the vector bundle $\mathbf{E}$ is provided with N--connection structure.%
\footnote{%
We shall use 'boldfaced' symbols for algebroids, bundles and manifolds and,
in general, for the geometrical objects defined on such spaces if it would
be necessary to emphasize that they are provided with (or adapted to) a
N--connection structure.}
\end{definition}

A section $\mathbf{X}$ of $\mathbf{E}$ has the h- and v--decompositions
\begin{equation*}
\mathbf{X=X}^{\alpha }\mathbf{e}_{\alpha }\mathbf{=}(X\equiv \
^{-}X=X^{i}e_{i},\ ^{\star }X=X^{b}v_{b})
\end{equation*}%
and 1-section of $\mathbf{E}^{\ast }$, $\mathbf{\Phi }\in Sec(\mathbf{E}%
^{\ast })$ has the h- and v--decomposition%
\begin{equation*}
\mathbf{\Phi }=(\Phi \equiv \ ^{-}\Phi =\Phi _{i}e^{i},\ ^{\star }\Phi =\Phi
_{b}v^{b}),
\end{equation*}%
where the rule of summing on repeating indices is used. Following the
convention of \cite{ma1,ma2,vnp}, we call \ respectively such objects to be
distinguished vectors and forms (in brief, $d$--vectors and $d$--forms)
being adapted to the global decomposition induced by the N--connection. In a
similar manner, we can introduce $d$--tensor, $d$--connection, d--spinor ...
objects if such objects are adapted to the N--connection structure. We
emphasize that the v--components, those labelled with indices $a,b,c,...$
are just those which would be subjected to the algebroid structure
conditions given by a couple $(\left[ \cdot ,\cdot \right] ,\mathbf{\rho }).$
$\ $The constructions for N--anholonomic algebroids have to be adapted to
the N--connection structure.

\subsubsection{Geometric structures induced by N--connections}

A N--connection may be described by its coefficients,%
\begin{equation*}
\mathbf{N}=N_{\ \underline{i}}^{\underline{a}}(u)dx^{\underline{i}}\otimes
\frac{\partial }{\partial y^{\underline{a}}}=N_{\ i}^{b}(u)e^{i}\otimes
v_{b},
\end{equation*}%
where we underlined the indices defining the coefficients with respect to a
local coordinate basis. The well known class of linear connections consists
on a particular subclasses with the coefficients being linear on $u^{%
\underline{a}},$ i. e. $N_{\underline{i}}^{\underline{a}}(x,u)=\Gamma _{%
\underline{b}\underline{i}}^{\underline{a}}(x)u^{\underline{b}}.$

On any Lie algebroid and its associated vector bundle we can can consider
'vielbein' transforms, \ stated by nondegenerated matrices $A_{\beta }^{\
\underline{\alpha }}(\mathbf{u})$ and their inverse ones $A_{\ \underline{%
\beta }}^{\alpha }(\mathbf{u})$, from local coordinate frame $e_{\underline{%
\alpha }}=\partial _{\underline{\alpha }}=(e_{\underline{i}}=\partial _{%
\underline{i}},v_{\underline{b}}=\partial _{\underline{a}})$ and,
respectively, coordinate co--frames, $e^{\underline{\alpha }}=du^{\underline{%
\alpha }}=(dx^{\underline{i}},du^{\underline{a}}))$ to any general ones $%
e_{\alpha }=(e_{i},v_{a})$ and, respectively, $e^{\alpha }=(e^{i},v^{a}),$
\begin{equation}
e_{\alpha }=A_{\alpha }^{\ \underline{\alpha }}(\mathbf{u})e_{\underline{%
\alpha }}\mbox{ and }e^{\alpha }=A_{\ \underline{\beta }}^{\alpha }(\mathbf{u%
})e^{\underline{\alpha }},  \label{vilebtr}
\end{equation}%
where the point $\mathbf{u=}(x,u)\in \mathbf{E}$ has the coordinates $%
u^{\alpha }=(x^{i},u^{a}).$\footnote{%
the symbols $\partial $ and $d$ are used correspondingly for usual partial
derivatives and differentials} In general, such frame transforms are not
adapted to the Lie algebroid and/or N--connection structure. Nevertheless,
by straightforward computations we can prove:

\begin{proposition}
The Lie algebroid and N--connection structures prescribe a subclass of local
frames related via a subclass of matrix transforms $\mathbf{A}_{\alpha }^{\
\underline{\alpha }}$ and $\mathbf{A}_{\ \underline{\beta }}^{\alpha }$ from
(\ref{vilebtr}) linearly depending on $N_{\ i}^{a}(x,u)$ and parametrized in
the form generating N--adapted frames%
\begin{equation}
\mathbf{e}_{\alpha }=(\mathbf{e}_{i}=\frac{\partial }{\partial x^{i}}-N_{\
i}^{b}v_{b},v_{b})  \label{dder}
\end{equation}%
\ and dual coframes
\begin{equation}
\mathbf{e}^{\alpha }=(e^{i},\ \mathbf{v}^{b}=v^{b}+N_{\ i}^{b}dx^{i}),
\label{ddif}
\end{equation}%
for any $v_{b}=A_{b}^{\ \underline{b}}\partial _{\underline{b}}$ satisfying
the condition $v_{c}\mathbf{\rfloor }v^{b}=\delta _{c}^{b}.$
\end{proposition}

The the Lie algebroid structure can be adapted to the N--connection and
resulting frame structures (\ref{dder}) \ and (\ref{ddif}). This can be done
following the procedure: Let us re--define the coefficients of the anchor
and structure functions with respect to the $\mathbf{e}_{\alpha }$ and $%
\mathbf{e}^{\alpha },$ when
\begin{eqnarray*}
\rho _{\underline{b}}^{\underline{i}}(x) &\rightarrow &\ \widehat{\mathbf{%
\rho }}_{b}^{i}(x,u)=\mathbf{A}_{\ \underline{i}}^{i}(x,u)\ \mathbf{A}%
_{b}^{\ \underline{b}}(x,u)\rho _{\underline{b}}^{\underline{i}}(x), \\
C_{\underline{d}\underline{b}}^{\underline{f}}(x) &\rightarrow &\mathbf{C}%
_{db}^{f}(x,u)=\mathbf{A}_{\ \underline{f}}^{f}(x,u)\ \mathbf{A}_{d}^{\
\underline{d}}(x,u)\ \mathbf{A}_{b}^{\ \underline{b}}(x,u)C_{\underline{d}%
\underline{b}}^{\underline{f}}(x),
\end{eqnarray*}%
where the transform $\mathbf{A}$--matrices are linear on coefficients $%
N_{i}^{a}$ as can be obtained from the formulas in the above presented
Proposition. In terms of N--adapted anchor $\ \widehat{\rho }_{b}^{i}(x,u)$
and structure functions $\mathbf{C}_{db}^{f}(x,u)$ (which depend also on
variables $u^{a}$), the structure equations of the Lie algebroids (\ref{anch}%
),\ (\ref{liea}) and (\ref{lasa}) transform respectively into
\begin{eqnarray}
{}\widehat{\mathbf{\rho }}(\mathbf{e}_{b}) &=&{}\widehat{\mathbf{\rho }}%
_{b}^{i}(x,u)\ \mathbf{e}_{i},  \label{anch1d} \\
\lbrack v_{d},v_{b}] &=&\mathbf{C}_{db}^{f}(x,u)\ v_{f}  \label{lie1d}
\end{eqnarray}%
and
\begin{eqnarray}
{}\widehat{\mathbf{\rho }}_{a}^{j}e_{j}({}\widehat{\mathbf{\rho }}%
_{b}^{i})-{}\widehat{\mathbf{\rho }}_{b}^{j}e_{j}({}\widehat{\mathbf{\rho }}%
_{a}^{i}) &=&{}\widehat{\mathbf{\rho }}_{e}^{j}\mathbf{C}_{ab}^{e},
\label{lased} \\
\sum\limits_{cyclic(a,b,e)}\left( {}\widehat{\mathbf{\rho }}_{a}^{j}e_{j}(%
\mathbf{C}_{be}^{f})+\mathbf{C}_{ag}^{f}\mathbf{C}_{be}^{g}-\mathbf{C}%
_{b^{\prime }e^{\prime }}^{f^{\prime }}{}\widehat{\mathbf{\rho }}_{a}^{j}%
\mathbf{Q}_{f^{\prime }bej}^{fb^{\prime }e^{\prime }}\right) &=&0,  \notag
\end{eqnarray}%
for $\mathbf{Q}_{f^{\prime }bej}^{fb^{\prime }e^{\prime }}=\mathbf{e}_{\
\underline{b}}^{b^{\prime }}\mathbf{e}_{\ \underline{e}}^{e^{\prime }}%
\mathbf{e}_{f^{\prime }}^{\ \underline{f}}~e_{j}(\mathbf{e}_{b}^{\
\underline{b}}\mathbf{e}_{e}^{\ \underline{e}}\mathbf{e}_{\ \underline{f}%
}^{f})$ computed for the values $\mathbf{e}_{\ \underline{b}}^{b^{\prime }}$
and $\mathbf{e}_{f^{\prime }}^{\ \underline{f}}$ taken \ from
\begin{eqnarray}
\mathbf{e}_{\alpha }^{\ \underline{\alpha }}(x,u) &=&\left[
\begin{array}{cc}
e_{i}^{\ \underline{i}}(x,u) & N_{i}^{b}(x,u)e_{b}^{\ \underline{a}}(x,u) \\
0 & e_{a}^{\ \underline{a}}(x,u)%
\end{array}%
\right] ,  \label{vt1a} \\
\mathbf{e}_{\ \underline{\beta }}^{\beta }(x,u) &=&\left[
\begin{array}{cc}
e_{\ \underline{i}}^{i\ }(x,u) & -N_{k}^{b}(x,u)e_{\ \underline{i}}^{k\
}(x,u) \\
0 & e_{\ \underline{a}}^{a\ }(x,u)%
\end{array}%
\right] ,  \label{vt2a}
\end{eqnarray}%
where $\mathbf{e}_{\alpha }=\mathbf{e}_{\alpha }^{\ \underline{\alpha }%
}\partial _{\underline{\alpha }}\mbox{ and }\mathbf{e}_{\ }^{\beta }=\mathbf{%
e}_{\ \underline{\beta }}^{\beta }du^{\underline{\beta }}$ depending
linearly on $N_{i}^{b}(x,u).$

We note that the operators $\mathbf{e}_{\alpha }$ and $\mathbf{e}^{\alpha }$
(respectively defined by (\ref{dder}) \ and (\ref{ddif})) are
''N--elongated'' partial derivatives and differentials defining a N--adapted
differential calculus on N--anholonomic manifolds. In the limit $N_{\
i}^{a}\rightarrow 0$ obtain the usual Lie algebroid constructions for
holonomic manifolds and bundles.

Following J.\ Grifone \cite{gr}, we introduce the curvature of a
N--connection $\mathbf{\Omega }$ as the Nijenhuis tensor
\begin{equation*}
N_{v}(\mathbf{X,Y})\doteqdot \lbrack \ ^{\star }X,\ \ ^{\star }Y]+\ ^{\star
\star }[\mathbf{X,Y}]-\ ^{\star }[\ ^{\star }X\mathbf{,Y}]-\ ^{\star }[%
\mathbf{X,}\ \ ^{\star }Y]
\end{equation*}%
for any $\mathbf{X,Y\in }\mathcal{X}(\mathbf{E})$ associated to the vertical
projection ''$\star "$ defined by this N--connection:%
\begin{equation*}
\mathbf{\Omega \doteqdot -}N_{v}
\end{equation*}%
written in the Lie algebroid and N--adapted form%
\begin{equation*}
\mathbf{\Omega =}\frac{1}{2}\Omega _{ij}^{b}\ e^{i}\wedge e^{j}\otimes v_{b}
\end{equation*}%
with coefficients%
\begin{equation*}
\Omega _{ij}^{a}=e_{[j}N_{\ i]}^{a\,}=e_{j}N_{\ i}^{a\,}-e_{i}N_{\
j}^{a\,}+N_{\ i}^{b\,}v_{b}N_{\ j}^{a\,}-N_{\ j}^{b\,}v_{b}N_{\ i}^{a\,}.
\end{equation*}

The vielbeins (\ref{dder}) satisfy the nonholonomy (equivalently,
anholonomy) relations
\begin{equation*}
\left[ \mathbf{e}_{\alpha },\mathbf{e}_{\beta }\right] =W_{\alpha \beta
}^{\gamma }\mathbf{e}_{\gamma }
\end{equation*}%
with nontrivial anholonomy coefficients $W_{jk}^{a}=\Omega _{jk}^{a}(x,u),$ $%
W_{ie}^{b}=v_{e}N_{\ i}^{b}(x,u)$ and $W_{ae}^{b}=C_{ae}^{b}(x)$ reflecting
the fact that the Lie algebroid is N--anholonom\-ic.

J.\ Vilms \cite{vilms} showed that any N--connection in a vector bundle $%
\mathcal{E}$ provides a linear connection in the vertical subbundle $\ v%
\mathcal{E}.$ Such linear connections are called Berwald connections after
the name of the geometer who introduced them originally in the Finsler
geometry, see details and historical remarks in \cite{ma1,ma2,bcs, v0408121}%
. The construction also holds for the Lie algebroids:

\begin{definition}
The Berwald connection with local coefficients
\begin{equation*}
\overline{N}_{bi}^{a}\doteqdot v_{b}N_{\ i}^{a}(x,y)\text{\mbox{ and }}%
\overline{N}_{be}^{a}\doteqdot 0
\end{equation*}%
is associated to a N--connection $\mathbf{N}=\{N_{\ i}^{a}\}$ and defines a
covariant derivative $\overline{\mathbf{D}}$\ on sections \ in the vertical
vector subbundle $v\mathbf{E.}$
\end{definition}

By using local expressions
\begin{equation*}
\overline{\mathbf{D}}_{e_{i}}v_{b}=v_{b}(N_{\ i}^{c})\ v_{c}\text{%
\mbox{ and
}}\overline{\mathbf{D}}_{v_{c}}v_{b}=0,
\end{equation*}%
a d--vector $\mathbf{X=}X^{i}e_{i}+X^{b}v_{b}\in \mathbf{E}$ and section $\
^{\star }B=B^{b}v_{b}\in v\mathbf{E,}$ we prove by direct component
computations:

\begin{proposition}
The Berwald covariant derivative $\overline{\mathbf{D}}$ has the local
expression%
\begin{equation*}
\overline{\mathbf{D}}_{X}\left( \ ^{\star }B\right) \doteqdot \mathbf{X\cdot
}\overline{\mathbf{D}}=\left[ X^{i}\left( \partial
_{i}B^{a}-(v_{b}N_{i}^{a})\ B^{b}\right) +X^{b}v_{b}B^{a}\right] v_{a.}
\end{equation*}
\end{proposition}

By definition, the Berwald connection is different from the notion of $E$%
--connection in a vector bundle (see, for instance, \cite{cr,cm1,dl1,can})
which was introduced for sections in holonomic Lie algebroids and vector
bundle maps. Roughly speaking, the standard Lie algebroid constructions
consider vielbeins, connections, metrics, maps... on sections of vector
bundles related to the structure functions $\rho _{a}^{i}(x)$ and $%
C_{ab}^{f}(x)$ depending on base coordinates $x^{i}.$

\subsection{N--connections on prolongations of Lie algebroids}

The aim of this subsection is to elaborate a general N--connection formalism
on any$\ \mathcal{L}^{\pi }E$ provided with a nonintegable distribution of
type (\ref{wsum1a}).

Let us consider the projection $pr:\mathcal{L}^{\pi }E\rightarrow E,$ when $%
(u,z)\rightarrow pr(u,z)=u,$ and introduce the vertical subbundle
\begin{equation*}
v\ \mathcal{L}^{\pi }E=\{(u,z)\in \mathcal{L}^{\pi }E;\ ^{\pi }\pi (u,z)=0\}.
\end{equation*}%
In this case, one has the projection $pr_{\mid v\ \mathcal{L}^{\pi }E}:v\
\mathcal{L}^{\pi }E\rightarrow E$ in a vector subbundle of $pr:\mathcal{L}%
^{\pi }E\rightarrow E$ allowing to define an exac sequence of vector
bundles: \
\begin{equation}
\xymatrix{ 0\ar[r] & v{\mathcal L}^{\pi}E \ar[dr]_{pr} \ar[r]^{\ \ ^{\circ
}i}& {\mathcal L}^{\pi} E \ar[d]^{pr} \ar[r]^{p}& \displaystyle{{\mathcal
L}^{\pi} E} / {v{\mathcal L}^{\pi} E}\ar[dl]^{pr}\ar[r]&0\\ &&E&& }
\label{exseq2}
\end{equation}%
We denote by $h:\mathcal{L}^{\pi }E\rightarrow z\mathcal{L}^{\pi }E$ the
complementary projection corresponding to $h:\mathcal{L}^{\pi }E\rightarrow v%
\mathcal{L}^{\pi }E$ when there are satisfied the conditions $%
h^{2}=h,v^{2}=v,vh=hv=0$ and $h+v=Id.$

\begin{definition}
A nonlinear connection (N--connection) $\ ^{\circ }\mathbf{N}$ on a Lie
algebroid $\mathcal{L}^{\pi }E$ \ is defined by an exact sequence (\ref%
{exseq2}), i.e. by a morphism of subspaces $\ ^{\circ }\mathbf{N}:\mathcal{L}%
^{\pi }E\rightarrow v\ \mathcal{L}^{\pi }E$ such that interior product $\
^{\circ }\mathbf{N\bullet }\ ^{\circ }i$ results in identity in the vertical
subbundle $v\ \mathcal{L}^{\pi }E$ (the kernel $\ker \ ^{\circ }\pi
^{\intercal }\doteqdot v\ \mathcal{L}^{\pi }E,$ for $\ ^{\circ }\pi
^{\intercal }:\ \mathcal{L}^{\pi }E\rightarrow TE)$ where $\ ^{\circ }i:v%
\mathcal{L}^{\pi }E\rightarrow \mathcal{L}^{\pi }E$ is an inclusion mapping.%
\footnote{%
The label ''$\ ^{\circ }$'' points that the geometric objects are defined
just for algebroid configurations and not for some nonholomic vectotr
bundles or nonholonomic manifolds.}
\end{definition}

We can say, equivalently, that a N--connection is defined by a Whitney sum
of type (\ref{wsum1a}) defining a general nonholonomic structure,\footnote{%
It should be emphasized that we can similarly define a N--connection by
using the splitting $T\mathcal{L}^{\pi }E=h\ T\mathcal{L}^{\pi }E\oplus v\ T%
\mathcal{L}^{\pi }E,$ but this would be a higher order structure if $E=TM,$
when $\mathcal{L}^{\pi }TM=TTM.$}
\begin{equation*}
\ \mathcal{L}^{\pi }E=h\ \mathcal{L}^{\pi }E\oplus v\ \mathcal{L}^{\pi }E.
\end{equation*}

\begin{definition}
A (prolongated) Lie algebroid $\mathcal{L}^{\pi }\mathbf{E}\doteqdot (%
\mathbf{E},\ ^{\pi }[\cdot ,\cdot ],\ ^{\pi }\rho )$ is N--anholonomic (in
brief, is a prolongation Lie N--algebroid) if it is provided with a
N--connection structure $\ ^{\circ }\mathbf{N}.$
\end{definition}

A N--connection $\ ^{\circ }\mathbf{N}$ may be described by its coefficients,%
\begin{equation*}
\ \ ^{\circ }\mathbf{N}=\ N_{\ a^{\prime }}^{b}\widetilde{z}^{a^{\prime
}}\otimes \widetilde{v}_{b},
\end{equation*}%
where \ $\widetilde{c}_{A}=(\widetilde{z}_{b},\widetilde{v}_{a})$ are
defined by (\ref{basis1a}) playing the role of local coordinate base
(nevertheless, being in general nonholonomic because of nontrivial structure
functions $C_{bc}^{a}(x))$ on $\mathcal{L}^{\pi }E.$ There are also
nonholonomic operators (local bases)
\begin{equation}
\mathbf{c}_{A}=(\mathbf{z}_{a}=\widetilde{z}_{a}-\ N_{\ a}^{b}v_{b},\
\mathbf{v}_{a}=\widetilde{v}_{a})  \label{dderalg}
\end{equation}%
and
\begin{equation}
\mathbf{c}^{A}=(\mathbf{z}^{a}=\widetilde{z}^{a},\ \mathbf{v}^{a}=v^{a}+\
N_{\ b^{\prime }}^{a}z^{b^{\prime }})  \label{ddifalg}
\end{equation}%
which for a general $\ ^{\circ }\mathbf{N}$ define respectively
''N--elongated'' partial derivatives and differentials stating a N--adapted
differential calculus on prolongated Lie N--algebroids.

\begin{theorem}
Any N--connection $\ ^{\circ }\mathbf{N}$ on a prolongated Lie algebroid $%
\mathcal{L}^{\pi }\mathbf{E}$ induces a N--connection $\mathbf{N}$ on the
associated vector bundle $\mathcal{E}.$ The inverse statement also holds
true.
\end{theorem}

\begin{proof}
Let us consider a Lie algebroid N--connection%
\begin{equation*}
\ ^{\circ }\mathbf{N=}\ N_{\ b^{\prime }}^{a}\ \widetilde{z}^{b^{\prime
}}\otimes \widetilde{v}_{a}=N_{\ b}^{a}\ \rho _{a}^{i}\frac{\partial }{%
\partial x^{i}}\otimes du^{b},
\end{equation*}%
where we have used the formulas (\ref{dder1a}), (\ref{bas2a}) \ and (\ref%
{basis1a}). Identifying $N_{\ b^{\prime }}^{i}\doteqdot \ N_{\ b^{\prime
}}^{a}\ \rho _{a}^{i},$ we obtain a N--connection on $\mathcal{E}.$ In order
to proof the inverse statement, we have consider a N--connection on the
vector bundle and to 'lift' it to $\mathcal{L}^{\pi }\mathbf{E}$ by using
similar inverse formulas. $\square $
\end{proof}

The curvature $^{\circ }\mathbf{\Omega }$ of a N--connection $\ ^{\circ }%
\mathbf{N}$ is just the Nijenhuis tensor on $\mathcal{L}^{\pi }\mathbf{E,}$%
\begin{equation*}
\ ^{\circ }\mathbf{\Omega \doteqdot -}\ ^{\circ }N_{v}=\frac{1}{2}\ \Omega
_{a^{\prime }e^{\prime }}^{b}\ \widetilde{z}^{a^{\prime }}\wedge \widetilde{z%
}^{e^{\prime }}\otimes \widetilde{v}_{b}
\end{equation*}%
with the coefficients
\begin{equation}
\ \Omega _{b^{\prime }c^{\prime }}^{a}=\rho _{b^{\prime }}^{j}\frac{\partial
\ N_{\ c^{\prime }}^{a}}{\partial x^{j}}-\rho _{c^{\prime }}^{i}\frac{%
\partial \ N_{\ b^{\prime }}^{a}}{\partial x^{i}}+\ N_{\ b^{\prime }}^{e}%
\frac{\partial N_{\ c^{\prime }}^{a}}{\partial u^{e}}-\ N_{\ c^{\prime }}^{e}%
\frac{\partial \ N_{\ b^{\prime }}^{a}}{\partial u^{e}}.  \label{ncurvla}
\end{equation}%
computed as those from (\ref{anhrel1a}) but for a general (not derived from
a Lagrangian) N--connection. We omitted the label ''$\ ^{\circ }$'' in
formulas (\ref{ncurv1a}) because the algebroid character of geometric
objects is identified already by ''primed'' \ indices of type $a^{\prime
},b^{\prime }...$ (we shall use this rule for our further considerations in
order to omit dubbing of labels).

\begin{definition}
The Lie algebroid Berwald connection with local coefficients
\begin{equation*}
\ \overline{N}_{be^{\prime }}^{a}\doteqdot \widetilde{v}_{b}(N_{\ e^{\prime
}}^{a})\text{\mbox{ and
}}\overline{N}_{be^{\prime }}^{a}\doteqdot 0
\end{equation*}%
is associated to a N--connection $\ ^{\circ }\mathbf{N}=\{N_{\ b^{\prime
}}^{a}\}$ and defines a covariant derivative $\overline{\mathcal{D}}$\ on
sections \ in the vertical vector subbundle $v\mathcal{L}^{\pi }E\mathbf{.}$
\end{definition}

One holds the

\begin{proposition}
\label{bcdl}The Berwald covariant derivative $\overline{\mathcal{D}}$ \ on $%
\mathcal{L}^{\pi }E$ has the local expression%
\begin{equation*}
\overline{\mathcal{D}}_{X}\left( \ ^{\star }B\right) \doteqdot \ ^{\circ }%
\mathbf{X\cdot }\overline{\mathcal{D}}=\left[ X^{b^{\prime }}\left(
\widetilde{z}_{b^{\prime }}B^{a}-(\widetilde{v}_{c}\ N_{\ b^{\prime }}^{c})\
B^{a}\right) +\ ^{\star }X^{e}\widetilde{v}_{e}B^{a}\right] \widetilde{v}%
_{a.}.
\end{equation*}
\end{proposition}

\begin{proof}
It is evident if we component computations with
\begin{equation*}
\overline{\mathcal{D}}_{\widetilde{z}_{a^{\prime }}}\widetilde{v}_{b}=%
\widetilde{v}_{b}(N_{\ a^{\prime }}^{c})\ \widetilde{v}_{c}\text{%
\mbox{
and }}\overline{\mathcal{D}}_{v_{c}}\widetilde{v}_{b}=0,
\end{equation*}%
for a d--vector $\ ^{\circ }\mathbf{X=}X^{A}\widetilde{c}_{A}=X^{a^{\prime }}%
\widetilde{z}_{a^{\prime }}+X^{b}\widetilde{v}_{b}\in \mathcal{L}^{\pi }E$
and section $\ ^{\star }B=B^{b}v_{b}\in v\mathbf{E}$ mapped into $B^{b}%
\widetilde{v}_{b}\in v\mathcal{L}^{\pi }E\mathbf{.\square }$
\end{proof}

We emphasize that the N--connection formalism is a natural one for
investigating physical systems with mixed sets of holonomic--anholonomic
variables. The imposed anholonomic constraints (anisotropies) are
characterized by the coefficients of N--connection defining a global
splitting of the components of geometrical objects with respect to some
'horizontal' (holonomic) and 'vertical' (anisotropic) directions. In brief,
we shall use respectively the terms h- and/or v--components, h- and/or
v--indices, and h- and/or v--subspaces which on Lie algebroids are
correspondingly substituted into z-- and v--components

A N--connection structure on $\mathcal{L}^{\pi }\mathbf{E}$ defines the
algebra of tensorial distinguished \ (by the N--connection structure) fields
$dT\left( \mathcal{L}^{\pi }\mathbf{E}\right) $ (d--fields, d--tensors,
d--objects, if to follow the terminology from \cite{ma1,ma2,vnp}) on $\
\mathcal{L}^{\pi }\mathbf{E}$ introduced as the tensor algebra $\mathcal{T}%
=\{\mathcal{T}_{qs}^{pr}\}$ of the distinguished tangent bundle $\mathcal{V}%
_{(d)},$ $p_{d}:\ h\mathcal{L}^{\pi }\mathbf{E}\oplus v\mathcal{L}^{\pi }%
\mathbf{E}\rightarrow \mathcal{L}^{\pi }\mathbf{E}.$ An element $\mathbf{t}%
\in \mathcal{T}_{qs}^{pr},$ a d--tensor field of type $\left(
\begin{array}{cc}
p & r \\
q & s%
\end{array}%
\right) ,$ can be written in local form as%
\begin{equation*}
\mathbf{t}=t_{e_{1}...e_{q}b_{1}^{\prime }...b_{r}^{\prime }}^{c_{1}^{\prime
}...c_{p}^{\prime }a_{1}...a_{r}}\left( u\right) \mathbf{z}_{c_{1}^{\prime
}}\otimes ...\otimes \mathbf{z}_{c_{p}^{\prime }}\otimes \mathbf{v}%
_{a_{1}}\otimes ...\otimes \mathbf{v}_{a_{r}}\otimes \mathbf{v}%
^{e_{1}}\otimes ...\otimes \mathbf{v}^{e_{q}}\otimes \mathbf{z}%
^{b_{1}^{\prime }}...\otimes \mathbf{z}^{b_{r}^{\prime }}.
\end{equation*}

\subsection{Algebroid d--connection and d--metric structures}

\label{dlcms} The Lie algebroid d--objects are defined in a coordinate free
form as geometric objects adapted to the N--connection structure on a
prolongated Lie algebroid $\mathcal{L}^{\pi }\mathbf{E.}$ In coordinate
form, we can characterize such objects (linear connections, metrics or any
tensor field) by certain group and coordinate transforms adapted to the
global space splitting (\ref{wsum1a}) into z- and v--subspaces
(z--projections on $\mathcal{L}^{\pi }\mathbf{E}$ play the role of
h--projections on $\mathbf{E}$).

\subsubsection{d--connections}

\label{dcons}

We analyze the general properties of the class of linear connections which
are adapted to the N--connection structure on $\mathcal{L}^{\pi }\mathbf{E.}$

\begin{definition}
\label{defdcon}A d--connection $\mathcal{D}$ on $\mathcal{L}^{\pi }\mathbf{E}
$ is defined as a linear connection conserving under a parallelism the
global decomposition (\ref{wsum1a}).
\end{definition}

A N-connection induces decompositions of d--tensor indices into sums of
horizontal and vertical parts, for example, for every d--vector $\ ^{\circ }%
\mathbf{X}\in \mathcal{V}_{(d)}$ and \ its dual, i. e. 1-form, $\ _{\circ }%
\mathbf{X}$ $\ $we have respectively
\begin{equation*}
\ ^{\circ }\mathbf{X}=\ ^{\circ }X+\ ^{\star }X\ \mbox{and \quad
}\ _{\circ }\mathbf{X}=\ _{\circ }X+\ _{\star }X.
\end{equation*}%
For simplicity, we shall not use boldface symbols for d--vectors and
d--forms as well we shall omit the Lie algebroid label ''$\circ $'' if this
will not result in ambiguities. We can associate to every d--covariant
derivation $\mathcal{D}_{X}=\ ^{\circ }\mathbf{X}\rfloor \mathcal{D}$ two
new operators of z- and v--covariant derivations, $\mathcal{D}_{X}=\ ^{\circ
}D_{X}+\ ^{\star }D_{X},$ defined respectively
\begin{equation*}
\ ^{\circ }D_{X}\ ^{\circ }\mathbf{Y}=\mathcal{D}_{\ ^{\circ }X}\ ^{\circ }%
\mathbf{Y}\quad \mbox{ and \quad }\ ^{\star }D_{X}\ ^{\circ }\mathbf{Y}=%
\mathcal{D}_{\ ^{\star }X}\ ^{\circ }\mathbf{Y},
\end{equation*}%
for which the following conditions hold:%
\begin{eqnarray}
\mathcal{D}_{X}\ ^{\circ }\mathbf{Y} &=&\ ^{\circ }D_{X}\ ^{\circ }\mathbf{Y}%
+\ ^{\star }D_{X}\ ^{\circ }\mathbf{Y},  \label{hvder} \\
\ ^{\circ }D_{X}f &=&(\ ^{\circ }X\mathbf{)}f\mbox{ \quad and\quad
}\ ^{\star }D_{X}f=(\ ^{\star }X)f.  \notag
\end{eqnarray}%
for any $\ ^{\circ }\mathbf{X,\ ^{\circ }Y}\in \mathcal{L}^{\pi }\mathbf{E}$
and any function $f$ on $\mathbf{E}.$

The N--adapted components $\ \mathbf{\Gamma }_{BC}^{A}$ of a d-connection $%
\mathcal{D}_{A}=\mathbf{c}_{A}\rfloor \mathcal{D}$ are defined by the
equations%
\begin{equation*}
\mathcal{D}_{A}\mathbf{c}_{B}=\ \mathbf{\Gamma }_{\ AB}^{E}\mathbf{c}_{E},
\end{equation*}%
from which one immediately follows
\begin{equation}
\mathbf{\Gamma }_{\ AB}^{E}=\left( \mathcal{D}_{A}\mathbf{c}_{B}\right)
\rfloor \mathbf{c}^{E}.  \label{dcon1}
\end{equation}%
The operations of c- and v-covariant derivations, $\ ^{\circ
}D_{c}=(L_{bc}^{a},\ L_{b^{\prime }c\;}^{a^{\prime }})$ and $\ ^{\star
}D_{c^{\prime }}=(K_{bc^{\prime }}^{a},\ K_{b^{\prime }c^{\prime
}}^{a^{\prime }})$ (see (\ref{hvder})) are introduced as the corresponding
c- and v--parametrizations of (\ref{dcon1}),%
\begin{eqnarray}
L_{b^{\prime }e^{\prime }}^{a^{\prime }} &=&\left( \mathcal{D}_{e^{\prime }}%
\mathbf{z}_{b^{\prime }}\right) \rfloor \mathbf{z}^{a^{\prime }},\quad
L_{be^{\prime }}^{a}=\left( \mathcal{D}_{e^{\prime }}\mathbf{v}_{b}\right)
\rfloor \mathbf{v}^{a}  \label{hcov} \\
K_{b^{\prime }c}^{a^{\prime }} &=&\left( \mathcal{D}_{c}\mathbf{z}%
_{b^{\prime }}\right) \rfloor \mathbf{z}^{a^{\prime }},\quad
K_{bc}^{a}=\left( \mathcal{D}_{c}\mathbf{v}_{b}\right) \rfloor \mathbf{v}%
^{a}.  \label{vcov}
\end{eqnarray}%
A set of h--components \ (\ref{hcov}) and v--components (\ref{vcov}),
distinguished in the form $\mathbf{\Gamma }_{\ AB}^{E}=(L_{b^{\prime
}e^{\prime }}^{a^{\prime }},\ L_{be^{\prime }\;}^{a};K_{b^{\prime
}c}^{a^{\prime }},\ K_{bc}^{a}),$ completely defines the local action of a
d--connection $\mathcal{D}$ in $\mathcal{L}^{\pi }\mathbf{E}.$ For instance,
having taken a Lie algebroid d--tensor field of type $\left(
\begin{array}{cc}
1 & 1 \\
1 & 1%
\end{array}%
\right) ,$ $\mathbf{s}=s_{e^{\prime }f}^{b^{\prime }a}\mathbf{z}_{b^{\prime
}}\otimes \mathbf{v}_{a}\otimes \mathbf{z}^{e^{\prime }}\otimes \mathbf{v}%
^{f},$ and a d--vector $\ ^{\circ }\mathbf{X}=X^{a^{\prime }}\mathbf{z}%
_{a^{\prime }}+X^{b}\mathbf{v}_{b}$ we can write%
\begin{equation*}
\mathcal{D}_{X}\mathbf{s=\ }^{\circ }D_{X}\mathbf{s+\ }^{\star }D_{X}\mathbf{%
s=}\left( X^{d^{\prime }}s_{e^{\prime }f|d^{\prime }}^{b^{\prime
}a}+X^{c}s_{e^{\prime }f\perp c}^{b^{\prime }a}\right) \mathbf{z}_{b^{\prime
}}\otimes \mathbf{v}_{a}\otimes \mathbf{z}^{e^{\prime }}\otimes \mathbf{v}%
^{f},
\end{equation*}%
where the z--covariant derivative is
\begin{equation*}
s_{e^{\prime }f|d^{\prime }}^{b^{\prime }a}=\mathbf{z}_{d^{\prime
}}(s_{e^{\prime }f}^{b^{\prime }a})+L_{h^{\prime }d^{\prime }}^{b^{\prime
}}s_{eb}^{h^{\prime }a}+\ L_{cd^{\prime }}^{a}s_{e^{\prime }f}^{b^{\prime
}c}-L_{e^{\prime }d^{\prime }}^{h^{\prime }}s_{h^{\prime
}f}^{ba}-L_{fd^{\prime }}^{c}s_{e^{\prime }c}^{b^{\prime }a}
\end{equation*}%
and the v--covariant derivative is
\begin{equation*}
s_{e^{\prime }f\perp c}^{b^{\prime }a}=\mathbf{v}_{c}s_{e^{\prime
}f}^{b^{\prime }a}+K_{h^{\prime }c}^{b^{\prime }}s_{e^{\prime }f}^{h^{\prime
}a}+K_{dc}^{a}s_{e^{\prime }f}^{b^{\prime }d}-K_{e^{\prime }c}^{h^{\prime
}}s_{h^{\prime }f}^{b^{\prime }a}-\ K_{fc}^{d}s_{e^{\prime }f}^{b^{\prime
}a}.
\end{equation*}%
For a scalar function $f$ we have
\begin{eqnarray*}
\ ^{\circ }D_{a^{\prime }}f &=&\mathbf{z}_{a^{\prime }}f=\widetilde{z}%
_{a^{\prime }}f-\ N_{\ a^{\prime }}^{b}\widetilde{v}_{b}f=z_{a^{\prime
}}f+C_{a^{\prime }e}^{b}u^{e}v_{b}f-\ N_{\ a^{\prime }}^{b}v_{b}f \\
&=&\rho _{a^{\prime }}^{i}\frac{\partial f}{\partial x^{i}}-\ N_{\ a^{\prime
}}^{b}\frac{\partial f}{\partial u^{b}}, \\
\mathbf{\ }^{\star }D_{c}f &=&\mathbf{v}_{c}f=v_{a}f=\frac{\partial f}{%
\partial u^{c}},
\end{eqnarray*}%
where the action of the N--elongated operators $\mathbf{c}_{A}=(\mathbf{z}%
_{a},\ \mathbf{v}_{a})$ is stated consequently by the formulas (\ref{dderalg}%
), (\ref{dder1a}), (\ref{bas2a}) \ and (\ref{basis1a}). We note that such
formulas are written in abstract index form and specify for d--connections
the covariant derivation rule.

\subsubsection{Metric structures and d--metrics}

We consider arbitrary metric structures on a Lie algebroid $\mathcal{L}^{\pi
}\mathbf{E}$ and state the possibility to adapt them to N--connection
structures.

\begin{definition}
A metric $^{\circ }\mathbf{g}$ on a Lie algebroid $\mathcal{L}^{\pi }\mathbf{%
E}$ is defined as a symmetric covariant tensor field of type $\left(
0,2\right) ,$ $g_{AB,}$ being nondegenerate and of constant signature.%
\footnote{%
Metric structures are considered also in the case of transitive Courant
algebroids when the Lie type bracket is changed into a a more general
product with deformations induced by the metric structure \cite{vais3}. In
our case, we shall also obtain certain deformations of the Lie structure by
the N--connections which can be treated as some off--diagonal metric terms.}
\end{definition}

We write a N--connection $^{\circ }\mathbf{N=}\{\ N_{\ b^{\prime }}^{b}\}$
and a metric structure%
\begin{equation}
\ ^{\circ }\mathbf{g}=g_{AB}\widetilde{c}^{A}\otimes \widetilde{c}^{B}
\label{mstr}
\end{equation}%
on $\mathcal{L}^{\pi }\mathbf{E,}$ where we underline the indices
considering that the basis $\widetilde{c}^{A}$ being dual to (\ref{basis1a}%
). The introduced geometric objects are mutually compatible if there are
satisfied the conditions
\begin{equation}
\ ^{\circ }\mathbf{g}\left( \widetilde{z}_{a},\widetilde{v}_{b}\right) =0,%
\mbox{ or equivalently,
}\ ^{\circ }g_{ab^{\prime }}-\ N_{\ b^{\prime }}^{c}h_{ac}=0,  \label{comp1}
\end{equation}%
where $\ h_{ab}\doteqdot \ ^{\circ }\mathbf{g}\left( \widetilde{v}_{a},%
\widetilde{v}_{b}\right) $ and $\ ^{\circ }g_{b^{\prime }a}\doteqdot \
^{\circ }\mathbf{g}\left( \widetilde{z}_{b^{\prime }},\widetilde{v}%
_{a}\right) \,$ resulting in
\begin{equation}
\ N_{\ c^{\prime }}^{b}=\ h^{ab}\ \ ^{\circ }g_{c^{\prime }a}
\label{nconstr}
\end{equation}%
(the matrix $\ h^{ab}$ is inverse to $h_{ab};$ for simplicity, we have not
underlined \ the indices in the last formula). We obtain a
z--v--decomposition of metric (in brief, d--metric)%
\begin{equation}
\ ^{\circ }\mathbf{g}(\ ^{\circ }\mathbf{X,\ ^{\circ }Y})\mathbf{=\ }z\
^{\circ }\mathbf{g}(\ ^{\circ }\mathbf{X,\ ^{\circ }Y})+v\ ^{\circ }\mathbf{g%
}(\ ^{\circ }\mathbf{X,\ ^{\circ }Y}),  \label{block1}
\end{equation}%
where the d-tensor $z\ ^{\circ }\mathbf{g}(\ ^{\circ }\mathbf{X,\ ^{\circ }Y}%
)=\mathbf{g}(X,\ ^{\star }Y)$ is of type $\left(
\begin{array}{cc}
0 & 0 \\
2 & 0%
\end{array}%
\right) $ and the d-tensor$\ v\ ^{\circ }\mathbf{g}(\ ^{\circ }\mathbf{X,\
^{\circ }Y})\mathbf{=\ h}(\ ^{\star }X,\ ^{\star }Y)$ is of type $\left(
\begin{array}{cc}
0 & 0 \\
0 & 2%
\end{array}%
\right) .$ With respect to a N--coframe (\ref{ddif1a}), the d--metric (\ref%
{block1}) is written
\begin{equation}
\ ^{\circ }\mathbf{g}=\mathbf{g}_{AB}\mathbf{c}^{A}\otimes \mathbf{c}^{B}=\
g_{a^{\prime }b^{\prime }}\mathbf{z}^{a^{\prime }}\otimes \mathbf{z}%
^{b^{\prime }}+\ h_{ab}\mathbf{v}^{a}\otimes \mathbf{v}^{b},  \label{block2}
\end{equation}%
where $\ g_{a^{\prime }b^{\prime }}\doteqdot \ ^{\circ }\mathbf{g}\left(
\mathbf{z}_{a^{\prime }},\mathbf{z}_{b^{\prime }}\right) .$ The d--metric (%
\ref{block2}) can be equivalently written in ''off--diagonal'' form (\ref%
{mstr}) if the basis of dual vectors consists from the coordinate
differentials,
\begin{equation}
\ ^{\circ }g_{AB}=\left[
\begin{array}{cc}
\ g_{a^{\prime }b^{\prime }}+N_{\ a^{\prime }}^{a}N_{\ b^{\prime }}^{b}h_{ab}
& N_{\ e^{\prime }}^{e}h_{ae} \\
N_{\ a^{\prime }}^{e}h_{be} & h_{ab}%
\end{array}%
\right] .  \label{ansatz}
\end{equation}%
It is easy to check that one holds the relations%
\begin{equation*}
\ ^{\circ }\mathbf{g}_{AB}=\mathbf{e}_{A}^{\ \underline{A}}\mathbf{e}_{B}^{\
\underline{B}}\ ^{\circ }\underline{g}_{\underline{A}\underline{B}}
\end{equation*}%
or, inversely,
\begin{equation*}
\ ^{\circ }\underline{g}_{\underline{A}\underline{B}}=\mathbf{e}_{\
\underline{A}}^{A}\ \mathbf{e}_{\ \underline{B}}^{B}\ ^{\circ }\mathbf{g}%
_{AB}
\end{equation*}%
for respective vielbein transforms which prove that a N--connection
structure can be associated to a prescribed ansatz of vielbein transforms%
\begin{eqnarray}
A_{A}^{\ \underline{A}} &=&\mathbf{e}_{A}^{\ \underline{A}}=\left[
\begin{array}{cc}
e_{a}^{\ \underline{a}} & N_{\ a}^{b}e_{b}^{\ \underline{a}} \\
0 & e_{a}^{\ \underline{a}}%
\end{array}%
\right] ,  \label{vt1} \\
A_{\ \underline{B}}^{B} &=&\mathbf{e}_{\ \underline{B}}^{B}=\left[
\begin{array}{cc}
e_{\ \underline{a}}^{a\ } & -N_{\ a}^{b}\ e_{\ \underline{a}}^{a\ } \\
0 & e_{\ \underline{a}}^{a\ }%
\end{array}%
\right] ,  \label{vt2}
\end{eqnarray}%
in a particular case $e_{a}^{\ \underline{a}}=\delta _{a}^{\underline{a}}$
with $\delta _{a}^{\underline{a}}$ being the Kronecker symbol, defining a
global splitting of $\mathcal{L}^{\pi }\mathbf{E}$ into z-- and v--subspaces
with the N--vielbein structure%
\begin{equation*}
\mathbf{c}_{A}=\mathbf{e}_{A}^{\ \underline{A}}\mathbf{c}_{\underline{A}}%
\mbox{ and }\mathbf{c}_{\ }^{B}=\mathbf{e}_{\ \underline{B}}^{B}\mathbf{c}^{%
\underline{B}}.
\end{equation*}

A metric, for instance, parametrized in the form (\ref{ansatz})\ is generic
off--diagonal if it can not be diagonalized by any coordinate transforms. If
the anholonomy coefficients (\ref{anhrel1a}) vanish for a such
parametrization, we can define certain coordinate transforms to diagonalize
both the off--diagonal form (\ref{ansatz}) and the equivalent d--metric (\ref%
{block2}).

\begin{definition}
The nonmetricity d--field
\begin{equation*}
\ \mathcal{Q}=\mathbf{Q}_{AB}\mathbf{c}^{A}\otimes \mathbf{c}^{B}
\end{equation*}%
on a Lie algebroid $^{\pi }\mathcal{L}\mathbf{E}$ provided with a
N--connection structure is defined by a d--tensor field with the
coefficients
\begin{equation}
\mathbf{Q}_{AB}\doteqdot -\mathcal{D\ }^{\circ }\mathbf{g}_{AB}  \label{nmf}
\end{equation}%
where the covariant derivative $\mathcal{D}$ is for a d--connection $\mathbf{%
\Gamma }_{\ A}^{E}=\mathbf{\Gamma }_{\ AB}^{E}\mathbf{c}^{B},$ see (\ref%
{dcon1}) with the respective splitting $\mathbf{\Gamma }_{\
AB}^{E}=(L_{b^{\prime }e^{\prime }}^{a^{\prime }},\ L_{be^{\prime
}\;}^{a};K_{b^{\prime }c}^{a^{\prime }},\ K_{bc}^{a}),$ in order to be
adapted to the N--connection structure.
\end{definition}

A linear connection $D_{X}$ is compatible\textbf{\ } with a d--metric $%
\mathcal{\ }^{\circ }\mathbf{g}$ if%
\begin{equation}
D_{X}\mathcal{\ }^{\circ }\mathbf{g}=0,  \label{mc}
\end{equation}
i. e. if $Q_{AB}\equiv 0.$ In a space provided with N--connection structure,
the metricity condition (\ref{mc}) may split into a set of compatibility
conditions on v- and v-- subspaces. We should consider separately which of
the conditions
\begin{equation}
\mathcal{\ }^{\circ }D(\mathcal{\ }^{\circ }g\mathbf{)}=0,\mathcal{\ }%
^{\star }D(\mathcal{\ }^{\circ }g\mathbf{)}=0,\mathcal{\ }^{\circ }D(%
\mathcal{\ }^{\star }g\mathbf{)}=0,\mathcal{\ }^{\star }D(\mathcal{\ }%
^{\star }g\mathbf{)}=0  \label{mca}
\end{equation}%
are satisfied, or not, for a given d--connection $\mathcal{D}=(\mathbf{%
\Gamma }_{\ AB}^{E}).$

\begin{definition}
A prolongated N--anholonomic algebroid $\mathcal{L}^{\pi }\mathbf{E}$ is
metric--affine if it is provided with a nontrivial nonmetricity structue $%
\mathcal{Q}=\mathbf{Q}_{AB}\mathbf{c}^{A}\otimes \mathbf{c}^{B}.$
\end{definition}

By acting on forms with the covariant derivative $\mathcal{D},$ on a
metric--affine N--anholonomic algebroid, we can also define another very
important geometric objects (the 'gravitational field potentials' on Lie
algebroids):

\begin{equation}
\mbox{ torsion }\ \mathcal{T}^{A}\doteqdot \mathcal{D}\mathbf{c}^{A}=d%
\mathbf{c}^{A}+\Gamma _{\ B}^{A}\wedge \mathbf{c}^{B},  \label{dta}
\end{equation}%
and
\begin{equation}
\mbox{ curvature }\ \mathcal{R}_{\ B}^{A}\doteqdot \mathcal{D}\Gamma _{\
B}^{A}=d\Gamma _{\ B}^{A}-\Gamma _{\ B}^{E}\wedge \Gamma _{\ E}^{A}.
\label{dra}
\end{equation}

The Bianchi identities are%
\begin{equation}
\mathcal{D}\mathbf{Q}_{AB}\equiv \mathcal{R}_{AB}+\mathcal{R}_{BA},\ D%
\mathcal{T}^{A}\equiv \mathcal{R}_{E}^{\ A}\wedge \mathbf{c}^{E}\mbox{
and }\mathcal{DR}_{E}^{\ A}\equiv 0,  \label{bi}
\end{equation}%
where we stress the fact that $\mathbf{Q}_{AB},\mathcal{T}^{A}$ and $%
\mathcal{R}_{\ B}^{A}$ are called also the strength fields of a
metric--affine theory on a N--anholonomic algebroids (we use the terms
considered in \cite{vmaf}).

\subsection{Torsions and Curvatures on Lie N--Algebroids}

\label{torscurv}We define and calculate the components of torsion and
curvature of a general d--connection $\mathcal{D}$ on a Lie N--algebroid \ $%
\mathcal{L}^{\pi }\mathbf{E}$.

\subsubsection{d--torsions and N--connections}

We give a definition being equivalent to (\ref{dta}) but in d--operator form:

\begin{definition}
The torsion $\mathcal{T}$ \ of a d--connection $\mathcal{D}=\left( \ ^{\circ
}D,\ ^{\star }D\right) \mathbf{\ }$ on $\mathcal{L}^{\pi }\mathbf{E}$ is
defined as an operator (d--tensor field) adapted to the N--connection
structure
\begin{equation}
\mathcal{T}\left( \ ^{\circ }\mathbf{X,\ ^{\circ }Y}\right) =\mathcal{D}_{\
^{\circ }\mathbf{X}}\mathbf{\ ^{\circ }Y-}\mathcal{D}_{\ ^{\circ }\mathbf{Y}%
}\ ^{\circ }\mathbf{X\ -}\left[ \ ^{\circ }\mathbf{X,\ ^{\circ }Y}\right]
\mathbf{.}  \label{torsion}
\end{equation}
\end{definition}

One holds the following c- and v--decompositions%
\begin{equation}
\mathcal{T}\left( \ ^{\circ }\mathbf{X,\ ^{\circ }Y}\right) \mathbf{=}%
\mathcal{T}\left( \mathbf{\ ^{\circ }}X,\mathbf{\ ^{\circ }}Y\right) \mathbf{%
+}\mathcal{T}\left( \mathbf{\ ^{\circ }}X,\mathbf{\ ^{\star }}Y\right)
\mathbf{+}\mathcal{T}\left( \mathbf{\ ^{\star }}X,\mathbf{\ ^{\circ }}%
Y\right) \mathbf{+}\mathcal{T}\left( \mathbf{\ ^{\star }}X,\mathbf{\ ^{\star
}}Y\right) \mathbf{.}  \label{hvtorsion}
\end{equation}%
We consider the projections:
\begin{eqnarray*}
c\mathcal{T}\left( \ ^{\circ }\mathbf{X,\ ^{\circ }Y}\right) &\doteqdot &%
\mathbf{\ ^{\circ }}T\left( \ ^{\circ }\mathbf{X,\ ^{\circ }Y}\right) ,v%
\mathcal{T}\left( \mathbf{\ ^{\circ }}X,\mathbf{\ ^{\circ }}Y\right)
\doteqdot \mathbf{\ ^{\star }}T\left( \mathbf{\ ^{\circ }}X,\mathbf{\
^{\circ }}Y\right) \mathbf{,} \\
c\mathcal{T}\left( \mathbf{\ ^{\circ }}X,\mathbf{\ ^{\circ }}Y\right)
&\doteqdot &\mathbf{\ ^{\circ }}T\left( \mathbf{\ ^{\circ }}X,\mathbf{\
^{\circ }}Y\right) \mathbf{,...}
\end{eqnarray*}%
and say that, for instance, $\mathbf{\ \ ^{\circ }}T\left( \mathbf{\ ^{\circ
}}X,\mathbf{\ ^{\circ }}Y\right) $ is the z(zz)-torsion of $\mathcal{D},$
\newline
$\mathbf{\ ^{\star }}T\left( \mathbf{\ ^{\circ }}X,\mathbf{\ ^{\circ }}%
Y\right) \mathbf{\ }$ is the v(zz)-torsion of $\mathcal{D}$ and so on.

The torsion (\ref{torsion}) is locally determined by five d--tensor fields,
d--torsions (N--adapted z--v--decompositions with respect to $\mathbf{c}%
_{A}=(\mathbf{z}_{a^{\prime }},\mathbf{v}_{a})$ and $\mathbf{c}^{A}=(\mathbf{%
z}^{a^{\prime }},\mathbf{v}^{a}),$ when it is convenient to use primed
abstract indices for the c--components of local bases) defined
\begin{eqnarray}
T_{b^{\prime }c^{\prime }}^{a^{\prime }} &\doteqdot &\mathbf{\ ^{\circ }}%
T\left( \mathbf{z}_{b^{\prime }},\mathbf{z}_{a^{\prime }}\right) \rfloor
\mathbf{z}^{a^{\prime }},\quad T_{b^{\prime }c^{\prime }}^{a}=\mathbf{\
^{\star }}\left( \mathbf{z}_{b^{\prime }},\mathbf{z}_{a^{\prime }}\right)
\rfloor \mathbf{v}^{a},   \notag \\
 T_{b^{\prime }b}^{a^{\prime }}&=&\mathbf{\
^{\circ }}T\left( \mathbf{v}_{b},\mathbf{z}_{b^{\prime }}\right) \rfloor
\mathbf{z}^{a^{\prime }},
\quad T_{b^{\prime }b}^{a} =\mathbf{\ ^{\circ }}T\left( \mathbf{v}_{b},%
\mathbf{z}_{b^{\prime }}\right) \rfloor \mathbf{v}^{a},  \notag \\
\ T_{bc}^{a} &=&
\mathbf{\ ^{\star }}T\left( \mathbf{v}_{c},\mathbf{v}_{b}\right) \rfloor
\mathbf{v}^{a}.  \notag
\end{eqnarray}%
Using the formulas (\ref{dderalg}), (\ref{ddifalg}), and (\ref{ncurvla}), we
can calculate the c- and v--components of torsion (\ref{hvtorsion}) for a
d--connection, i. e. we  prove

\begin{theorem}
\label{tdtors}The torsion $\mathcal{T}_{.BE}^{A}=(T_{.b^{\prime }c^{\prime
}}^{a^{\prime }},T_{b^{\prime }a}^{a^{\prime }},T_{.a^{\prime }b^{\prime
}}^{a},T_{.bb^{\prime }}^{a},T_{.bc}^{a})$ of a given d--connec\-ti\-on $\ \mathbf{%
\Gamma }_{BE}^{A}=\left( L_{b^{\prime }e^{\prime }}^{a^{\prime
}},L_{be^{\prime }}^{a},K_{b^{\prime }c}^{a^{\prime }},K_{bc}^{a}\right) $(%
\ref{dcon1}) is defined by the corresponding z- and v--components (d--torsions)
\begin{eqnarray}
T_{.b^{\prime }e^{\prime }}^{a^{\prime }} &=&-T_{e^{\prime }b^{\prime
}}^{a^{\prime }}=L_{.b^{\prime }e^{\prime }}^{a^{\prime }}-L_{e^{\prime
}b^{\prime }}^{a^{\prime }},\quad T_{b^{\prime }a}^{a^{\prime
}}=-T_{ab^{\prime }}^{a^{\prime }}=K_{b^{\prime }a}^{a^{\prime }},\   \notag
\\
T_{.b^{\prime }a^{\prime }}^{a} &=&-T_{.a^{\prime }b^{\prime }}^{a}=\Omega
_{.b^{\prime }a^{\prime }}^{a},\
T_{.bc}^{a}=-T_{.cb}^{a}=K_{bc}^{a}-K_{cb}^{a},  \notag \\
T_{.bb^{\prime }}^{a} &=&-T_{.b^{\prime }b}^{a}=\frac{\partial \ N_{\
a^{\prime }}^{a}}{\partial u^{b}}-L_{.ba^{\prime }}^{a}.\   \label{dtorsb}
\end{eqnarray}
\end{theorem}

We note that for (pseudo) Riemannian structures on Lie N--algebroids the
d--torsions can be induced by the N--connection coefficients and reflect the
nonholonomic character of the the corresponding manifold provided with a
nonintegrable distribution. Such objects vanishes when we transfer our
considerations with respect to holonomic bases for a trivial N--connection
and zero ''vertical'' dimension.

\subsubsection{d--curvatures and N--connections}

In operator form, the curvature (\ref{dra}) is stated from the

\begin{definition}
The curvature $\mathcal{R}$ of a d--connection $\mathcal{D}=\left( \ ^{\circ
}D,\ ^{\star }D\right) \mathbf{\ }$ on $\mathcal{L}^{\pi }\mathbf{E}$ is
defined as an operator (d--tensor field) adapted to the N--connection
structure
\begin{equation}
\mathcal{R}\left( \ ^{\circ }\mathbf{X,\ ^{\circ }Y}\right) \mathbf{\
^{\circ }Z}=\left( \mathcal{D}_{\ ^{\circ }\mathbf{X}}\mathcal{D}_{\ ^{\circ
}\mathbf{Y}}\mathbf{\ }-\mathcal{D}_{\ ^{\circ }\mathbf{Y}}\mathcal{D}_{\
^{\circ }\mathbf{X}}-\mathcal{D}_{\left[ \ ^{\circ }\mathbf{X,\ ^{\circ }Y}%
\right] }\right) \ \mathbf{^{\circ }Z.}  \label{curvaturea}
\end{equation}
\end{definition}

One holds certain properties for the z- and v--decompositions of curvature, $%
\mathcal{R}()\mathcal{=}\left( \mathbf{\ ^{\circ }}R(),\mathbf{\ ^{\star }}%
R()\right) ,$ when
\begin{eqnarray*}
\mathbf{\ ^{\star }}R\left( \ ^{\circ }\mathbf{X,\ ^{\circ }Y}\right)
\mathbf{\ ^{\circ }}Z &=&0,\ \mathbf{\ ^{\circ }}R\left( \ ^{\circ }\mathbf{%
X,\ ^{\circ }Y}\right) \ ^{\circ }Z\mathbf{=}0,\  \\
\mathcal{R}\left( \ ^{\circ }\mathbf{X,\ ^{\circ }Y}\right) \mathbf{\
^{\circ }Z} &=&\mathbf{\ ^{\circ }}R\left( \ ^{\circ }\mathbf{X,\ ^{\circ }Y}%
\right) \ ^{\circ }Z\mathbf{+\ ^{\star }}R\left( \ ^{\circ }\mathbf{X,\
^{\circ }Y}\right) \ ^{\star }Z\mathbf{,}
\end{eqnarray*}%
where, for instance, $\mathbf{^{\circ }Z=(\ }^{\circ }Z,\ ^{\star }Z\mathbf{%
).}$ From (\ref{curvaturea}) and the equation
\begin{equation*}
\mathcal{R}\left( \ ^{\circ }\mathbf{X,\ ^{\circ }Y}\right) \mathbf{=-}%
\mathcal{R}\left( \ \mathbf{^{\circ }Y,\ }^{\circ }\mathbf{X}\right) ,
\end{equation*}%
we get that the curvature of a d-con\-necti\-on $\mathcal{D}$ in $\mathcal{L}%
^{\pi }\mathbf{E}$ is completely determined by the following six d--tensor
fields (d--curvatures):%
\begin{eqnarray}
R_{\ e^{\prime }b^{\prime }c^{\prime }}^{a^{\prime }} &=&\mathbf{z}%
^{a^{\prime }}\rfloor \mathcal{R}\left( \mathbf{z}_{c^{\prime }},\mathbf{z}%
_{b^{\prime }}\right) \mathbf{z}_{e^{\prime }},~R_{\ bb^{\prime }e^{\prime
}}^{a}=v^{a}\rfloor \mathcal{R}\left( \mathbf{z}_{e^{\prime }},\mathbf{z}%
_{b^{\prime }}\right) \mathbf{v}_{b},  \label{curvaturehv} \\
P_{\ b^{\prime }c^{\prime }c}^{a^{\prime }} &=&\mathbf{z}^{a^{\prime
}}\rfloor \mathcal{R}\left( \mathbf{v}_{c},\mathbf{z}_{c^{\prime }}\right)
\mathbf{z}_{b^{\prime }},~P_{\ bc^{\prime }c}^{a}=\mathbf{v}^{a}\rfloor
\mathcal{R}\left( \mathbf{v}_{c},\mathbf{z}_{c^{\prime }}\right) \mathbf{v}%
_{b},  \notag \\
S_{\ b^{\prime }bc}^{a^{\prime }} &=&\mathbf{z}^{a^{\prime }}\rfloor
\mathcal{R}\left( \mathbf{v}_{c},\mathbf{v}_{b}\right) \mathbf{z}_{b^{\prime
}},~S_{\ bcd}^{a}=\mathbf{v}^{a}\rfloor \mathcal{R}\left( \mathbf{v}_{d},%
\mathbf{v}_{c}\right) \mathbf{v}_{b}.  \notag
\end{eqnarray}%
By a direct computation, using (\ref{dderalg}), (\ref{ddifalg}), (\ref{hcov}%
), (\ref{vcov}) and (\ref{curvaturehv}), we prove

\begin{theorem}
The curvature
\begin{equation*}
\mathcal{R}_{.BEM}^{A}=(R_{\ e^{\prime }b^{\prime }c^{\prime }}^{a^{\prime
}},R_{\ bb^{\prime }e^{\prime }}^{a},P_{\ b^{\prime }c^{\prime
}c}^{a^{\prime }},P_{\ bc^{\prime }c}^{a},S_{\ b^{\prime }bc}^{a^{\prime
}},S_{\ bcd}^{a})
\end{equation*}%
of a d--con\-nec\-ti\-on $\mathbf{\Gamma }_{BC}^{A}$ (\ref{dcon1}) is
defined by the corresponding z- v--components (d--curvatures)
\begin{eqnarray}
R_{\ e^{\prime }b^{\prime }c^{\prime }}^{a^{\prime }} &=&\mathbf{z}%
_{c^{\prime }}(L_{.e^{\prime }b^{\prime }}^{a^{\prime }})-\mathbf{z}%
_{b^{\prime }}(L_{.e^{\prime }c^{\prime }}^{a^{\prime }})+L_{.e^{\prime
}b^{\prime }}^{d^{\prime }}L_{d^{\prime }c^{\prime }}^{a^{\prime
}}-L_{.e^{\prime }c^{\prime }}^{d^{\prime }}L_{d^{\prime }b^{\prime
}}^{a^{\prime }}-K_{.e^{\prime }a}^{a^{\prime }}\Omega _{.b^{\prime
}c^{\prime }}^{a},  \label{dcurv} \\
R_{\ bb^{\prime }e^{\prime }}^{a} &=&\mathbf{z}_{e^{\prime }}(L_{.bb^{\prime
}}^{a})-\mathbf{z}_{b^{\prime }}(L_{.be^{\prime }}^{a})+L_{.bb^{\prime
}}^{c}L_{.ce^{\prime }}^{a}-L_{.be^{\prime }}^{c}L_{.cb^{\prime
}}^{a}-K_{.bc}^{a}\ \Omega _{.b^{\prime }e^{\prime }}^{c},  \notag \\
P_{\ e^{\prime }b^{\prime }a}^{a^{\prime }} &=&\mathbf{v}_{a}(L_{.e^{\prime
}b^{\prime }}^{a^{\prime }})-\left( \mathbf{z}_{b^{\prime }}(K_{.e^{\prime
}a}^{a^{\prime }})+L_{.d^{\prime }b^{\prime }}^{a^{\prime }}K_{.e^{\prime
}a}^{d^{\prime }}-L_{.e^{\prime }b^{\prime }}^{d^{\prime }}K_{.d^{\prime
}a}^{a^{\prime }}-L_{.ab^{\prime }}^{c}K_{.e^{\prime }c}^{a^{\prime }}\right)
\notag \\
&&+K_{.e^{\prime }b}^{a^{\prime }}T_{.b^{\prime }a}^{b},  \notag \\
P_{\ ba^{\prime }a}^{c} &=&\mathbf{v}_{a}(L_{.ba^{\prime }}^{c})-\left(
\mathbf{z}_{a^{\prime }}(K_{.ba}^{c})+L_{.da^{\prime
}}^{c\,}K_{.ba}^{d}-L_{.ba^{\prime }}^{d}K_{.da}^{c}-L_{.aa^{\prime
}}^{d}K_{.bd}^{c}\right)  \notag \\
&&+K_{.bd}^{c}T_{.a^{\prime }a}^{d},  \notag \\
S_{\ b^{\prime }bc}^{a^{\prime }} &=&S_{\ jbc}^{i}=\mathbf{v}%
_{c}(K_{.b^{\prime }b}^{a^{\prime }})-\mathbf{v}_{b}(K_{.b^{\prime
}c}^{a^{\prime }})+K_{.b^{\prime }b}^{e^{\prime }}K_{.e^{\prime
}c}^{a^{\prime }}-K_{.b^{\prime }c}^{e^{\prime }}K_{e^{\prime }b}^{a^{\prime
}},  \notag \\
S_{\ bcd}^{a} &=&\mathbf{v}_{d}(K_{.bc}^{a})-\mathbf{v}%
_{c}(K_{.bd}^{a})+K_{.bc}^{e}K_{.ed}^{a}-K_{.bd}^{e}K_{.ec}^{a}.  \notag
\end{eqnarray}
\end{theorem}

The components of the Ricci d-tensor
\begin{equation*}
\mathbf{R}_{AB}\doteqdot \mathcal{R}_{\ ABE}^{E}
\end{equation*}%
with respect to a locally adapted frame (\ref{dderalg}) has four h-
v--components, $\mathbf{R}_{AB}=\left( R_{a^{\prime }b^{\prime
}},R_{a^{\prime }a},R_{aa^{\prime }},S_{ab}\right) ,$ where%
\begin{eqnarray}
R_{a^{\prime }b^{\prime }} &=&R_{\ a^{\prime }b^{\prime }c^{\prime
}}^{c^{\prime }},\quad R_{a^{\prime }a}=-\ ^{2}P_{a^{\prime }a}=-P_{\
a^{\prime }b^{\prime }a}^{b^{\prime }},  \label{dricci} \\
R_{aa^{\prime }} &=&\ ^{1}P_{aa^{\prime }}=P_{\ aa^{\prime }b}^{b},\quad
S_{ab}=S_{\ abc}^{c}.  \notag
\end{eqnarray}%
We point out that because, in general, $^{1}P_{aa^{\prime }}\neq
~^{2}P_{a^{\prime }a}$ the Ricci d--tensor is non symmetric.\footnote{%
We note that we consider such h- and v--splitting which are adapted to the
N--connection decomposition into subspaces as the Whitney sum:\ a
h--component can not be transformed into a v--term and inversely.}

Having defined a d--metric of type (\ref{block2}) in $\mathcal{L}^{\pi }%
\mathbf{E},$ we can introduce the scalar curvature ${\overleftarrow{\mathcal{%
R}}}$ of a d--connection $\mathcal{D}\mathbf{,}$
\begin{equation}
{\overleftarrow{\mathcal{R}}}=\ \mathbf{g}^{AB}\mathbf{R}_{AB}=\ ^{\circ
}R+\ ^{\star }S,  \label{dscal}
\end{equation}%
where $\ ^{\circ }R=g^{a^{\prime }b^{\prime }}R_{a^{\prime }b^{\prime }}$
and $\ ^{\star }S=h^{ab}S_{ab}$ and define the distinguished form of the
Einstein tensor (the Einstein d--tensor),
\begin{equation}
\mathbf{G}_{AB}\doteqdot \mathbf{R}_{AB}-\frac{1}{2}\ \mathbf{g}_{AB}{%
\overleftarrow{\mathcal{R}}.}  \label{deinst}
\end{equation}

The Ricci and Bianchi identities (\ref{bi}) of d--connections are formulated
in terms of z- and v--forms on vector bundle \cite{ma1,ma2}. In a similar
form, by using operators on $\mathcal{L}^{\pi }\mathbf{E,}$ we can formulate
them in component form on Lie N--algebroids (we omit in this work such
cumbersome formulas).

\subsection{Classes of d--connections}

In this section, we analyze a set of linear connections and associated
covariant derivations being important for modelling of mechanics and
gravitational field interactions on Lie N--algebroids provided with
anholonomic frame structure and generic off--diagonal metrics.

\subsubsection{The Levi--Civita connection and N--connections}

By definition, the Levi--Civita connection $\ ^{\circ }\nabla =\{\ ^{\nabla }%
\mathbf{\Gamma }_{\ BE}^{A}\}$ on $\mathcal{L}^{\pi }\mathbf{E,}$ with
coefficients
\begin{equation}
\ ^{\nabla }\mathbf{\Gamma }_{ABE}=\ ^{\circ }\mathbf{g}\left( \mathbf{c}%
_{A},\nabla _{E}\mathbf{c}_{B}\right) =\ \mathbf{g}_{AF}\ ^{\nabla }\mathbf{%
\Gamma }_{\ E}^{F},\,  \label{lccon1}
\end{equation}%
is torsionless,
\begin{equation*}
\ ^{\nabla }\mathcal{T}^{A}\doteqdot \ ^{\circ }\nabla \mathbf{c}^{A}=d%
\mathbf{c}^{A}+\ ^{\nabla }\mathbf{\Gamma }_{\ BE}^{A}\wedge \mathbf{c}%
^{B}=0,
\end{equation*}%
and metric compatible, $\ ^{\circ }\nabla (\ ^{\circ }\mathbf{g)}=0.$ The
formula (\ref{lccon1}) states that the operator $\ ^{\circ }\nabla $ can be
considerid on spaces provided with N--connection structure but this linear
connection is not adapted to the N--connection splitting \ (\ref{wsum1a}),
i. e. it is not a d--connection, see Definition \ref{defdcon} (so, we do not
use a 'boldfaced' symbol for the Levi--Civita connection). One holds

\begin{theorem}
If a Lie N--algebroid $\mathcal{L}^{\pi }\mathbf{E}$ is provided with both
N--connection $\ ^{\circ }\mathbf{N}$\ and d--metric $\ ^{\circ }\mathbf{g=}$
$\left\{ \mathbf{g}_{AB}\right\} $ structures, there is a unique linear
symmetric and torsionless connection $\ ^{\circ }\nabla \mathbf{=}\left\{
\nabla _{E}\right\} ,$ being metric compatible such that $\ \nabla _{E}\
\mathbf{g}_{AB}=0$ $\ $for $\mathbf{g}_{AB}=\left( g_{a^{\prime }b^{\prime
}},h_{ab}\right) ,$ see (\ref{block2}), with the coefficients
\begin{equation*}
\ ^{\nabla }\mathbf{\Gamma }_{ABE}=\ ^{\circ }\mathbf{g}\left( \mathbf{c}%
_{A},\nabla _{E}\mathbf{c}_{B}\right) =\mathbf{g}_{AD}\ ^{\nabla }\mathbf{%
\Gamma }_{\ BE}^{D},\,
\end{equation*}%
where
\begin{eqnarray}
\ ^{\nabla }\mathbf{\Gamma }_{ABE} &=&\frac{1}{2}[\mathbf{c}_{B}\mathbf{g}%
_{AE}+\mathbf{c}_{E}\mathbf{g}_{BA}-\mathbf{c}_{A}\mathbf{g}_{EB}
\label{lcsym} \\
&&+\mathbf{g}_{AK}\ W_{EB}^{K}+\mathbf{g}_{BK}\ W_{AE}^{K}-\mathbf{g}_{EK}\
W_{BA}^{K}]  \notag
\end{eqnarray}%
with respect to N--adapted frames $\mathbf{c}_{A}$ (\ref{dderalg}) and
N--coframes $\mathbf{c}_{\ }^{A}$ (\ref{ddifalg}), when the nonholonomy
coefficients $\ W_{AB}^{K}$ are defined by formulas (\ref{anhrel1a}).
\end{theorem}

\begin{proof}
It is a straightforward calculation in order to verify the metricity and
torsionless conditions with respect to nonholonomic frames, like it is
given, for instance, in \cite{stw,mtw} for the case of (pseudo)\ Riemannian
spaces. In our case the geometric objects and abstract indices are defined
for a prolongated Lie N--algebroid $\ \mathcal{L}^{\pi }\mathbf{E.}$ $%
\square $
\end{proof}

We can introduce the 1-form formalism and express
\begin{equation*}
\ ^{\nabla }\mathbf{\Gamma }_{\ EA}=\ ^{\nabla }\mathbf{\Gamma }_{EAB}%
\mathbf{c}^{B}
\end{equation*}%
where
\begin{equation}
\ ^{\nabla }\mathbf{\Gamma }_{\ EA}=\frac{1}{2}\left[ \mathbf{c}_{E}\rfloor
\ d\mathbf{c}_{A}-\mathbf{c}_{A}\rfloor \ d\mathbf{c}_{E}-\left( \mathbf{c}%
_{E}\rfloor \ \mathbf{c}_{A}\rfloor \ d\mathbf{c}_{B}\right) \wedge \mathbf{c%
}^{B}\right] ,  \label{christa}
\end{equation}%
contains z- v-components, $\ ^{\nabla }\mathbf{\Gamma }_{\ AB}^{E}=\left( \
^{\nabla }L_{\ b^{\prime }c^{\prime }}^{a^{\prime }},\ ^{\nabla }L_{\
bc^{\prime }}^{a},\ ^{\nabla }K_{\ b^{\prime }c}^{a^{\prime }},\ ^{\nabla
}K_{\ bc}^{a}\right) ,$ defined similarly to (\ref{hcov}) and (\ref{vcov})
but using the operator $\nabla ,$
\begin{eqnarray*}
\ ^{\nabla }L_{\ b^{\prime }c^{\prime }}^{a^{\prime }} &=&\left( \nabla
_{c^{\prime }}\mathbf{z}_{b^{\prime }}\right) \rfloor \mathbf{z}^{a^{\prime
}},\quad \ ^{\nabla }L_{\ ba^{\prime }}^{a}=\left( \nabla _{a^{\prime }}%
\mathbf{v}_{b}\right) \rfloor \mathbf{v}^{a}, \\
\ \ ^{\nabla }K_{\ b^{\prime }c}^{a^{\prime }} &=&\left( \nabla _{c}\mathbf{z%
}_{b^{\prime }}\right) \rfloor \mathbf{z}^{a^{\prime }},\quad \ ^{\nabla
}K_{\ bc}^{a}=\left( \nabla _{c}\mathbf{v}_{b}\right) \rfloor \mathbf{v}^{a}.
\end{eqnarray*}%
In explicit form, the components $L_{\bigtriangledown b^{\prime }c^{\prime
}}^{a^{\prime }},L_{\bigtriangledown ba^{\prime }}^{a},K_{\bigtriangledown
b^{\prime }c}^{a^{\prime }}$ and $K_{\bigtriangledown bc}^{a}$ are defined
by the formula (\ref{christa}), the N--adapted frame $\mathbf{c}_{A}\ $ and
coframe $\mathbf{c}^{B}$ and a d--metric $\ ^{\circ }\mathbf{g}=\left(
g_{a^{\prime }b^{\prime },}h_{ab}\right) .$

\subsubsection{The canonical d--connection and the Levi--Civita connection}

We search a d--connection which is similar to the Levi--Civita connection
satisfying the metricity conditions adapted to the N--connection and
possessing some flexibility on existing of nontrivial d--torsion components.

\begin{proposition}
There are metric d--connections $\mathcal{D}\mathbf{=}\left( \ ^{\circ }D,\
^{\star }D\right) $ in a Lie N--algebroid \ $\mathcal{L}^{\pi }\mathbf{E},$
see (\ref{hvder}), satisfying the metricity conditions if and only if
\begin{equation}
\ ^{\circ }D_{a^{\prime }}g_{b^{\prime }c^{\prime }}=0,\ \ ^{\star
}D_{a}g_{b^{\prime }c^{\prime }}=0,\ \ ^{\circ }D_{a^{\prime }}h_{ab}=0,\ \
^{\star }D_{a}h_{ab}=0.  \label{mcas}
\end{equation}
\end{proposition}

A proof consists from an explicit example:

\begin{definition}
The canonical d--connection $\widehat{\mathcal{D}}\mathbf{=}\left( \ ^{\circ
}\widehat{D},\ ^{\star }\widehat{D}\right) ,$ equivalently $\widehat{\mathbf{%
\Gamma }}_{\ A}^{E}=\widehat{\mathbf{\Gamma }}_{\ AB}^{E}\mathbf{c}^{B},$\
is defined by the h-- v--components
\begin{equation*}
\widehat{\mathbf{\Gamma }}_{\ AB}^{E}=\left( \widehat{L}_{b^{\prime
}c^{\prime }}^{a^{\prime }},\widehat{L}_{bc^{\prime }}^{a},\widehat{K}%
_{b^{\prime }c}^{a^{\prime }},\widehat{K}_{bc}^{a}\right) ,
\end{equation*}%
where%
\begin{eqnarray}
\widehat{L}_{b^{\prime }c^{\prime }}^{a^{\prime }} &=&\frac{1}{2}%
g^{a^{\prime }e^{\prime }}\left( \mathbf{z}_{c^{\prime }}g_{b^{\prime
}e^{\prime }}+\mathbf{z}_{b^{\prime }}g_{c^{\prime }e^{\prime }}-\mathbf{z}%
_{e^{\prime }}g_{b^{\prime }c^{\prime }}\right) ,  \label{candcon} \\
\widehat{L}_{bc^{\prime }}^{a} &=&\mathbf{v}_{b}\ \left( N_{\ c^{\prime
}}^{a}\right) +\frac{1}{2}h^{ac}\left( \mathbf{z}_{c^{\prime }}h_{bc}-%
\mathbf{v}_{b}\ \left( N_{\ c^{\prime }}^{d}\right) h_{dc}-\mathbf{v}_{c}\
\left( N_{\ c^{\prime }}^{d}\right) h_{db}\right) ,  \notag \\
\widehat{K}_{b^{\prime }c}^{a^{\prime }} &=&\frac{1}{2}g^{a^{\prime
}e^{\prime }}\mathbf{v}_{c}g_{e^{\prime }b^{\prime }},  \notag \\
\widehat{K}_{bc}^{a} &=&\frac{1}{2}h^{ad}\left( \mathbf{v}_{c}h_{bd}+\mathbf{%
v}_{b}h_{cd}-\mathbf{v}_{d}h_{bc}\right) .  \notag
\end{eqnarray}%
satisfying the torsionless conditions for the c--subspace and v--subspace,
respectively, $\widehat{T}_{b^{\prime }c^{\prime }}^{a^{\prime }}=0$ and $%
\widehat{T}_{bc}^{a}=0.$
\end{definition}

By straightforward calculations with (\ref{candcon}) we can verify that the
conditions (\ref{mcas}) are satisfied and that the d--torsions are really
subjected to the conditions $\widehat{T}_{b^{\prime }c^{\prime }}^{a^{\prime
}}=0$ and $\widehat{T}_{bc}^{a}=0$ (see section \ref{torscurv})). We
emphasize that the canonical d--torsion posses nonvanishing torsion
components,%
\begin{eqnarray*}
\widehat{T}_{.b^{\prime }c^{\prime }}^{a}&=&-\widehat{T}_{.c^{\prime
}b^{\prime }}^{a}=\Omega _{.b^{\prime }c^{\prime }}^{a},\ ~\widehat{T}%
_{a^{\prime }a}^{b^{\prime }}=-\widehat{T}_{aa^{\prime }}^{b^{\prime }}=%
\widehat{K}_{.a^{\prime }a}^{b^{\prime }},   \\
~\widehat{T}_{.bb^{\prime }}^{a} &=&-%
\widehat{T}_{.b^{\prime }b}^{a}=\mathbf{v}_{b}\ \left( N_{\ b^{\prime
}}^{a}\right) -\widehat{L}_{.bb^{\prime }}^{a}
\end{eqnarray*}%
which reflects a nontrivial anholonmic frame structure on Lie N--algebroids.

\begin{proposition}
\label{lccdc}The components of the Levi--Civita connection $\mathbf{\Gamma }%
_{\bigtriangledown BA}^{E}$ and the components of the canonical
d--connection \ $\widehat{\mathbf{\Gamma }}_{\ BA}^{E}$\ are related by
formulas%
\begin{equation}
\mathbf{\Gamma }_{\bigtriangledown BA}^{E}=\left( \widehat{L}_{b^{\prime
}c^{\prime }}^{a^{\prime }},\widehat{L}_{bb^{\prime }}^{a}-\mathbf{v}_{b}\
\left( N_{\ b^{\prime }}^{a}\right) ,\widehat{K}_{b^{\prime }c}^{a^{\prime
}}+\frac{1}{2}g^{a^{\prime }e^{\prime }}\Omega _{b^{\prime }e^{\prime
}}^{a}h_{ca},\widehat{K}_{bc}^{a}\right) ,  \label{lcsyma}
\end{equation}%
where $\Omega _{a^{\prime }b^{\prime }}^{a}$\ \ is the N--connection
curvature\ (\ref{cncla}).
\end{proposition}

The proof follows from an explicit \ calculus with respect to the N--adapted
frame (\ref{dderalg}) and N--adapted coframe (\ref{ddifalg}) in (\ref{lcsym}%
) (equivalently, in (\ref{christa})) and re--groupation of the components as
to distinguish the z- and v--components (\ref{candcon}) for $\mathbf{g}%
_{AB}=\left( g_{a^{\prime }a^{\prime }},h_{ab}\right) .$

\subsubsection{The set of metric d--connections}

Let us define the set of all possible metric d--connections, satisfying the
conditions (\ref{mcas}) and being constructed only form $g_{a^{\prime
}b^{\prime }},h_{ab}$ and $\ N_{\ a^{\prime }}^{a}$ and their partial
derivatives. Such d--connections satisfy the conditions$\ $for d--torsions
that$\ T_{~b^{\prime }c^{\prime }}^{a^{\prime }}=0$ and $T_{~bc}^{a}=0$ and
can be generated by two procedures of deformation of the connection
\begin{eqnarray*}
\widehat{\mathbf{\Gamma }}_{\ AB}^{E} &\rightarrow &\ ^{[K]}\mathbf{\Gamma }%
_{\ EAB}^{E}=\mathbf{\Gamma }_{\ AB}^{E}+\ \mathbf{Z}_{\ AB}^{E}, \\
\mbox{ or } &\rightarrow &^{[M]}\mathbf{\Gamma }_{\ \alpha \beta }^{\gamma }=%
\widehat{\mathbf{\Gamma }}_{\ \alpha \beta }^{\gamma }+\ ^{[M]}\mathbf{Z}_{\
\alpha \beta }^{\gamma }.
\end{eqnarray*}

\begin{theorem}
\ \label{kmp}Every deformation d--tensor (equivalently, distorsion, or
deflection) \
\begin{eqnarray*}
\mathbf{Z}_{\ AB}^{E} &=&\{\ Z_{\ b^{\prime }c^{\prime }}^{a^{\prime }}=%
\frac{1}{2}g^{a^{\prime }e^{\prime }}\ ^{\circ }D_{b^{\prime }}g_{e^{\prime
}c^{\prime }},\ Z_{\ bb^{\prime }}^{a}=\frac{1}{2}h^{ac}\ ^{\circ
}D_{b^{\prime }}h_{cb},\  \\
&&\ Z_{\ a^{\prime }a}^{b^{\prime }}=\frac{1}{2}g^{b^{\prime }e^{\prime }}\
^{\star }D_{a}g_{e^{\prime }a^{\prime }},\ Z_{\ bc}^{a}=\frac{1}{2}h^{ad}\
^{\star }D_{c}h_{db}\}
\end{eqnarray*}%
\ transforms a d--connection $\mathbf{\Gamma }_{\ AB}^{E}=\left(
L_{b^{\prime }c^{\prime }}^{a^{\prime }},L_{bc^{\prime }}^{a},K_{b^{\prime
}c}^{a^{\prime }},K_{bc}^{a}\right) $\ (\ref{dcon1}) into a metric
d--connection%
\begin{equation*}
\ ^{[K]}\mathbf{\Gamma }_{\ AB}^{E}=\left( L_{b^{\prime }c^{\prime
}}^{a^{\prime }}+\ Z_{b^{\prime }c^{\prime }}^{a^{\prime }},L_{bc^{\prime
}}^{a}+\ Z_{\ bc^{\prime }}^{a},K_{b^{\prime }c}^{a^{\prime }}+\ Z_{\
b^{\prime }a}^{a^{\prime }},C_{bc}^{a}+\ Z_{\ bc}^{a}\right) .
\end{equation*}
\end{theorem}

\begin{proof}
The proof consists from a straightforward verification that the conditions (%
\ref{mcas}) are satisfied for $\ ^{[K]}\mathcal{D}\mathbf{=\{^{[K]}\mathbf{%
\Gamma }_{AB}^{E}\}}$ and $\mathbf{g}_{AB}=\left( g_{a^{\prime }b^{\prime
}},h_{ab}\right) .$ We note that this metrization procedure contains
additional covariant derivations of the d--metric coefficients, defined by
arbitrary d--connection, not only N--adapted derivatives of the d--metric
and N--connection coefficients as in the case of the canonical d--connection.%
$\square $
\end{proof}

\begin{theorem}
\label{mconnections}For a fixed d--metric structure \ (\ref{block2}),\ $%
\mathbf{g}_{AB}=\left( g_{a^{\prime }b^{\prime }},h_{ab}\right) ,$ on a Lie
N--algebroid $\mathcal{L}^{\pi }\mathbf{E},$ the set of metric
d--connections \ \
\begin{equation*}
^{\lbrack M]}\mathbf{\Gamma }_{\ AB}^{E}=\widehat{\mathbf{\Gamma }}_{\
AB}^{E}+\ ^{[M]}\mathbf{Z}_{\ AB}^{E}\ \ \
\end{equation*}%
is defined by the deformation d--tensor \
\begin{eqnarray*}
^{\lbrack M]}\mathbf{Z}_{AB}^{E} &=&\{\ ^{[M]}Z_{\ b^{\prime }c^{\prime
}}^{a^{\prime }}=\ ^{[-]}O_{c^{\prime }m^{\prime }}^{l^{\prime }a^{\prime
}}Y_{l^{\prime }b^{\prime }}^{m^{\prime }},\ ^{[M]}Z_{\ bc^{\prime }}^{a}=\
^{[-]}O_{bd}^{ea}Y_{ec^{\prime }}^{m},\  \\
&&\ ^{[M]}Z_{\ b^{\prime }a}^{a^{\prime }}=\ ^{[+]}O_{b^{\prime }k^{\prime
}}^{m^{\prime }a^{\prime }}Y_{m^{\prime }a}^{k^{\prime }},\ ^{[M]}Z_{\
bc}^{a}=\ ^{[+]}O_{bd}^{ea}Y_{ec}^{d}\}
\end{eqnarray*}%
where the so--called Obata operators are
\begin{equation*}
\ ^{[\pm ]}O_{c^{\prime }m^{\prime }}^{l^{\prime }a^{\prime }}=\frac{1}{2}%
\left( \delta _{c^{\prime }}^{l^{\prime }}\delta _{m^{\prime }}^{a^{\prime
}}\pm g_{c^{\prime }m^{\prime }}g^{l^{\prime }a^{\prime }}\right)
\mbox{ and
}\ ^{[\pm ]}O_{bd}^{ea}=\frac{1}{2}\left( \delta _{b}^{e}\delta _{d}^{a}\pm
h_{bd}h^{ea}\right)
\end{equation*}%
and \ $Y_{l^{\prime }b^{\prime }}^{m^{\prime }},$\ $Y_{ec^{\prime
}}^{m},Y_{m^{\prime }c}^{k^{\prime }},$\ $Y_{ec}^{d}$ are arbitrary
d--tensor fields.
\end{theorem}

\begin{proof}
The proof consists from a direct verification of the fact that the
conditions (\ref{mcas}) are satisfied on $\mathcal{L}^{\pi }\mathbf{E}$ for $%
\ ^{[M]}\mathcal{D}\mathbf{=\{^{[M]}\mathbf{\Gamma }_{\ AB}^{E}\}.}$ We note
that the relation (\ref{lcsyma}) \ between the Levi--Civita and the
canonical d--connection is a particular case of $^{[M]}\mathbf{Z}_{AB}^{E},$
when $Y_{l^{\prime }b^{\prime }}^{m^{\prime }},$\ $Y_{ec^{\prime }}^{m}$ and
$Y_{ec}^{d}$ are zero, but $Y_{m^{\prime }c}^{k^{\prime }}$ is taken to have
$\ ^{[+]}O_{a^{\prime }b^{\prime }}^{m^{\prime }c^{\prime }}Y_{m^{\prime
}c}^{b^{\prime }}=\frac{1}{2}g^{c^{\prime }b^{\prime }}\Omega _{a^{\prime
}b^{\prime }}^{a}h_{ca}.\square $
\end{proof}

\section{N--Connections, Geometric Mechanics and Lie Algebroids}

The general idea on geometrization of mechanics on the tangent /cotangent
bundle and/or on Lie algebroids is that a regular (for simplicity)
Lagrangian, or Hamiltonian, define the fundamental geometric objects of the
model. There were elaboratede two general approaches: In the first one, the\
basic geometric constructions are derived from the so--called Poincar\'{e}%
--Cartan 1-form, the Poincar\'{e}--Cartan 2-form and the energy function
(see \cite{AM,LR3}) permitting us to geometrize the Euler--Lagrange
equations in terms of the (pre-) sympletic geometry. In the second approach,
there are emphasized the (semi) spray configuration and associated
N--connection, canonical metric and linear connection, almost
complex/sympletic ... structures, all adapted to a N--connection (see \cite%
{mhss}). In this Section, we state the main results and outline the proofs
for Lie algebroid constructions in the second approach to geometrization of
mechanics and classical field theory. We refer to \cite{weins1,cm1,dl1,ml}
and \cite{ma1,ma2,v0408121} for respective details and proofs on algebroid
geometrization of the Euler--Lagrange equations and N--connections on the
tangent bundle, i.e. to details on the first approach.

\subsection{Lie algebroids, vector bundles, and the Lagrange formalism}

Let $\widetilde{TM}\doteqdot TM\backslash \{0\},$ \ $\dim M=n,$ where $\{0\}$
means the null section of the tangent bundle $\tau _{M}:TM\rightarrow M,$
and $L:TM\rightarrow \R$ be a Lagrangian function. Fixing the local
coordinates $(x^{i},y^{i}),$ the elements of the Hessian are defined%
\begin{equation}
\ ^{L}g_{ij}\doteqdot \frac{1}{2}\frac{\partial ^{2}L}{\partial
y^{i}\partial y^{j}},  \label{lmf}
\end{equation}%
when, for simplicity, we consider that the regularity condition is
satisfied, i. e. $rank\left( \ ^{L}g_{ij}\right) =n.$ We shall also use the
matrix $(\ ^{L}g^{ij})$ inverse to $(\ ^{L}g_{ij})$.

\begin{definition}
A Lagrange space is a pair $L^{n}=(M,L)$ defined by a regular Lagrangian $%
L(x^{i},y^{k})$ for which $\ ^{L}g_{ik}$ is of constant signature on $%
\widetilde{TM}.$
\end{definition}

We can elaborate a similar construction on a Lie algebroid $\mathcal{A}$
with $\pi :E\rightarrow M$ and $\pi _{\pi }:\ \mathcal{L}^{\pi }E\rightarrow
E$ for a Lagrangian $\ l:E\rightarrow \R$ which defines a Lagrange
fundamental function $\ l(x,u)$ on $\widetilde{E}\doteqdot E/\{0\}_{E},$
where $\{0\}_{E}$ is the null section of the vector bundle $\pi
:E\rightarrow M,$ with%
\begin{equation}
\ ^{l}g_{ab}\doteqdot \frac{1}{2}\frac{\partial ^{2}\ l}{\partial
u^{a}\partial u^{b}},  \label{lamf}
\end{equation}%
being nondegenerate, i. e. $rank\left( \ ^{l}g_{ab}\right) =m,$ where $m$ is
the dimension of the typical fiber of $E,$ and of constant signature.

\begin{definition}
\label{defls}A Lagrange algebroid is a pair $L\mathcal{A}=(E,\ l)$ defined
by a regular Lagrangian $\ l(x^{i},u^{a})$ for which $\ ^{l}g_{ab}$ is of
constant signature on $\widetilde{E}.$
\end{definition}

Let us define the basic geometric objects necessary for a geometrization of
the Euler--Lagrange equations in the usual context (see, for instance, \cite%
{ml,LR3}):

\begin{enumerate}
\item The Poincar\'e--Cartan 1--form
\begin{equation*}
\theta _{L}\doteqdot S^*(dL)=\hat{p}_{i}dx^{i},
\end{equation*}%
where $S$ is the vertical endomorphism on $TM$, and the generalized momenta
is
\begin{equation*}
\hat{p}_{i}=\frac{\partial L}{\partial y^{i}}.
\end{equation*}

\item The Poincar\'e--Cartan 2-form $\omega _{L}\doteqdot -d\theta _{L},$%
\begin{equation*}
\omega _{L}=2\ ^{L}g_{ij}dy^{i}\wedge dx^{j}+\frac{\partial ^{2}L}{\partial
y^{i}\partial x^{j}}dx^{j}\wedge dx^{i}=dx^{i}\wedge d\hat{p}_{i}.
\end{equation*}

\item The energy function
\begin{equation*}
E_{L}\doteqdot C_{M}L-L=y^{i}\hat{p}_{i}-L,
\end{equation*}%
where $\Delta=y^{i}\partial /\partial y^{i}$ is the Liouville vector field
on $TM.$
\end{enumerate}

A vector field $\xi $ on $TM\;$ is said to be a second order differential
equation (SODE, or a semispray) if $S\xi =\Delta .$ This allows us to
express
\begin{equation*}
\xi =y^{i}\frac{\partial }{\partial x^{i}}+\xi ^{i}(x,y)\frac{\partial }{%
\partial y^{i}}.
\end{equation*}%
A curve $\gamma :\R\rightarrow M,$ parametrized $\gamma (t)=\{x^{i}(t)\},$
with the canonical extension to $TM,$ $\dot{\gamma}(t)\doteqdot d\gamma
/dt=\{x^{i}(t),y^{i}(t)\},$ is a solution of the SODE\ $\xi $ if and only if
it is satisfied the equation%
\begin{equation*}
\frac{d^{2}x^{i}}{dt^{2}}=\xi ^{i}(t,x^{i},\frac{dx^{i}}{dt})
\end{equation*}%
for $y^{i}=dx^{i}/dt,$ that is, if $\dot{\gamma}$ is an integral curve of $%
\xi .$

The introduced geometric objects can be redefined for a Lie algebroid $%
\mathcal{L}^{\pi }E$ and its dual $\mathcal{L}^{\pi }E^{\ast }$ (see \cite%
{mart,dl1} for an intrinsic definition).

\begin{enumerate}
\item The Poincar\'{e}--Cartan 1--section
\begin{equation}
\theta _{l}\doteqdot \ ^{\pi }S^{\ast }(d^{E}l)=\ ^{l}\hat{p}_{a}\mathbf{c}%
^{a}\in Sec\left( (\mathcal{L}^{\pi }E)^{\ast }\right) ,  \label{pc1fa}
\end{equation}%
where the general momenta is
\begin{equation*}
\ ^{l}\hat{p}_{a}\doteqdot \frac{\partial l}{\partial u^{a}}.
\end{equation*}

\item The Poincar\'{e}--Cartan 2-section $\omega _{l}\doteqdot -d^{\mathcal{L%
}}\theta _{l},$%
\begin{equation}
\omega _{l}=2\ ^{l}g_{ab}\mathbf{z}^{a}\wedge \mathbf{v}^{b}+\left( \frac{1}{%
2}\ ^{l}\hat{p}_{e}C_{ab}^{e}-\rho _{a}^{j}\frac{\partial ^{2}l}{\partial
u^{b}\partial x^{j}}\right) \mathbf{z}^{a}\wedge \mathbf{z}^{b}.
\label{pc2fa}
\end{equation}

\item The energy function
\begin{equation}
E_{l}\doteqdot \ ^{\pi }\Delta (l)-l=\ ^{l}\hat{p}_{a}\ u^{a}-l.
\label{efunca}
\end{equation}
\end{enumerate}

We shall use additional ``algebroid'' labels like ``$\circ $'', ''$l$'',...
(on the left and right, upper or lower ones, for convenience) for certain
algebroid constructions if it would be necessary to distinguish them from
some geometric objects on the vector/tangent bundle spaces.

In an intrinsic way, the variational Euler--Lagrange equations \cite{mart}
can be geometrized in terms of the introduced three geometrical objects,
respectively, on $\widetilde{TM}$ or on $\mathcal{L}^{\pi }E.$ One holds:

\begin{theorem}
\label{theleq}

\begin{itemize}
\item[$a)$] For any regular Lagrangian $L$, there is a unique SODE, which is
called the Euler--Lagran\-ge vector field:
\begin{equation*}
\xi _{L}=y^{i}\frac{\partial }{\partial x^{i}}+\frac{1}{2}\ \ ^{L}g^{ik}(%
\frac{\partial L}{\partial x^{k}}-y^{j}\frac{\partial ^{2}L}{\partial
x^{j}\partial y^{k}})\frac{\partial }{\partial y^{i}}
\end{equation*}%
on $TM$ such that
\begin{equation*}
i_{\Gamma _{L}}\omega _{L}=dE_{L}
\end{equation*}%
and its solutions are solutions of the Euler--Lagrange equations%
\begin{equation}
\frac{d}{dt}\left( \frac{\partial L}{\partial y^{i}}\right) -\frac{\partial L%
}{\partial x^{i}}=0  \label{eleq1a}
\end{equation}%
and $y^{i}=\dot{x}^{i};$\

\item[$b)$] For any regular Lagrangian $l,$ there is a unique SODE, which is
called the Euler--Lagran\-ge section:
\begin{equation*}
\xi _{l}=u^{a}\mathbf{c}_{a}+\frac{1}{2}\ ^{l}g^{ab}(\rho _{b}^{i}\frac{%
\partial l}{\partial x^{i}}-\rho _{c}^{i}u^{c}\frac{\partial ^{2}l}{\partial
x^{i}\partial u^{b}}+u^{c}C_{cb}^{e}\frac{\partial l}{\partial u^{e}})%
\mathbf{v}_{a}
\end{equation*}%
on $\ \mathcal{L}^{\pi }E$ such that
\begin{equation*}
i_{\Gamma _{l}}\omega _{l}=d^{\mathcal{L}}E_{l}
\end{equation*}%
and its solutions are solutions of the Euler--Lagrange equations%
\begin{equation}
\frac{d}{dt}\left( \frac{\partial l}{\partial u^{a}}\right) -\rho _{a}^{i}%
\frac{\partial l}{\partial x^{i}}+C_{ab}^{e}u^{a}\frac{\partial l}{\partial
u^{e}}=0  \label{eleq1b}
\end{equation}%
for
\begin{equation}
\dot{x}^{i}=\rho _{b}^{i}u^{b}.  \label{rel2c}
\end{equation}
\end{itemize}
\end{theorem}

\begin{proof}
The proof of the results equivalent to the part b) of the Theorem is
considered in \cite{dl1,mart}. It transforms into a usual one for the
geometric mechanics if the trivial Lie algebroid structures are considered
on $TM.$
\end{proof}

\subsection{Geometric structures defined by Lagrangians}

A Lagrange space also defines another important geometric objects and
structures (see \cite{mhss} and references therein, for more details).

\subsubsection{The Euler--Lagrange equations as 'nonlinear' geodesic
equations}

For certain purposes of geometric mechanics, it is enough to consider that
the solutions of the Euler--Lagrange equations are defined by a set of
nonlinear geodesic equations.

\begin{theorem}
\label{thnge}The Euler--Lagrange equations a) (\ref{eleq1a}) and b) (\ref%
{eleq1b}) are equivalent to the corresponding `nonlinear' geodesic equations

\begin{itemize}
\item[$a)$] on Lagrange spaces,%
\begin{equation}
\frac{dy^{i}}{dt}+2\ ^{L}G^{i}(x^{k},y^{j})=0  \label{ngeq}
\end{equation}%
where
\begin{equation}
2\ ^{L}G^{i}(x^{k},y^{k})=\frac{1}{2}\ ^{L}g^{ij}\left( \frac{\partial ^{2}L%
}{\partial y^{i}\partial x^{k}}y^{k}-\frac{\partial L}{\partial x^{i}}%
\right) ,  \label{gcoeff}
\end{equation}%
and

\item[$b)$] on Lagrange algebroids,%
\begin{equation}
\frac{du^{a}}{dt}+2\ ^{l}G^{a}(x^{k},u^{a})=0  \label{ngeqa}
\end{equation}%
where
\begin{equation}
2\ ^{l}G^{a}(x^{k},u^{b})=\frac{1}{2}\ ^{l}g^{ab}\left( \frac{\partial ^{2}l%
}{\partial u^{b}\partial x^{i}}\rho _{c}^{i}u^{c}-C_{bc}^{e}u^{c}\frac{%
\partial l}{\partial u^{e}}-\rho _{b}^{i}\frac{\partial l}{\partial x^{i}}%
\right)  \label{gcoeffa}
\end{equation}
\end{itemize}
\end{theorem}

\begin{proof}
The proof of part $b)$ of this theorem follows from a straightforward
computation
\begin{eqnarray}
\frac{d}{dt}\left( \frac{\partial l}{\partial u^{a}}\right) &=&\frac{%
\partial ^{2}l}{\partial u^{a}\partial x^{i}}\frac{dx^{i}}{dt}+\frac{%
\partial ^{2}l}{\partial u^{a}\partial u^{b}}\frac{du^{b}}{dt},  \notag \\
&=&\frac{\partial ^{2}l}{\partial u^{a}\partial x^{i}}\rho _{b}^{i}u^{b}+2\
^{l}g_{ab}\frac{du^{b}}{dt},  \label{aux2a}
\end{eqnarray}%
where we have taken into account the formulas (\ref{lamf}) and (\ref{rel2c}%
). Introducing the (\ref{aux2a}) into the Euler--Lagrange equations (\ref%
{eleq1b}) and re--grouping the terms in order to emphasize the value (\ref%
{gcoeffa}), we obtain the formula (\ref{ngeqa}). The proof of the part $a)$
may be considered as a trivial limit for the Lie algebroid structures on $%
TM, $ which is outlined in explicit form in Refs. \cite{ma1,mhss}. $\square $
\end{proof}

\subsubsection{Canonical semispray and N--connection}

The semisprays of the previous nonlinear geodesic equations are related to a
fundamental geometric object, the N--connection defined canonically by the
Lagrangian.

\begin{theorem}
\label{r2} The coefficients a) $\ ^{L}G^{i}(x^{k},y^{k})$ (\ref{gcoeff}) and
\ b) $\ ^{l}G^{a}(x^{k},u^{b})$ (\ref{gcoeffa}) define respectively:

\begin{itemize}
\item[$a)$] the solutions of both type equations (\ref{eleq1a}) and (\ref%
{ngeq}) as paths of the canonical semispray%
\begin{equation*}
\xi _{L}=y^{i}\frac{\partial }{\partial x^{i}}-2\ ^{L}G^{i}\frac{\partial }{%
\partial y^{i}}
\end{equation*}%
and the canonical N--connection structure on Lagrange space,
\begin{equation}
^{L}N_{\ j}^{i}\doteqdot \frac{\partial \ ^{L}G^{i}(x,y)}{\partial y^{i}},
\label{cncl}
\end{equation}%
and

\item[$b)$] the solutions of both type equations (\ref{eleq1b}) and (\ref%
{ngeqa}) as paths of the canonical semispray%
\begin{equation*}
\xi _{l}=u^{a}\rho _{a}^{i}\frac{\partial }{\partial x^{i}}-2\ ^{l}G^{a}%
\frac{\partial }{\partial u^{a}}
\end{equation*}%
and the canonical N--connection structure on Lagrange algebroid,
\begin{equation}
^{l}N_{\ b}^{a}\doteqdot \frac{\partial \ ^{l}G^{a}(x,u)}{\partial u^{b}}.
\label{cncla}
\end{equation}
\end{itemize}
\end{theorem}

\begin{proof}
The idea to proof the part $a)$ of this theorem is to show that the
coefficients $\ ^{L}N_{j}^{i}$ define a local distribution of h- and
v--subspaces, $hT_{x}M$ and $vT_{x}M,$ for any point $x\in M.$ Unifying the
construction on all points, $\bigcup_{x},$ we get a global splitting on $%
TTM, $ i. e. a Whitney sum,
\begin{equation}
TTM=hTM\oplus vTM,  \label{wsum1}
\end{equation}%
as a nonintegrable distribution (nonholonomic structure) into horizontal (h)
and vertical (v) subspaces. This is equivalent to the definition of
N--connecti\-on, see, for instance, \cite{ma1,ma2} and the discussion in
next Section, related to the formula (\ref{withns}) when $E=TM.$ The proof
of the part $b)$ is related to a similar proof (via local distributions and
their globalization) of existence of a Whitney sum decomposition
\begin{equation}
\ \mathcal{L}^{\pi }E=h\ \mathcal{L}^{\pi }E\oplus v\ \mathcal{L}^{\pi }E
\label{wsum1a}
\end{equation}%
defined just by $^{l}N_{\ b}^{a}(x,u)$ and $l(x,u).$

For various geometric applications it is enough to show that such
N--coefficients prescribe a canonical nonholonomic frame structure (on the
Lagrange spaces or on the Lagrange algebroid) as we shall do in the next
subsection.$\square $
\end{proof}

We note that in the presented Proof we consider a nonholonomic
(nonintegrable) distribution just for $\ \mathcal{L}^{\pi }E$ because for $%
E=TM$ we get $\ \mathcal{L}^{\pi }E=TTM$ and (\ref{wsum1a}) transforms into (%
\ref{wsum1}). As a matter of principle, for instance, by considering such
splitting on sets of sections with the attempt to define the Ehressmann
connection like in the usual approach \cite{ml,LR3}, it is more useful to
considers splitting of $T\mathcal{L}^{\pi }E.$ In this work, we shall not
consider such type of higher order N--connections which for $E=TM$ are
defined for $TTTM.$

\subsubsection{ Canonical nonholonomic frames}

Any regular Lagrangians $a)$ $L(x,y)$ and $b)$ $l(x,u)$ prescribe
respectively clas\-ses of local (co)frames defined by the canonical
N--connection.

\begin{proposition}
\label{resws}There are preferred local nonholonomic (co) bases
(equivalently, vielbeins\footnote{%
this term is largely used in modern physical literature}) induced linearly
by the coefficients of the N--connection structure:

\begin{itemize}
\item[$a)$] on Lagrange spaces,
\begin{equation}
\mathbf{e}_{\alpha }=(e_{i},\ v_{i})=(e_{i}=\frac{\partial }{\partial x^{i}}%
-\ ^{L}N_{\ i}^{j}\ \frac{\partial }{\partial y^{j}},\ v_{i}=\frac{\partial
}{\partial y^{i}})  \label{dder1}
\end{equation}%
and
\begin{equation}
\mathbf{e}^{\beta }=(e^{i},\ v^{i})=(e^{i}=dx^{i},v^{i}=dy^{i}+\ ^{L}N_{\
j}^{i}\ dx^{j}),  \label{ddif1}
\end{equation}%
and

\item[$b)$] on Lagrange algebroids,
\begin{equation}
\mathbf{c}_{A}=(\mathbf{z}_{a^{\prime }}=\widetilde{z}_{a^{\prime }}-\ \
^{l}N_{\ a^{\prime }}^{b}\widetilde{v}_{b},\ \mathbf{v}_{a}=\widetilde{v}%
_{a})  \label{dder1a}
\end{equation}%
and
\begin{equation}
\mathbf{c}^{A}=(\mathbf{z}^{a^{\prime }}=\widetilde{z}^{a^{\prime }},\
\mathbf{v}^{a}=\widetilde{v}^{a}+\ \ ^{l}N_{\ b}^{a}\widetilde{z}^{b})
\label{ddif1a}
\end{equation}%
where $1\leq a,b\leq m$.
\end{itemize}
\end{proposition}

\begin{proof}
The proof follows from the presented formulas for $\mathbf{e}_{\alpha },%
\mathbf{e}^{\beta }$ and $\mathbf{c}_{A},\mathbf{c}^{A}$ depending linearly
on $N_{\ i}^{j}\ $\ and, respectively, $N_{\ b}^{a}.$ For instance, the
vielbeins (\ref{dder1}) satisfy the nonholonomy relations
\begin{equation}
\lbrack \mathbf{e}_{\alpha },\mathbf{e}_{\beta }]=\mathbf{e}_{\alpha }%
\mathbf{e}_{\beta }-\mathbf{e}_{\beta }\mathbf{e}_{\alpha }=W_{\alpha \beta
}^{\gamma }\mathbf{e}_{\gamma }  \label{anhrel1}
\end{equation}%
with (antisymmetric) nontrivial anholonomy coefficients
\begin{equation*}
\ ^{1}W_{ij}^{k}=\frac{\partial N_{\ i}^{k}}{\partial y^{j}}\mbox{ and }\
^{2}W_{ji}^{k}=\Omega _{ij}^{k}
\end{equation*}%
where
\begin{equation}
\Omega _{ij}^{k}=\frac{\partial N_{\ i}^{k}}{\partial x^{j}}-\frac{\partial
N_{\ j}^{k}}{\partial x^{i}}+N_{\ i}^{p}\frac{\partial N_{\ j}^{k}}{\partial
y^{p}}-N_{\ j}^{p}\frac{\partial N_{\ i}^{k}}{\partial y^{p}}.
\label{ncurv1}
\end{equation}%
Similar nonholonomy relations hold for the algebroid vielbein $\mathbf{c}%
_{A}=(\mathbf{z}_{a},\ \mathbf{v}_{a}),$
\begin{equation}
\lbrack \mathbf{c}_{A},\mathbf{c}_{B}]=\mathbf{c}_{A}\mathbf{c}_{B}-\mathbf{c%
}_{B}\mathbf{c}_{A}=\ W_{AB}^{D}\mathbf{c}_{D}  \label{anhrel1a}
\end{equation}%
but there are additional nontrivial anholonomy coefficients determined by
the nontrivial Lie algebroid structure, $\ ^{\circ }W_{AB}^{D}=(\
^{1}W_{ca}^{b},\ ^{2}W_{ca}^{b},\ ^{3}W_{c^{\prime }a^{\prime }}^{b}),$ with
\begin{equation*}
\ ^{1}W_{c^{\prime }a}^{b}=\frac{\partial \ N_{\ c^{\prime }}^{b}}{\partial
u^{a}},\ ^{2}W_{ca}^{b}=C_{bc}^{a},\ \mbox{ and }\ ^{3}W_{ca}^{b}=\Omega
_{ca}^{b}
\end{equation*}%
where
\begin{equation}
\ \ \Omega _{c^{\prime }a^{\prime }}^{b}=\rho _{c^{\prime }}^{j}\frac{%
\partial \ N_{\ a^{\prime }}^{b}}{\partial x^{j}}-\rho _{a^{\prime }}^{i}%
\frac{\partial \ N_{\ c^{\prime }}^{b}}{\partial x^{i}}+\ N_{\ c^{\prime
}}^{e}\frac{\partial \ N_{\ a^{\prime }}^{b}}{\partial u^{e}}-\ N_{\
a^{\prime }}^{e}\frac{\partial \ N_{\ c^{\prime }}^{b}}{\partial u^{e}}.
\label{ncurv1a}
\end{equation}%
So, we conclude that the nonholonomy coefficients are defined by the
N--connection structure. In the trivial case, we obtain holonomic bases.$%
\square $
\end{proof}

We note that we omitted, for simplicity, in this Proposition, the labels ''$%
L",$ ''$l$'' and ''$\circ $'' for the canonical Lagrange N--connection
coefficients and another objects because the abstract indices label already
the space for which the geometric objects are considered. Such formulas hold
true for arbitrary N--connections (see next Section).

In order to preserve a relation with our previous denotations \cite%
{vnp,v0408121}, we note that $\mathbf{e}_{\alpha }=(e_{i},v_{i})$ and $%
\mathbf{e}^{\alpha }=(e^{i},v^{i})$ are, respectively, the former $\delta
_{\alpha }=\delta /\partial u^{\alpha }=(\delta _{i},\partial _{i})$ and $%
\delta ^{\alpha }=\delta u^{\alpha }=(dx^{i},\delta y^{i})$ which emphasize
that the operators (\ref{dder1}) and (\ref{ddif1}) define, correspondingly,
certain 'N--elongated' partial derivatives and differentials which are more
convenient for calculations on nonholonomic manifolds. In a similar manner
we can argue that the N--connection elongates certain Lie algebroid local
frames to generate vielbeins both adapted to the N--connection and algebroid
structure.

\subsection{Almost Hermitian Mechanics and Lie Algebroids}

Any Lagrange space can be ``lifted'' to an almost Kahlerian structure over $%
\widetilde{TM},$ see Refs. \cite{ma1,ma2}. Similar lifts can be defined for
Lagrange algebroids:

\subsubsection{Canonical almost complex structures}

By explicit constructions, one proves

\begin{proposition}
\label{r5} The canonical N--connections a)$\ ^{L}N_{\ j}^{i}$ (\ref{cncl})
and b)$\ ^{l}N_{\ b}^{a}$ (\ref{cncla}) naturally induce, respectively,

\begin{itemize}
\item[$a)$] an almost complex structure $\mathbf{F}:\mathcal{X}(\widetilde{TM%
})\rightarrow \mathcal{X}(\widetilde{TM}),$ $\ $defined for a tangent bundle
$TM,$
\begin{equation*}
\mathbf{F}(e_{i})=\ v_{i}\mbox{ and  }\mathbf{F}(v_{i})=-e_{i},
\end{equation*}%
and
\begin{equation}
\mathbf{F}=\ -v_{i}\otimes e^{i}+e_{i}\otimes v^{i}  \label{acs1}
\end{equation}%
satisfies the condition $\mathbf{F\bullet F=-I,}$ i. e. $F_{\ \ \beta
}^{\alpha }F_{\ \ \gamma }^{\beta }=-\delta _{\gamma }^{\alpha },$ where $%
\delta _{\gamma }^{\alpha }$ is the Kronecker symbol and $\mathcal{X}(%
\widetilde{TM})$\ denotes the module of vector fields on $\widetilde{TM};$

\item[$b)$] an almost complex (algebroid) structure $\ ^{\circ }\mathbf{F}:%
\mathcal{X}(\mathcal{L}^{\pi }E)\rightarrow \mathcal{X}(\mathcal{L}^{\pi
}E), $
\begin{equation*}
\ ^{\circ }\mathbf{F}(\mathbf{z}_{a})=\ \mathbf{v}_{a}\mbox{ and }\ ^{\circ }%
\mathbf{F}(\ \mathbf{v}_{a})=-\mathbf{z}_{a},
\end{equation*}%
such that
\begin{equation}
\ ^{\circ }\mathbf{F}=\ -\mathbf{v}_{a}\otimes \mathbf{z}^{a}+\mathbf{z}%
_{a}\otimes \mathbf{\ v}^{a}  \label{acs1a}
\end{equation}%
satisfies the condition $\ ^{\circ }\mathbf{F\bullet \ \ ^{\circ }F=-I,}$ i.
e. $F_{\ B}^{A}F_{\ D}^{B}=-\delta _{D}^{A}.$
\end{itemize}
\end{proposition}

\subsubsection{Canonical metric structures}

A regular Lagrangian defines also the canonical metric on Lagrange spaces
(algebroids) \ constructed by using Sasaki type lifts from $M$ to $%
\widetilde{TM}$ (to $\mathcal{L}^{\pi }E)$ where the metric tensor is $\
^{L}g_{ab}$ (\ref{lmf}) (or $^{l}g_{ab}$ (\ref{lamf})).

\begin{theorem}
There are canonical metric structures

\begin{itemize}
\item[$a)$] on $\widetilde{TM},$ i. e.
\begin{equation}
\ ^{L}\mathbf{g}=\ ^{L}\mathbf{g}_{\alpha \beta }\mathbf{e}^{\alpha }\otimes
\mathbf{e}^{\beta }=\ ^{L}g_{ij}\ e^{i}\otimes e^{j}+\ ^{L}g_{ij}\
v^{i}\otimes \ v^{j}  \label{slm}
\end{equation}%
and

\item[$b)$] on $\ \mathcal{L}^{\pi }E,$%
%\ ^{l}\mathbf{g}=[\ ^{\circ }g,\^{\star }g],$ i. e.
\begin{equation}
\ ^{l}\mathbf{g}=\ ^{l}\mathbf{g}_{AB}\mathbf{c}^{A}\otimes \mathbf{c}^{B}=\
^{l}g_{a^{\prime }b^{\prime }}\ \mathbf{z}^{a^{\prime }}\otimes \mathbf{z}%
^{b^{\prime }}+\ ^{l}g_{ab}\ \mathbf{v}^{a}\otimes \mathbf{v}^{b}
\label{slma}
\end{equation}%
called distinguished metrics (d--metrics) defined by the corresponding
Lagrangians. % and, induced by such Lagrangians,canonical N--connecti\-ons.
\end{itemize}
\end{theorem}

It is possible to prove that $\ ^{L}\mathbf{g}$ and $\ ^{l}\mathbf{g}$ does
not depend on local tranformations of coordinates but, respectively, on $L$
and $l.$

In a standard manner, by using the metric (\ref{slm}) (or (\ref{slma})) it
is possible to construct the Levi--Civita connection $^{L}\nabla $ \ (or $%
^{l}\nabla )$ on a Lagrange space (algebroid) which, by definition,
satisfies both the metricity, $^{L}\nabla (\ ^{L}\mathbf{g)}=0,$ (or $%
^{l}\nabla \ (^{l}\mathbf{g})=0)$ and the torsionless conditions. From a
formal point of view, this geometrizes the Lagrange mechanics in terms of a
(pseudo) Riemannian model on the Lagrange space (or algebroid). But a such
approach would consider a linear connection structure which is not adapted
to the N--connection (i. e. to the nonholonomic distribution) which is
defined canonically by the Lagrangian, see section \ref{ahlm}.

The nonholonomic frames (\ref{dder1}) \ (respectively, (\ref{dder1a}))
induce naturally a torsion structure via the anholonomy coefficients (\ref%
{anhrel1}) (respectively, (\ref{anhrel1a})). We can elaborate a covariant
N--adapted differential calculus by considering a class of linear
connections preserving the global splitting (\ref{wsum1}) (respectively, (%
\ref{wsum1a})), called distinguished (by the N--connection) connections, in
brief, d--connections.\footnote{%
for details, see Refs \cite{ma1,ma2,vnp} and the subsection \ref{dcons} in
this work.} Such connections may be chosen to satisfy the metricity
condition, but contain a nontrivial torsion component which is defined by
the Lagrangian and, in general, another geometric/physical terms.

\subsubsection{Canonical almost sympletic structures}

A regular Lagrangian defines a canonical almost sympletic structure via the
canonical N--connection, almost complex structure and metric constructed on
tangent bundles and/or on Lie algebroids.

\begin{theorem}
There are almost Kahlerian models of the a) Lagrange spaces and b) Lagrange
algebroids defined respectively by

\begin{itemize}
\item[$a)$] triads $K^{2n}=(\widetilde{TM},\ ^{L}\mathbf{g,F})$ with the
induced almost sympletic 2--form
\begin{equation}
\ ^{L}\mathbf{\omega }=\ ^{L}\mathbf{\omega }_{\alpha \beta }\mathbf{e}%
^{\alpha }\wedge \mathbf{e}^{\beta }=\ ^{L}g_{ij}\ v^{i}\wedge \ e^{j}
\label{s2f}
\end{equation}%
and

\item[$b)$] triads $\ ^{\circ }K^{2m}=(\mathcal{L}^{\pi }E,\ ^{l}\mathbf{g,\
}^{\circ }\mathbf{F})$ with the induced almost sympletic 2--section
\begin{equation}
\ ^{l}\mathbf{\omega }=\ ^{l}\mathbf{\omega }_{AB}\mathbf{c}^{A}\wedge
\mathbf{c}^{B}=\ ^{l}g_{ab}\ \mathbf{v}^{a}\wedge \mathbf{z}^{b}.
\label{s2fa}
\end{equation}
\end{itemize}
\end{theorem}

\begin{proof}
It is evident if we define, correspondingly,%
\begin{equation*}
a)\ ^{L}\mathbf{\omega (e}_{\alpha },\mathbf{e}_{\beta }\mathbf{)\doteqdot }%
\ ^{L}\mathbf{g(Fe}_{\alpha },\mathbf{e}_{\beta }\mathbf{)}\mbox{ and }\ b)\
^{l}\mathbf{\omega (c}_{A},\mathbf{c}_{B}\mathbf{)\doteqdot }\ ^{l}\mathbf{%
g(\ }^{\circ }\mathbf{Fc}_{A},\mathbf{c}_{B}\mathbf{)}
\end{equation*}%
and consider the components of formulas (\ref{slm}), (\ref{acs1}) and (\ref%
{slma}), (\ref{acs1a}). $\square $
\end{proof}

\subsubsection{Canonical d--connection structures}

\label{ahlm} It should noted that the almost Kahler manifolds $K^{2n}$ and $%
\ ^{\circ }K^{2m}$ transform into Kahlerian spaces if the N--connection
structure is integrable for the corresponding Lagrange space and Lagrange
algebroid.

\begin{definition}
A linear connection $\widetilde{\mathbf{D}}$ on $TM$ ($\widetilde{\mathcal{D}%
}$ on $\mathcal{L}^{\pi }E)$ is said to be a distinguished connection
(d--connection) if it preserves by parallelism (i. e. by parallel transports
defined by the corresponding covariant derivative)the vertical and
horizontal distributions (\ref{wsum1}) on $TM$ ((\ref{wsum1a}) on $\mathcal{L%
}^{\pi }E$).
\end{definition}

We consider a particular class of d--connections:

\begin{definition}
A normal (or natural) $d$--connection $\ \mathbf{D,}$ or$\ \mathcal{D},$ is
adapted to the almost sympletic structure $\mathbf{F}$ for Lagrange spaces,
or $\ ^{\circ }\mathbf{F}$ $\ $for Lagrange algebroids, when (respectively)%
\begin{equation*}
\ \mathbf{D}_{X}\mathbf{F=0,\mbox{ or }\ }\mathcal{D}_{\ ^{\circ }X}\
^{\circ }\mathbf{F=0,}
\end{equation*}%
for any vector field $\mathbf{X}$ on $TM,$ or $\ ^{\circ }\mathbf{X}$ on $%
TE. $
\end{definition}

A normal d--connection $\mathbf{D}$ is characterized by its coefficients,
\begin{equation*}
\Gamma _{\ \beta \gamma }^{\alpha }\mathbf{=}(L_{jk}^{i}(x,y),\
K_{jk}^{i}(x,y))
\end{equation*}%
on $TM,$ where
\begin{equation*}
\mathbf{D}_{e_{k}}e_{j}\doteqdot L_{jk}^{i}e_{i},~\mathbf{D}_{e_{k}}\
v_{j}\doteqdot L_{jk}^{i}\ v_{i},\mathbf{D}_{v_{k}}e_{j}\doteqdot \
K_{jk}^{i}e_{i},~\mathbf{D}_{v_{k}}\ v_{j}\doteqdot \ K_{jk}^{i}\ v_{i},
\end{equation*}%
and (for Lie algebroids)
\begin{equation*}
\mathcal{D}\mathbf{=}(L_{ab}^{e}(x,u),\ K_{ab}^{e}(x,u))
\end{equation*}%
on $TE,$ denoted%
\begin{equation*}
\mathcal{D}_{\mathbf{c}_{a}}\mathbf{c}_{b}\doteqdot L_{ab}^{e}\mathbf{c}%
_{e},~\mathcal{D}_{\mathbf{c}_{a}}\mathbf{v}_{b}\doteqdot L_{ab}^{e}\mathbf{v%
}_{e},\mathcal{D}_{\mathbf{v}_{a}}\mathbf{c}_{b}\doteqdot \ K_{ab}^{e}%
\mathbf{c}_{e},~\mathcal{D}_{\mathbf{v}_{a}}\mathbf{v}_{b}\doteqdot \
K_{ab}^{e}\mathbf{v}_{e}.
\end{equation*}

\begin{definition}
A d--connection $a)$ $\widetilde{\mathbf{D}},$ or $b)$ $\widetilde{\mathcal{D%
}},$ is a) h-- and/or v--metric, and b) c-- and/or v--metric, respectively,
if there are satisfied the conditions:

\begin{itemize}
\item[$a)$] for Lagrange spaces,%
\begin{equation*}
\widetilde{D}_{k}\ ^{L}g_{ij}=0\mbox{ and/or }\ ^{\star }\widetilde{D}_{k}\
^{L}g_{ij}=0,
\end{equation*}%
and

\item[$b)$] for Lagrange algebroids,%
\begin{equation*}
\widetilde{\mathcal{D}}_{a}\ ^{l}g_{bc}=0\mbox{ and/or }\ ^{\star }%
\widetilde{\mathcal{D}}_{a}\ ^{l}g_{bc}=0;
\end{equation*}%
a such connection is metric (compatible) if it satisfies both h- and
v--metricity conditions.
\end{itemize}
\end{definition}

The torsion of a $d$--connection, for instance of $\widetilde{\mathcal{D}}$,
can be defined in component free form%
\begin{equation*}
\widetilde{\mathcal{T}}(\ ^{\circ }\mathbf{X},\ ^{\circ }\mathbf{Y}%
)\doteqdot \widetilde{\mathcal{D}}_{\ ^{\circ }\mathbf{X}}\ ^{\circ }\mathbf{%
Y}-\widetilde{\mathcal{D}}_{\ ^{\circ }\mathbf{Y}}\ ^{\circ }\mathbf{X}-[\
^{\circ }\mathbf{X},\ ^{\circ }\mathbf{Y}].
\end{equation*}%
Any d--vector decompose in its $z$-- and $v$--components, $\ ^{\circ }%
\mathbf{Y}=zY+vY,$ for $zY=Y_{a}\mathbf{z}^{a}$ and $vY=Y_{a}\mathbf{v}^{a}.$
Considering such projections, we can decompose the torsion $\widetilde{%
\mathcal{T}}(\mathbf{c}_{A},\mathbf{c}_{B})$ into N--adapted components
\begin{equation*}
z\widetilde{\mathcal{T}}(\mathbf{z}_{a},\mathbf{z}_{b}),z\widetilde{\mathcal{%
T}}(\mathbf{z}_{a},\mathbf{v}_{b}),z\widetilde{\mathcal{T}}(\mathbf{v}_{a},%
\mathbf{v}_{b}),v\widetilde{\mathcal{T}}(\mathbf{z}_{a},\mathbf{z}_{b}),v%
\widetilde{\mathcal{T}}(\mathbf{z}_{a},\mathbf{v}_{b}),v\widetilde{\mathcal{T%
}}(\mathbf{v}_{a},\mathbf{v}_{b}).
\end{equation*}

\begin{theorem}
\label{alhmla}There are almost Kahlerian models of the a) Lagrange spaces
and b) Lagrange algebroids defined by respective unique (canonical) almost
Kahlerian d--connections a) $\widehat{\mathbf{D}}$ on $TM$ and \ b) $%
\widehat{\mathcal{D}}$ on $\ \mathcal{L}^{\pi }E$ which preserve by
parallelism the vertical distributions and satisfy the conditions:

\begin{itemize}
\item[$a)$] one holds the compatibility with the almost Kahlerian structure
\begin{equation*}
\widehat{\mathbf{D}}_{X}\ ^{L}\mathbf{g}=\widehat{\mathbf{D}}_{X}\ ^{L}%
\mathbf{\omega =0\mbox{ and }D}_{X}\mathbf{F=0,}
\end{equation*}%
for any vector field $\mathbf{X}$ on $TM,$ and the propery of vanishing of
the complete ``horizontal'' and ``vertical'' torsions, i. e.
\begin{equation*}
h\widehat{\mathbf{T}}(e_{i},e_{j})=0\mathbf{\mbox{ and }}v\widehat{\mathbf{T}%
}(v_{i},v_{j})=0;
\end{equation*}

\item[$b)$] one holds the compatibility with the almost Kahlerian structure
\begin{equation*}
\widehat{\mathcal{D}}_{\ ^{\circ }\mathbf{X}}\ ^{l}\mathbf{g}=\widehat{%
\mathcal{D}}_{\ ^{\circ }\mathbf{X}}\ ^{l}\mathbf{\omega }^{\circ }\mathbf{=0%
\mbox{ and }}\widehat{\mathcal{D}}_{\ ^{\circ }\mathbf{X}}\ ^{\circ }\mathbf{%
F}\mathbf{=0,}
\end{equation*}%
for any section $\ ^{\circ }\mathbf{X}$ on $\ \mathcal{L}^{\pi }E,$ and the
property of vanishing of \ the complete ''horizontal'' and ''vertical''
torsions, i. e.
\begin{equation*}
z\widehat{\mathcal{T}}(\mathbf{z}_{a},\mathbf{z}_{b})=0\mathbf{\mbox{ and }}v%
\widehat{\mathcal{T}}(\mathbf{v}_{a},\mathbf{v}_{b})=0;
\end{equation*}
\end{itemize}
\end{theorem}

\begin{proof}
Let us state that the almost Kahlerian d--connections $a)$ $\widehat{\mathbf{%
D}}=(\widehat{L}_{jk}^{i},\ \widehat{K}_{jk}^{i})$ and \ $b)$ $\widehat{%
\mathcal{D}}=(\widehat{L}_{ab}^{e},\widehat{K}_{ab}^{e})$ are defined by
respective coefficients of the canonical d--connections ,%
\begin{eqnarray}
a)\ \widehat{L}_{jk}^{i} &=&\frac{1}{2}\ ^{L}g^{ih}\left( e_{k}\
^{L}g_{hj}+e_{j}\ ^{L}g_{hk}-e_{h}\ ^{L}g_{jk}\right) ,  \label{dccanls} \\
\widehat{K}_{jk}^{i} &=&\frac{1}{2}\ ^{L}g^{ih}\left( v_{k}\
^{L}g_{hj}+v_{j}\ ^{L}g_{hk}-v_{h}\ ^{L}g_{jk}\right) ,  \notag
\end{eqnarray}%
and
\begin{eqnarray}
b)\ \widehat{L}_{ab}^{e} &=&\frac{1}{2}\ ^{l}g^{ac}\left( \mathbf{z}_{b}\
^{l}g_{ca}+\mathbf{z}_{a}\ ^{l}g_{cb}-\mathbf{z}_{c}\ ^{l}g_{ab}\right) ,
\label{dccanla} \\
\widehat{K}_{bc}^{a} &=&\frac{1}{2}\ ^{l}g^{ac}\left( \ \mathbf{v}_{b}\
^{l}g_{ca}+\ \mathbf{v}_{a}\ ^{l}g_{cb}-\ \mathbf{v}_{c}\ ^{l}g_{ab}\right) .
\notag
\end{eqnarray}%
By straightforward calculations with covariant derivatives defined by the
the coefficients (\ref{dccanls}) and (\ref{dccanla}) we can verify that one
holds true all conditions of the theorem.$\square $
\end{proof}

The existence of canonical almost complex and almost sympletic structures
defined by Lagrangian and/or N--connection is very important for elaborating
an approach to geometric quantization of mechanical systems modelled on
nonholonomic manifolds \cite{v0407495} as well for a rigorous definition of
nonholonomic (anisotropic) Clifford structures and spinors in commutative
and noncommutative spaces \cite{vnp,v0408121}.

\section{Finsler and Hamilton Algebroids}

The theory of N--connections and adapted metric and linear connection
structures on prolongated Lie algebroids, elagborated in the previous
Section, gives rise to a number of possibilities to construct geometric and
physical models on Lie N--anholonomic algebroids and their duals. The aim of
this Section is to consider two such geometries (the Finsler and generalized
Lagrange algebroids) realized on Lie algebroids and two dual algebroid
geometries (the Hamilton and Cartan algebroids).

\subsection{Finsler and generalized Lagrange algebroids}

We analyze in brief two possibilities for the algebroid constructions
presented in Section 3: 1) to constrain them to a Finsler like configuration
and 2) to extend them for generalized Lagrange configurations.

\subsubsection{Finsler algebroids}

The Finsler geometry was modelled on tangent bundles (the canonical ways
with metric compatible, or not, d--connections, see \cite%
{mats,bej,ma1,ma2,bcs}) and, more recently, as embedding of N--anholonomic
spaces of even dimensions into Riemann--Cartan and metric--affine manifolds,
superbundles and Clifford bundles and on finite projective modules for
noncommutative spaces, see references and discussion from \cite%
{vnp,vmaf,v0408121}. There is a possibility to define a new class of
''singular'' Finsler geometries on Lie algebroids with associated vector
bundles (which are not tangent bundles, or their mimics on manifolds of odd
dimensions). Such algebroid Finsler configurations can be modelled on
prolongated Lie algebroids.

\begin{definition}
A Finsler algebroid $F\mathcal{A}=(E,\mathbf{f}^{2})$ is a Lie algebroid $%
\mathcal{L}^{\pi }\mathbf{E}$ provided with a fundamental Finsler function $%
\mathbf{f:}E\rightarrow \R$ satisfying the conditions:

\begin{enumerate}
\item $\mathbf{f}$ is a scalar differentiable function on the manifold $%
\widetilde{E}=E/\{0\}$ and continuous on the null section of $\pi
:E\rightarrow M;$

\item $\mathbf{f}$ is a positive function, homogeneous on the fibers of $%
\mathcal{E},$ i. e. $\mathbf{f(}x,\lambda u\mathbf{)=}$ $\lambda \mathbf{f(}%
x,u\mathbf{),}\lambda \in \R;$

\item the Hessian of $\mathbf{f}^{2}$ with elements%
\begin{equation}
\ ^{f}g_{ab}=\frac{1}{2}\frac{\partial ^{2}\mathbf{f}^{2}}{\partial
u^{a}\partial u^{b}}  \label{famf}
\end{equation}
\end{enumerate}

is positively defined on $\widetilde{E}.$
\end{definition}

There are two general ways to model Finsler algebroids on $\mathcal{L}^{\pi }%
\mathbf{E:}$

The first one is to say that they consist a particular case of Lagrange
algebroids (see Definition \ref{defls}) when $l=\mathbf{f}^{2}$ and the
Hessian (\ref{lamf}) transforms into (\ref{famf}). In this case, we can
define an almost Hermitian model of Finsler algebroids and restrict all
results of the Sections 3 and 4 for metric compatible canonical
d--connections derived from $l=\mathbf{f}^{2}$ and respective canonical
N--connection $^{l}N_{\ b}^{a}\rightarrow \ ^{f}N_{\ b}^{a},$ see (\ref%
{cncla}), when%
\begin{equation*}
\ ^{f}N_{\ b}^{a}=\frac{\partial \ ^{f}G^{a}}{\partial u^{b}}=\frac{1}{4}%
\frac{\partial }{\partial u^{b}}\left\{ \ ^{f}g^{ac}\left( u^{e}\rho _{e}^{i}%
\frac{\partial ^{2}f^{2}}{\partial u^{c}\partial x^{i}}-\rho _{c}^{k}\frac{%
\partial f^{2}}{\partial x^{k}}\right) \right\} ,
\end{equation*}%
the Sasaki type of Finsler d--metric, on $\mathcal{L}^{\pi }E,\ ^{l}\mathbf{g%
}=[\ ^{\circ }g,\ ^{\star }g],$ see (\ref{slma}), i. e.
\begin{equation}
\ ^{f}\mathbf{g}=\ ^{f}\mathbf{g}_{AB}\mathbf{c}^{A}\otimes \mathbf{c}^{A}=\
^{f}g_{ab}\ \mathbf{z}^{a}\otimes \mathbf{z}^{b}+\ ^{f}g_{ab}\ \mathbf{v}%
^{a}\otimes \mathbf{v}^{b},  \label{fsdm}
\end{equation}%
and canonical d--connection computed similarly to (\ref{dccanla})
\begin{eqnarray}
\widehat{L}_{ab}^{e} &=&\frac{1}{2}\ ^{f}g^{ac}\left( \mathbf{z}_{b}\
^{f}g_{ca}+\mathbf{z}_{a}\ ^{f}g_{cb}-\mathbf{z}_{c}\ ^{f}g_{ab}\right) ,
\label{candconf} \\
\widehat{K}_{bc}^{a} &=&\frac{1}{2}\ ^{f}g^{ac}\left( \ \mathbf{v}_{b}\
^{f}g_{ca}+\ \mathbf{v}_{a}\ ^{f}g_{cb}-\ \mathbf{v}_{c}\ ^{f}g_{ab}\right) ,
\notag
\end{eqnarray}%
where the N--elongation of the operators $\mathbf{z}_{b}$ is defined by $\
^{f}N_{\ b}^{a}.$

In the second way of elaborating Finsler geometries on Lie algebroids, one
may consider that a Lagrange structure is a singular Finsler structure on
higher dimension and to follow the idea that such geometries possess
nontrivial nonmetricity d--tensors (of Berwald or Chern type\footnote{%
see discussions in \cite{bcs} and \cite{v0408121}; the models satisfying the
metricity conditions admit a more simple geometric and physical
interpretation of interactions with spinor and gravity fields, but in
another turn the ''nonmetricity'' physics also presents certain interest.}).
Here, we present some details on Berwald type nonmetricity for Finsler
algebroids:

A d--connection of Berwald type (see, for instance, Ref. \cite{ma1,ma2,bcs}
on such configurations in Finsler and Lagrange geometry) and denoted $\
_{[B]}\mathcal{D}=\left( \ _{[B]}^{\circ }D,\ \ _{[B]}^{\star }D\right) $
\begin{equation*}
\ _{[B]}\mathbf{\Gamma }_{\ A}^{E}=\ \ _{[B]}\mathbf{\Gamma }_{\ AB}^{E}%
\mathbf{c}^{B},
\end{equation*}%
with c- and v--components
\begin{equation}
\ \ \ _{[B]}\mathbf{\Gamma }_{AB}^{E}=\left( \widehat{L}_{\ b^{\prime
}c^{\prime }}^{a^{\prime }},\mathbf{v}_{b}\left( \ ^{f}N_{\ a^{\prime
}}^{a}\right) ,0,\widehat{K}_{\ bc}^{a}\right) ,  \label{berw}
\end{equation}%
with $\widehat{L}_{\ jk}^{i}$ and $\widehat{K}_{\ bc}^{a}$ taken as in (\ref%
{candconf}), satisfying only partial metric compatibility conditions for a
d--metric \ (\ref{fsdm}),\ $\ \ ^{f}\mathbf{g}_{AB}=\left( \
^{f}g_{a^{\prime }b^{\prime }},\ ^{f}g_{ab}\right) $ on $\mathcal{L}^{\pi }E$
\begin{equation*}
\ \ _{[B]}^{\circ }D_{c^{\prime }}\ ^{f}g_{a^{\prime }b^{\prime }}=0%
\mbox{
and }\ \ _{[B]}^{\star }D_{c}\ ^{f}g_{ab}=0.
\end{equation*}%
This is an example of d--connections which may possess nontrivial
nonmetricity components, $\ \ _{[B]}\mathbf{Q}_{ABC}=\left( \
_{[B]}Q_{ca^{\prime }b^{\prime }},\ \ _{[B]}Q_{a^{\prime }ab}\right) $ with
\begin{equation}
\ \ _{[B]}Q_{ca^{\prime }b^{\prime }}=\ _{[B]}^{\star }D_{c}\
^{f}g_{a^{\prime }b^{\prime }}\mbox{ and }\ \ _{[B]}Q_{a^{\prime }ab}=\ \
_{[B]}^{\circ }D_{c^{\prime }}\ ^{f}g_{ab}.  \label{berwnm}
\end{equation}%
So, the Berwald d--connection defines a metric--affine algebroid $\mathcal{L}%
^{\pi }E$ provided with N--connection structure of Finsler type.

If $\widehat{L}_{\ jk}^{i}$ and $\widehat{K}_{\ bc}^{a}$ vanish, we obtain a
Berwald type connection $\ $%
\begin{equation*}
^{\lbrack N]}\mathbf{\Gamma }_{\alpha \beta }^{\gamma }=\left( 0,\mathbf{v}%
_{b}\left( \ ^{f}N_{\ a^{\prime }}^{a}\right) ,0,0\right)
\end{equation*}%
induced only by the canonical Finsler N--connection structure. It defines a
vertical covariant derivation $\ _{[N]}^{\star }D_{c}$ acting in the
v--subspace of $\mathcal{L}^{\pi }E,$ with the coefficients being partial
derivatives on v--coordinates $u^{a}$ of the N--connection coefficients $\
^{f}N_{\ b}^{a}.$

We can generalize the Berwald connection (\ref{berw}) to contain any
prescribed values of d--torsions $T_{.b^{\prime }c^{\prime }}^{a^{\prime }}$
and $T_{.bc}^{a}$ from the c- v--decomposition (\ref{dtorsb}), but redefined
with respect to the canonical Finsler d--connection (\ref{candconf}). We can
check by a straightforward calculations that the d--connection%
\begin{equation}
\ _{[B\tau ]}\mathbf{\Gamma }_{\ AB}^{E}=\left( \widehat{L}_{\ b^{\prime
}c^{\prime }}^{a^{\prime }}+\tau _{\ b^{\prime }c^{\prime }}^{a^{\prime }},%
\mathbf{v}_{b}\ \left( ^{f}N_{\ a^{\prime }}^{a}\right) ,0,\widehat{K}_{\
bc}^{a}+\tau _{\ bc}^{a}\right)  \label{bct}
\end{equation}%
with
\begin{eqnarray}
\tau _{\ b^{\prime }c^{\prime }}^{a^{\prime }} &=&\frac{1}{2}\
^{f}g^{a^{\prime }d^{\prime }}\left( \ ^{f}g_{bf}T_{\ dc}^{f}+\
^{f}g_{cf}T_{\ db}^{f}-\ ^{f}g_{df}T_{\ bc}^{f}\right)  \label{tauformulas}
\\
\tau _{\ bc}^{a} &=&\frac{1}{2}\ ^{f}g^{ad}\left( \ ^{f}g_{bf}T_{\ dc}^{f}+\
^{f}g_{cf}T_{\ db}^{f}-\ ^{f}g_{df}T_{\ bc}^{f}\right)  \notag
\end{eqnarray}%
results in $\ _{[B\tau ]}\mathbf{T}_{.b^{\prime }c^{\prime }}^{a^{\prime
}}=T_{.b^{\prime }c^{\prime }}^{a^{\prime }}$ and $\ _{[B\tau ]}\mathbf{T}%
_{bc}^{a}=T_{.bc}^{a}.$ The d--connection (\ref{bct}) has nonvanishing
nonmetricity components, $\ _{[B\tau ]}\mathbf{Q}_{\alpha \beta \gamma
}=t(\ _{[B\tau ]}Q_{ca^{\prime }b^{\prime }}, \ _{[B\tau
]}Q_{a^{\prime }ab}).$

In general, by using the metrization procedure (see Theorem \ref{kmp}) we
can also construct metric d--connections with prescribed values of
d--torsions $T_{.b^{\prime }c^{\prime }}^{a^{\prime }}$ and $T_{.bc}^{a},$
or to express, for instance, the Levi--Civita connection via the
coefficients of an arbitrary metric d--connection.

We can express a general affine Berwald d--connection$\ $%
\begin{equation*}
\ _{[B\tau ]}\mathbf{\Gamma }_{\ \ A}^{E}=\ \ _{[B\tau ]}\mathbf{\Gamma }_{\
AB}^{E}\mathbf{c}^{B}
\end{equation*}%
via its deformations from the Levi--Civita connection $\mathbf{\Gamma }%
_{\bigtriangledown \ B}^{A},$
\begin{equation}
\ _{[B\tau ]}\mathbf{\Gamma }_{\ \ A}^{E}=\ ^{\nabla }\mathbf{\Gamma }_{\
A}^{E}+\ _{[B\tau ]}\mathbf{Z}_{\ \ A}^{E},  \label{accnb}
\end{equation}%
$\ ^{\nabla }\mathbf{\Gamma }_{\ E}^{A}$ being expressed as (\ref{christa})
(equivalently, defined by (\ref{lcsym})) and
\begin{eqnarray}
\ _{[B\tau ]}\mathbf{Z}_{AB} &=&\mathbf{c}_{B}\rfloor \ \ _{[B\tau ]}\mathbf{%
T}_{A}-\mathbf{c}_{A}\rfloor \ \ _{[B\tau ]}\mathbf{T}_{B}+\frac{1}{2}\left(
\mathbf{c}_{A}\rfloor \mathbf{c}_{B}\rfloor \ \ _{[B\tau ]}\mathbf{T}%
_{E}\right) \mathbf{c}^{E}  \label{distanb} \\
&&+\left( \mathbf{c}_{A}\rfloor \ \ _{[B\tau ]}\mathbf{Q}_{BE}\right)
\mathbf{c}^{E}-\left( \mathbf{c}_{B}\rfloor \ \ _{[B\tau ]}\mathbf{Q}%
_{AE}\right) \mathbf{c}^{E}+\frac{1}{2}\ \ _{[B\tau ]}\mathbf{Q}_{AB}  \notag
\end{eqnarray}%
defined with prescribed d--torsions $_{[B\tau ]}\mathbf{T}_{b^{\prime
}c^{\prime }}^{a^{\prime }}=T_{.b^{\prime }c^{\prime }}^{a^{\prime }}$ and $%
^{[B\tau ]}\mathbf{T}_{bc}^{a}=T_{.bc}^{a},$ where, for simplicity, we have
omitted the label ''$\mathbf{f}$''. Such formulas hold true for any
d--connection expressed via deformations of a metric compatible
d--connection. The Berwald d--connection defines a particular subclass of
metric--affine connections being adapted to the N--connection structure and
with prescribed values of d--torsions.

If the deformations of d--metrics in (\ref{accnb}) are considered with
respect to the canonical d--connection $\widehat{\mathbf{\Gamma }}_{\
BC}^{A} $ with z- v-- coefficients (\ref{candcon}), we can construct a set
of canonical metric--affine d--connections. Such d--connections $\mathbf{%
\Gamma }_{\ \ A}^{E}=\ \mathbf{\Gamma }_{\ AB}^{E}\mathbf{c}^{B}$ are
defined via deformations
\begin{equation}
\ \mathbf{\Gamma }_{\ B}^{A}=\widehat{\mathbf{\Gamma }}_{\ B}^{A}+\ \widehat{%
\mathbf{Z}}_{\ \ B}^{A},  \label{mafdc}
\end{equation}%
$\widehat{\mathbf{\Gamma }}_{\ B}^{A}$ being the canonical d--connection (%
\ref{dcon1}) and
\begin{eqnarray}
\ \widehat{\mathbf{Z}}_{AB} &=&\mathbf{c}_{B}\rfloor \ \mathbf{T}_{A}-%
\mathbf{c}_{A}\rfloor \ \mathbf{T}_{A}+\frac{1}{2}\left( \mathbf{c}%
_{A}\rfloor \mathbf{c}_{B}\rfloor \ \mathbf{T}_{A}\right) \mathbf{c}^{E}
\label{dmafdc} \\
&&+\left( \mathbf{c}_{A}\rfloor \ \ _{[B\tau ]}\mathbf{Q}_{BE}\right)
\mathbf{c}^{E}-\left( \mathbf{c}_{B}\rfloor \ \mathbf{Q}_{AE}\right) \mathbf{%
c}^{E}+\frac{1}{2}\ \ _{[B\tau ]}\mathbf{Q}_{AB}  \notag
\end{eqnarray}%
where $\mathbf{T}_{A}$ and $\mathbf{Q}_{AB}$ are arbitrary torsion and
nonmetricity structures.

A metric--affine d--connection $\mathbf{\Gamma }_{\ \ A}^{E}$ can be also
considered as a deformation from the Berwald connection $\ _{[B\tau ]}%
\mathbf{\Gamma }_{\ AB}^{E}$
\begin{equation}
\ \mathbf{\Gamma }_{\ B}^{A}=\ \ _{[B\tau ]}\mathbf{\Gamma }_{\ B}^{A}+\ \
_{[B\tau ]}\ \widehat{\mathbf{Z}}_{\ \ B}^{A},  \label{mafbc}
\end{equation}%
$\ \ _{[B\tau ]}\mathbf{\Gamma }_{\ AB}^{E}$ being the Berwald d--connection
(\ref{bct}) and
\begin{eqnarray}
\ \ _{[B\tau ]}\ \widehat{\mathbf{Z}}_{\ \ B}^{A} &=&\mathbf{c}_{B}\rfloor \
\mathbf{T}_{A}-\mathbf{c}_{A}\rfloor \ \mathbf{T}_{B}+\frac{1}{2}\left(
\mathbf{c}_{A}\rfloor \mathbf{c}_{B}\rfloor \ \mathbf{T}_{E}\right) \mathbf{c%
}^{E}  \label{dmafbc} \\
&&+\left( \mathbf{c}_{A}\rfloor \ \ _{[B\tau ]}\mathbf{Q}_{BE}\right)
\mathbf{c}^{E}-\left( \mathbf{c}_{B}\rfloor \ \mathbf{Q}_{AE}\right) \mathbf{%
c}^{E}+\frac{1}{2}\ _{[B\tau ]}\mathbf{Q}_{AB}.  \notag
\end{eqnarray}

The z- and v--splitting of formulas can be computed by introducing the
adapted N--frames (\ref{dder1a}) and (\ref{ddif1a}) and d--metric $\ ^{\circ
}\mathbf{g}=\left( g_{a^{\prime }b^{\prime },}h_{ab}\right) $ into (\ref%
{christa}), (\ref{accnb}) and (\ref{distanb}) for the general Berwald
d--connections. In a similar form, we can compute splittings of connections
by introducing the N--frames and d--metric into \ (\ref{dcon1}), (\ref{mafdc}%
) and (\ref{dmafdc}) \ for the metric affine canonic d--connections and,
respectively, into (\ref{bct}), (\ref{mafbc}) and (\ref{dmafbc}) for the
metric--affine Berwald d--connections.

Finally, we note that for the respective classes of d--connections, the
components of the torsion and curvature tensors may be defined by
introducing the corresponding connections (\ref{christa}), (\ref{candcon}), (%
\ref{berw}), (\ref{bct}), (\ref{accnb}), (\ref{mafdc}) and (\ref{mafbc})
into the general formulas for torsion (\ref{dta}) and curvature (\ref{dra})
on spaces provided with N--connection structure.

\subsubsection{Generalized Lagrange algebroids}

\label{ssgla}The d--metric (\ref{slma}) was introduced for the canonical
geometric modelling of Lagrange mechanics on Lie N--algebroids. There are
physical arguments to consider more general configurations than those for
the Lagrange algebroids. For instance, J. L. Synge \cite{synge} considered a
metric of type
\begin{equation*}
g_{ij}(x,V(x))=\ ^{0}g_{ij}(x)+(1-u^{2}(x,V(x))V_{i}\
\end{equation*}%
in order to study the propagation of electromagnetic waves in a medium with
the index of refraction $n^{i}(x,V(x))=1/u^{i}(x,V(x)),$ $V^{2}=V_{i}V_{j}\
^{0}g_{ij}$ where $\ ^{0}g_{ij}(x)$ is the background (pseudo) Riemannian
metric of the medium and $V_{i}(x)$ is the velocity of the medium. There
were also considered metrics of type
\begin{equation*}
g_{ij}=e^{\sigma (x,y)}\ ^{l}g_{ij}(x,y),\mbox{ or }g_{ij}=e^{\sigma (x,y)}\
^{0}g_{ij}(x)
\end{equation*}%
related to physical processes in dispersive media or in general relativity:
a detailed study of the relativitstic optic and mechanical and
electromagnetic models resulting in generalizations of the Finsler, Lagrange
and (pseudo) Riemannian geometries is contained in Chapters XI and XII of
the monograph \cite{ma2} (see \cite{vnp} on such generalizations suggested
from higher energy physics). In order to model such processes on Lie
algebroids, we have to introduce into consideration classes of d--metrics
with more general coefficients and N--connections than $\ ^{l}g_{ab}$ (\ref%
{lamf}) and $\ ^{l}N_{\ b}^{a}$ (\ref{cncl}) on $\mathcal{L}^{\pi }E,$ i. e.
d--metrics of type
\begin{equation}
\ ^{gl}\mathbf{g}=\ \mathbf{g}_{AB}\mathbf{c}^{A}\otimes \mathbf{c}^{A}=\
g_{a^{\prime }b^{\prime }}(x,u)\ \mathbf{z}^{a^{\prime }}\otimes \mathbf{z}%
^{b^{\prime }}+\ g_{ab}(x,u)\ \mathbf{v}^{a}\otimes \mathbf{v}^{b}
\label{dmgl}
\end{equation}%
with arbitrary $g_{ab}(x,u)$ and $v^{a}$ elongated by arbitrary $N_{\
b}^{a}(x,u),$ see (\ref{ddif1a}).

\begin{definition}
A generalized Lagrange algebroid is a pair $GL\mathcal{A}=(E,\ g_{ab})$
defined by a nongenerated and constant signature$\ g_{ab}$ on $\widetilde{E}%
. $
\end{definition}

On $GL\mathcal{A},$ it is also possible to define a canonical N--connection
defined only by $g_{ab}$ and to associate a semispray configuration. To do
this, we introduce the absolute energy
\begin{equation}
\varepsilon (x,u)\doteqdot g_{ab}(x,u)u^{a}u^{b}  \label{dabse}
\end{equation}%
and consider the action integral
\begin{equation*}
I=\int\limits_{0}^{1}\varepsilon (x^{k}(t),\dot{x}^{i}(t)=\rho
_{b}^{i}u^{b}(t))dt.
\end{equation*}%
For a regular system, when the auxiliary d--tensor
\begin{equation*}
\ ^{\varepsilon }g_{ab}\doteqdot \frac{1}{2}\frac{\partial ^{2}\varepsilon }{%
\partial u^{a}\partial u^{b}}
\end{equation*}%
is nondegenerated, we can compute the Euler--Lagrange equations (\ref{eleq1b}%
), where $\varepsilon (x,u)$ is considered instead of $l(x,u),$ which are
equivalent to the nonlinear geodesic equations%
\begin{equation*}
\frac{du^{a}}{dt}+2\ ^{\varepsilon }G^{a}(x^{k},u^{b})=0
\end{equation*}%
where
\begin{equation*}
2\ ^{\varepsilon }G^{a}(x^{k},u^{b})=\frac{1}{2}\ ^{\varepsilon
}g^{ab}\left( \frac{\partial ^{2}\varepsilon }{\partial u^{b}\partial x^{i}}%
\rho _{c}^{i}u^{c}+C_{bc}^{e}u^{c}\frac{\partial \varepsilon }{\partial u^{e}%
}-\rho _{b}^{i}\frac{\partial \varepsilon }{\partial x^{i}}\right) .
\end{equation*}%
This follows just from the Theorem \ref{thnge} b) if $l\rightarrow
\varepsilon .$ $\ $The canonical N--connection is
\begin{equation*}
\ ^{\varepsilon }N_{\ b}^{a}\doteqdot \frac{\partial \ ^{\varepsilon }G^{a}}{%
\partial u^{b}}.
\end{equation*}

We may prove all results of the Section 3 (to define the canonical
d--connection, the almost complex and almost sympletic structure, ....) for
the generalized Lagrange algebroids, satisfying the regularity condition, by
substituting the absolute energy instead of the Lagrange function. This
results in

\begin{corollary}
Any generalized Lagrange algebroid defined by a metric tensor $g_{ab}$ can
be modelled equivalently as a Lagrange algebroid provided with a
corresponding absolute energy function $\varepsilon (x,u)$ if the regularity
conditions are satisfied.
\end{corollary}

For some explicit purposes, it could be more convenient to work directly
with $g_{ab},$ instead of $\ ^{\varepsilon }g_{ab},$ and with the d--metric (%
\ref{dmgl}) which is a particular case of (\ref{block2}) (when $%
h_{ab}=g_{ab}).$ In this case, we can introduce the canonical d--connection (%
\ref{candcon}) and compute the respective d--torsions (\ref{dtorsb}) and
d--curvatures (\ref{dcurv}).

\subsection{Hamilton--Cartan algebroids}

The Hamilton mechanics can be geometrized on cotangent bundles\\ $(T^{\ast
}M,\ ^{\ast }\pi ,$ $M)$ where $T^{\ast }M$ is dual to $TM$ (see, for
instance, a summary of approaches in Refs. \cite{mhss,ml}). There were
elaborated Lie algebroid geometrizations of the Hamilton equations in terms
of sympletic and Poisson structures on algebroids (see details and
references in \cite{acsw,dl1}). The aim of this section is to outline the
main features of the Hamilton mechanics realized in terms of prolongations
to Lie algebroids (with associated covector, dual, bundles) provided with
N--connection structure, defined by a Lagrangian and/or Hamiltonian. It
should be noted here that the Hamilton algebroids have been considered also
in Ref. \cite{lpopescu} following the formalism from \cite{cm1}.

We shall use the concept a covector bundle $\mathcal{E}^{\ast }=(E^{\ast },\
^{\ast }\pi ,M)$ where $E^{\ast }$ is dual to $E.$ When $E=TM,$ we obtain
the particular case of (co) tangent bundle. The local coordinates of a point
$\mathbf{\breve{u}=}(x,p)\in \mathcal{E}^{\ast }$ are denoted $\breve{u}%
^{\alpha }=(x^{i},p_{a})$ where the coordinates $p_{a}$ are dual to $u^{a}.$%
\footnote{%
We preserve all conventions and denotations introduced in the previous
sections with respect to Lie algebroids, vector bundles and manifolds, in
general, provided with N--connection structure. If would be necessary, we
shall use the labels '' $\breve{...}"$ and/or ''$\ast "$ in order to
emphasize that some geometric objects refer just to certain dual spaces or,
equivalently, co--spaces, and call such objects like co--vectors, but we
shall omit the term co--algebroid in order to avoid confusions with the
concept of coalageras. It should be noted that under dualization the
uper/lower indices will be transformed into the corresponding inverse ones,
lower/uper, indices. Following such formal rules, we can re-derive all
formulas for the dual spaces directly from the similar formulas for
'non--dual' objects. Nevertheless, in some cases, the relation between the
objects of Lagrange and Hamilton geometry in not a complete formal duality:
we have to take into account the Legendre transforms which introduce more
sophisticate constructions for the Hamilton spaces and geometries.} The
local bases and cobases on $\mathcal{E}^{\ast }$ are denoted, for instance, $%
\ \mathbf{\breve{e}}_{\alpha }=(e_{i},\breve{e}^{a})\ $and $\mathbf{\breve{e}%
}^{\alpha }=(e^{i},\breve{e}_{a}).$ A dual Lie algebroid (prolongated to a
Lie algebroid) is defined as a usual one but associated to a covector bundle
$\mathcal{E}^{\ast }.$ It should be emphasized here that we use the term
''coalgebroid'' induced from covector/ cotangent bundle but not from
''co--algebra''.

\subsubsection{Prolongations to Lie algebroids on dual vector bundles}

Let $\mathcal{L}^{\pi }\mathbf{E}\doteqdot (\mathbf{E},\ ^{\pi }[\cdot
,\cdot ],\ ^{\pi }\rho )$ be a prolongation Lie N--algebroid derived for a
Lie N--algebroid $(\mathbf{E},\ [\cdot ,\cdot ],\ \rho )$ with associated
vector bundle $\pi :\mathbf{E\rightarrow }M$ when the algebroids and related
vector bundles, in general, are provided with mutually compatible
N--connection structures. We also consider \textbf{\ }$^{\ast }\pi :\mathbf{E%
}^{\ast }\mathbf{\rightarrow }M$ to be the vector bundle projection of the
dual bundle $\mathbf{E}^{\ast }$ to $\mathbf{E}.$ For trivial N--connection
structures, we can write respectively $E^{\ast }$ and $E$ (we shall also use
''not--boldfaced'' symbols if the constructions will not depend on existence
of the N-connection structure).

The prolongation $\ \mathcal{L}^{\ast \pi }\mathbf{E}$ of $\mathbf{E}$ over $%
^{\ast }\pi $ is defined by the set of elements satisfying the conditions%
\begin{equation*}
\mathcal{L}^{\ast \pi }\mathbf{E=}\left\{ (u,\breve{z})\in \mathbf{E}\times T%
\mathbf{E}^{\ast }/\rho (b)=(T\mathbf{\ }^{\ast }\pi )(\breve{z})\right\} .
\end{equation*}%
We call $\ \mathcal{L}^{\ast \pi }\mathbf{E=}(\mathbf{E}^{\ast },\ ^{\ast
\pi }[\cdot ,\cdot ],\ ^{\ast \pi }\rho )$ a prolongation N--algebroid over $%
\mathbf{E}^{\ast },$ of rank $2m,$ with fibers isomorphic to $(E,E^{\ast })$
with the Lie algebroid structure $(\ ^{\ast \pi }[\cdot ,\cdot ],\ ^{\ast
\pi }\rho )$ defined for $\ast \pi $ and $\mathbf{E}^{\ast }$ instead of $%
\pi $ and $\mathbf{E}.$\footnote{%
see Ref. \cite{dl1} for details on such constructions; in this subsection we
outline the basic results in a form adapted to the N--connection structure}
\ We note that if $E=TM$ the Lie algebroid $\mathcal{L}^{\ast \pi }E$
transforms into the standard Lie algebroid $(T(T^{\ast }M),\ [\cdot ,\cdot
],\ ^{\ast \pi }\rho =Id).$

The space $\mathcal{L}^{\ast \pi }\mathbf{E}$ is fibred over $\mathbf{E}%
^{\ast }$ by the projection $\ ^{\ast \pi }\pi :\mathcal{L}^{\pi }\mathbf{E}%
\longrightarrow \mathbf{E}^{\ast },$ given by $\ \ ^{\pi }\pi (u,\breve{z}%
)=\ ^{\ast }\tau _{E}(\breve{z})$ where $^{\ast }\tau _{E}:T\mathbf{E}^{\ast
}\longrightarrow \mathbf{E}^{\ast }$ is the tangent projection. \ It is also
interesting to define the projection into the second factor: $\ ^{\ast \pi
}\rho :\mathcal{L}^{\ast \pi }\mathbf{E}\longrightarrow T\mathbf{E},$ \
given by $\ ^{\ast \pi }\rho (u,\breve{z})=\breve{z}.$ Stating a local basis
$\{\breve{v}^{a}\}$ of $Sec(\mathbf{E}^{\ast }),$ for $p\in E^{\ast },\pi
(p)=x\in M,$ and $x^{i}$ are local coordinats around $x,$ we have $p=p_{a}%
\breve{v}^{a}$ and the bundle coordinates on $E^{\ast }$ are $(x^{i},p_{a}).$

We denote respectively the section $\breve{s}\in Sec(\mathbf{E}^{\ast })$
and the sections of the modules of vector fields $\ ^{v}\breve{s}\in {%
\hbox
{\fr X}}(\mathbf{E}^{\ast }),\ ^{c}\breve{s}\in {\hbox
{\fr X}}(\mathbf{E}^{\ast }),$ and $\ ^{\mathbf{v}}\breve{s}\in {%
\hbox {\fr
X}}(\mathcal{L}^{\ast \pi }\mathbf{E}),\ ^{\mathbf{c}}\breve{s}\in {Sec}(%
\mathcal{L}^{\ast \pi }\mathbf{E}),$ and define the corresponding vertical
and complete lifts of sections of $\mathbf{E}$ into sections of $\mathcal{L}%
^{\ast \pi }\mathbf{E}.$ \ One holds the expressions
\begin{equation}
\ ^{\mathbf{c}}\breve{s}(u)=\left( \breve{s}(\ast \pi (p)),\ ^{c}\breve{s}%
(p)\right) \mbox{ and }\ ^{\mathbf{v}}\breve{s}(p)=\left( 0,\ ^{v}\breve{s}%
(p)\right) .  \label{rul1b}
\end{equation}%
There is an unique Lie algebroid structure $(\ ^{\ast \pi }[\cdot ,\cdot ],\
^{\ast \pi }\rho )$ on $\ \mathcal{L}^{\ast \pi }\mathbf{E}$ \ which defined
by
\begin{equation*}
~\ ^{\ast \pi }[\ ^{\mathbf{v}}\breve{s},\ ^{\mathbf{v}}\breve{s}]=0,~\
^{\ast \pi }[\ ^{\mathbf{c}}\breve{s},\ ^{\mathbf{v}}\breve{s}]=\ ^{\mathbf{v%
}}[\breve{s},\breve{s}],\ ^{\ast \pi }[\ ^{\mathbf{c}}\breve{s},\ ^{\mathbf{c%
}}\breve{s}]=\ ^{\mathbf{c}}[\breve{s},\breve{s}].
\end{equation*}%
For the lifts of functions we write
\begin{eqnarray*}
\ ^{\pi }\rho (\ ^{\mathbf{c}}\breve{s})(\ ^{c}f) &=&\ ^{c}\left( \rho (%
\breve{s})(f)\right) ,~\ ^{\pi }\rho (\ ^{\mathbf{c}}\breve{s})(\ ^{v}f)=\
^{v}\left( \rho (\breve{s})(f)\right) , \\
\ ^{\pi }\rho (\ ^{\mathbf{v}}\breve{s})(\ ^{c}f) &=&\ ^{v}\left( \rho (%
\breve{s})(f)\right) ,~\ ^{\pi }\rho (\breve{s}^{\mathbf{v}})(\ ^{v}f)=0.
\end{eqnarray*}

We denote the local coordinates on $M$ and $\mathcal{E}^{\ast }$
respectively by $x^{i}$ and $(x^{i},p_{a})$ and consider the Lie algebroid
structure functions $\rho _{a}^{i}(x)$ and $C_{be}^{a}(x).$ The local bases
for the considered vertical and complete lifts are written%
\begin{equation}
\ ^{c}\breve{e}_{a}=\rho _{a}^{i}\frac{\partial }{\partial x^{i}}%
-C_{ae}^{b}p_{b}\frac{\partial }{\partial p_{e}}\mbox{ and }\ ^{v}\breve{e}%
^{a}=\frac{\partial }{\partial p_{a}}  \label{bas2b}
\end{equation}
transforming any section $\breve{s}$ $=s_{a}\breve{v}^{a}$ of $E^{\ast },$
respectively, into the vector fields $\ ^{v}\breve{s}$ and $\ ^{c}\breve{s},$
when
\begin{equation*}
\ ^{c}\breve{s}=s^{a}\rho _{a}^{i}\frac{\partial }{\partial x^{i}}-\left(
\rho _{a}^{i}\frac{\partial s^{b}}{\partial x^{i}}-s^{d}C_{da}^{b}\right)
p_{b}\frac{\partial }{\partial p_{a}}\mbox{ and }\ ^{v}\breve{s}=s_{a}\frac{%
\partial }{\partial p_{a}}.
\end{equation*}%
These are local expressions, for a complete definition see Ref. \cite{dl1}.

The relations (\ref{rul1b}) for $z_{a^{\prime }}=\ ^{c}\breve{e}_{a^{\prime
}}$ and $\breve{v}^{a}=\ ^{v}\breve{e}^{a},$ in termis of the basis $\{z_{a%
%TCIMACRO{\U{b4}}%
%BeginExpansion
{\acute{}}%
%EndExpansion
},\breve{v}^{a}\}$ of sections of $\mathcal{L}^{\ast \pi }\mathbf{%
E\rightarrow E}^{\ast },$ we may transform the local frame (\ref{bas2b}) into%
\begin{equation*}
\ ^{\mathbf{c}}\breve{s}=s^{a^{\prime }}z_{a^{\prime }}-(\rho _{a}^{i}\frac{%
\partial s^{b}}{\partial x^{i}}p_{b}+s^{d}C_{ad}^{b})\breve{v}^{a}%
\mbox{ and
}\ ^{\mathbf{v}}s=s_{a}\breve{v}^{a}.
\end{equation*}%
It is convenient to introduce a new local basis on sections of $\ \mathcal{L}%
^{\ast \pi }\mathbf{E}$ $\ $over $\mathbf{E}^{\ast },$

\begin{equation}
\mathring{c}_{A}=(\mathring{z}_{a^{\prime }}=z_{a}+C_{ae}^{b}p_{b}\breve{v}%
^{e},\mathring{v}^{a}=\breve{v}^{a})  \label{basis1b}
\end{equation}%
with the components satisfying the typical Lie algebroid structure relations
(\ref{anch}) and (\ref{liea}). Defining
\begin{equation*}
\ ^{\pi }\rho (\mathring{z}_{a})=\rho _{a}^{i}\frac{\partial }{\partial x^{i}%
},~\ ^{\pi }\rho (\breve{v}^{a})=\frac{\partial }{\partial p_{a}},
\end{equation*}%
one obtains
\begin{equation*}
\ ^{\ast \pi }\left[ \mathring{z}_{a^{\prime }},\mathring{z}_{b^{\prime }}%
\right] =C_{ab}^{e}\mathring{z}_{e^{\prime }},~\ ^{\ast \pi }\left[
\mathring{z}_{a^{\prime }},\breve{v}^{b}\right] =0,~\ ^{\ast \pi }\left[
\breve{v}^{a},\breve{v}^{a}\right] =0.
\end{equation*}

With respect to the (\ref{basis1b}) for an element $\omega =\gamma
^{a^{\prime }}\mathring{z}_{a^{\prime }}+\zeta _{a}\mathring{v}^{a}\in \
\mathcal{L}^{\ast \pi }\mathbf{E},$ we can define the natural local
coordinates $(x^{i},p_{a},\gamma ^{a^{\prime }},\zeta _{a})$ on $\ \mathcal{L%
}^{\ast \pi }\mathbf{E},$ when the point $\omega $ $\in \ ^{\ast \pi }\pi
(\ast \pi ^{-1}(x))$ [for a vector bundle projection $\ ^{\ast \pi }\pi :\
\mathcal{L}^{\ast \pi }\mathbf{E}\rightarrow \mathbf{E}^{\ast }$ and $x\in
M, $ and $(x^{i},p_{a})$ considered also as the coordinates of the point $\
^{\ast \pi }\pi (\omega )\in \ast \pi ^{-1}(x)]$ may be expressed in
coordinate form
\begin{equation*}
\omega =\gamma ^{a^{\prime }}\mathring{z}_{a^{\prime }}(\ ^{\ast \pi }\pi
(\omega ))+\zeta _{a}\mathring{v}^{a}(\ ^{\ast \pi }\pi (\omega )).
\end{equation*}%
In coordinate form, the anchor map is defined
\begin{equation*}
\ ^{\pi }\rho (x^{i},p_{a},\gamma ^{a},\zeta _{a})=(x^{i},p_{a},\rho
_{a}^{i}\gamma ^{a},\zeta _{a}).
\end{equation*}

We can elaborate a differential form calculus by stating an abstract
differential operator $d^{\ast \mathcal{L}}\equiv d^{\ \mathcal{L}^{\ast \pi
}E}$ acting in the form
\begin{eqnarray}
d^{\ast \mathcal{L}}f &=&\rho _{a}^{i}\frac{\partial f}{\partial x^{i}}%
\mathring{z}^{a}+\frac{\partial f}{\partial p_{a}}\mathring{v}_{a},
\label{form1c} \\
d^{\ast \mathcal{L}}\mathring{z}^{a} &=&-\frac{1}{2}C_{be}^{a}\mathring{z}%
^{b}\wedge \mathring{z}^{e},~d^{\ast \mathcal{L}}\mathring{v}_{a}=0,  \notag
\end{eqnarray}%
where the local basis $\mathring{c}^{A}=(\mathring{z}^{a^{\prime }},%
\mathring{v}_{a})$ is the dual to $\mathring{c}_{A}=(\mathring{z}_{a^{\prime
}},\mathring{v}^{a}).$ Such formulas generalize on $\mathcal{L}^{\ast \pi }E$%
\ the similar ones (\ref{form1a}) defined by $f\in C^{\infty }(M)$ and $%
\theta =\theta _{b}\mathring{v}^{a}\in Sec(E^{\ast }),$ compare also with
the formulas (\ref{form1b}) for$\ \mathcal{L}^{\pi }E.$

\subsubsection{Hamilton equations and Poisson brackets on Lie algebroids}

We introduce the Liouville section (1--form) of $\mathcal{L}^{\ast \pi }E,$%
\begin{equation}
~^{h}\breve{\theta}\doteqdot p_{a}\breve{v}^{a}  \label{om1}
\end{equation}%
and, following the rules (\ref{form1c}), we can derive that the 2--form
\begin{equation}
~^{h}\breve{\omega}\doteqdot -d^{\ast \mathcal{L}}~^{h}\breve{\theta}=%
\mathring{z}^{a}\wedge \breve{v}_{a}+\frac{1}{2}C_{be}^{a}p_{a}\mathring{z}%
^{b}\wedge \mathring{z}^{e}  \label{om1a}
\end{equation}%
defines a canonical sympletic structure which is nondegenerate and and
satisfies the condition $d^{\ast \mathcal{L}}~^{h}\breve{\omega}=0.$ For the
standard Lie algebroid with $E=TM$ the $\breve{\theta}_{TE}$ and $\breve{%
\omega}_{TM}$ are respectively the usual Liouville 1-form and the canonical
sympletic 2--form on $T^{\ast }M.$

Let $\breve{h}:E^{\ast }\rightarrow \R$ be a map defining a Hamiltonian
function $\breve{h}(x^{i},p_{a})$ which, for simplicity, satisfies the
regularity condition when the Hessian
\begin{equation}
~^{h}\breve{g}^{ab}\doteqdot \frac{1}{2}\frac{\partial ^{2}\breve{h}}{%
\partial p_{a}\partial p_{b}}  \label{hehs}
\end{equation}%
is nondegenerated and of constant signature. There is a unique section
\begin{equation}
\breve{\xi}_{h}=\frac{\partial \breve{h}}{\partial p_{a}}\mathring{z}%
_{a}-\left( C_{ab}^{e}p_{e}\frac{\partial \breve{h}}{\partial p_{b}}+\rho
_{a}^{i}\frac{\partial \breve{h}}{\partial x^{i}}\right) \breve{v}^{a}\in
Sec(\mathcal{L}^{\ast \pi }E),  \label{aux6a}
\end{equation}%
inducing a vector filed $\rho ^{\ast \pi }(\breve{\xi}_{h})$ on $E^{\ast },$%
\footnote{%
for a standard Lie algebroid, $\breve{\xi}$ transforms into a usual Hamilton
vector field}%
\begin{equation*}
\rho ^{\ast \pi }(\breve{\xi}_{h})=\rho _{a}^{i}\frac{\partial \breve{h}}{%
\partial p_{a}}\frac{\partial }{\partial x^{i}}-\left( C_{ab}^{e}p_{e}\frac{%
\partial \breve{h}}{\partial p_{b}}+\rho _{a}^{i}\frac{\partial \breve{h}}{%
\partial x^{i}}\right) \frac{\partial }{\partial p_{a}},
\end{equation*}%
satisfying the equation
\begin{equation}
\breve{\xi}_{h}\rfloor \breve{\omega}_{E}=d^{\ast \mathcal{L}}\breve{h}
\label{aux7a}
\end{equation}%
for a given $d\breve{h}\in Sec(\mathcal{L}^{\ast \pi }E)^{\ast }.$ Such
formulas sketch the proof of a theorem (the ''dual'' of Theorem \ref{theleq}%
, for Hamilton structures on Lie algebroids):

\begin{theorem}
The integral curves of the section $\breve{\xi}_{h}$ (\ref{aux6a}) (with
induced vector field $\rho ^{\ast \pi }(\breve{\xi}_{h}))$ defining the
solution of (\ref{aux7a})) satisfy the Hamilton equations for $\breve{h}%
(x^{i},p_{a}),$%
\begin{eqnarray}
\frac{dx^{i}}{dt} &=&\rho _{a}^{i}\frac{\partial \breve{h}}{\partial p_{a}},
\label{hameq} \\
\frac{dp_{a}}{dt} &=&-\left( C_{ab}^{e}p_{e}\frac{\partial \breve{h}}{%
\partial p_{b}}+\rho _{a}^{i}\frac{\partial \breve{h}}{\partial x^{i}}%
\right) .  \notag
\end{eqnarray}
\end{theorem}

The dual bundle $E^{\ast }$ admits a linear Poisson structure, a 2--vector
field, $\Lambda _{E\ast }$ such that%
\begin{equation*}
\left[ \Lambda _{E\ast },\Lambda _{E\ast }\right] =0
\end{equation*}%
and $\Lambda _{E\ast }(df,df^{\prime })$ is a linear function for any linear
functions $f,f^{\prime }$ on $E^{\ast }.$ The local coordinate expression is
\begin{equation}
\Lambda _{E\ast }=\rho _{a}^{i}\frac{\partial }{p_{a}}\wedge \frac{\partial
}{\partial x^{i}}+\frac{1}{2}p_{e}C_{ab}^{e}\frac{\partial }{\partial p_{a}}%
\wedge \frac{\partial }{\partial p_{b}}.  \label{linpostr}
\end{equation}%
This structure induces a linear Poisson bracket%
\begin{equation*}
\left\{ f,w\right\} _{E\ast }\doteqdot \Lambda _{E\ast }(d^{TE\ast
}f,d^{TE\ast }w)
\end{equation*}%
where the operator $d^{TE\ast }$ is defined by the rules (\ref{form1c}) but
on $TE^{\ast }.$ In a particular case of local coordinates, we have
\begin{equation*}
\left\{ x^{i},x^{j}\right\} _{E\ast }=0,\ \left\{ p_{a},x^{j}\right\}
_{E\ast }=\rho _{a}^{i}\mbox{\ and \ }\left\{ p_{a},p_{b}\right\} _{E\ast
}=p_{e}C_{ab}^{e},
\end{equation*}%
see details in \cite{mart,dl1}.

For Hamiltonian sections, we can naturally use $~^{h}\breve{\omega}$ (\ref%
{om1a}) satisfying (\ref{aux7a}), instead of $\Lambda _{E\ast },$ in order
to define
\begin{equation*}
\left\{ f,w\right\} _{E\ast }\doteqdot -~^{h}\breve{\omega}(\breve{\xi}_{f},%
\breve{\xi}_{w}).
\end{equation*}

The formula for energy $E_{l}$ on a Lagrange algebroid (\ref{efunca}) can be
rewritten in the form
\begin{equation*}
\ \breve{h}=p_{a}u^{a}-\frac{l}{2}
\end{equation*}%
which relates the integral curves of the Euler--Lagrange equations to the
integral curves of the Hamilton equations if the regularity conditions are
satisfied. This relation stated in a more rigorous geometric form by the
Legendre transform associated to $l$ which is a smooth map%
\begin{equation*}
Leg_{l}:~E\rightarrow E^{\ast };\ Leg_{l}(a)(b)=~^{l}\theta (a)(z)
\end{equation*}%
defined with respect to the Poincare--Cartan 1--form $~^{l}\theta \in
Sec\left( (\mathcal{L}^{\pi }E)^{\ast }\right) $ (\ref{pc1fa}), for any $%
a,b\in E_{x},\ x\in M$ and $z\in $ $\mathcal{L}^{\pi }E_{\mid a}$ such that $%
pr_{1}(z)=b$ when $pr_{1}:\ \mathcal{L}^{\pi }E\rightarrow E$ is the
restriction to $\ \mathcal{L}^{\pi }E$ of the first canonical projection $%
pr_{1}:E\times TE\rightarrow E.$ The transform $Leg_{l}$ induces a map for
the prolongated algebroids,
\begin{equation*}
\mathcal{L}Leg_{l}:~\mathcal{L}^{\pi }E\rightarrow \ \mathcal{L}^{\ast \pi
}E;~(\mathcal{L}Leg_{l})(b,\mathbf{X}_{a})=\left( b,(T_{a}Leg_{l})(\mathbf{X}%
_{a})\right) ,
\end{equation*}%
where $a,b\in E$ and $(b,\mathbf{X}_{a})\in \left( \ \mathcal{L}^{\pi
}E\right) _{a}\subseteq E_{\pi (a)}\times T_{a}E$ and $TLeg_{l}:~TE%
\rightarrow TE^{\ast }$ being the tangent map of $Leg_{l}.$ The map $%
\mathcal{L}Leg_{l}$ is well defined because $\pi ^{\ast }\circ Leg_{l}=\pi .$

In local coordinates, the introduced Legendre maps are parametrized
respectively
\begin{equation*}
Leg_{l}(x^{i},u^{a})\rightarrow \left( x^{i},p_{a}=\frac{\partial l}{%
\partial u^{a}}\right)
\end{equation*}%
and%
\begin{equation*}
\mathcal{L}Leg_{l}(x^{i},u^{a};z^{a^{\prime }},v^{a})\rightarrow \left(
x^{i},p_{a};z^{a^{\prime }},\rho _{b}^{i}z^{b}\frac{\partial p_{a}}{\partial
x^{i}}+v^{b}~^{l}g_{ab}\right) .
\end{equation*}%
Using such coordinate maps, we can prove by straightforword computations
that under Legendre transforms the Poincare forms$~^{l}\theta $ (\ref{pc1fa}%
) and $~^{l}\omega $ (\ref{pc2fa}) transform respectively into $~^{h}\breve{%
\theta}$ (\ref{om1}) and $~~^{h}\breve{\omega}$ (\ref{om1a}). This deduces
that the pair $\left( Leg_{l},\mathcal{L}Leg_{l}\right) $ defines a morphism
between the prolongated Lie algebroids $\mathcal{L}^{\pi }E$ and $\ \mathcal{%
L}^{\ast \pi }E$ with compatible Lie algebroid structure functions.

\subsubsection{Hamilton algebroids and Cartan algebroids}

The Hamilton algebroids can be introduced as geometric mechanics structures
on $\ \mathcal{L}^{\ast \pi }E$ and $\mathcal{L}^{\pi }E$ related by
Legendre transforms:

\begin{definition}
A Hamilton algebroid is a pair $H\mathcal{A}=(E^{\ast },\ \ \breve{h})$
defined by a regular Hamiltonian $\ \breve{h}(x^{i},p_{a})$ being
differentiable on $\widetilde{E^{\ast }}\doteqdot E^{\ast }/\{0^{\ast }\}$
and continuous on the null sections of $\ 0^{\ast }:E^{\ast }\rightarrow M,$
with a nondegenerated and constant signature Hessian $\ ^{h}\breve{g}^{ab}$(%
\ref{hehs}) on $\widetilde{E^{\ast }}.$
\end{definition}

There are possibilities to restrict and/or to generalize this definition:

\begin{definition}
A Cartan algebroid $C\mathcal{A}=(E^{\ast },\ \ ^{c}\breve{h})$ is a
Hamilton algebroid with positive $\ \breve{h}=\ ^{c}\breve{h}(x^{i},p_{a})$
on $E^{\ast }$ and 1--homogeneous with respect to the momenta $p_{a},$ i. e.
$\ ^{c}\breve{h}(x^{i},\lambda p_{a})=\lambda \ ^{c}\breve{h}(x^{i},p_{a}).$
\end{definition}

Roughly speaking, the Cartan algebroids are Finsler algebroids but defined
on $\mathcal{L}^{\ast \pi }E$ and $E^{\ast }$ (with some additional
geometric structures related to the Legendre transforms).

In a more general case, we can consider prolongated Lie algebroid structures
defined by an arbitrary nondegeneratd d--tensor field $\breve{g}^{ab},$
co--metric on $\widetilde{E^{\ast }},$ not obligatory defined as the second
derivative on the momenta from a Hamiltonian. Nevertheless, the geometry of
such generalized Hamilton algebroids $GH\mathcal{A}=(E^{\ast },\ \ \breve{g}%
^{ab})$ can be modelled similarly to that of the usual Hamilton algebroids
by introducing an additional dual global ''energy'' function
\begin{equation*}
\breve{\varepsilon}(x,u)\doteqdot \breve{g}^{ab}(x,p)p_{a}p_{b}
\end{equation*}%
like for the generalized Lagrange algebroids.

\ For simplicity, in this work we shall consider the main geometric
constructions only for the Hamilton algebroids. We shall sketch the idea of
such proofs.

\paragraph{Canonical N--connections on Hamilton algebroids\newline
}

The results of Theorems \ref{thnge} and \ref{r2} reformulated for the
Hamilton algebroids are stated by

\begin{theorem}
The set of coefficients
\begin{equation}
~^{h}N_{ab}=\frac{1}{4}\left\{ ~^{h}g_{ab},\breve{h}\right\} _{E\ast }-\frac{%
1}{4}\left( g_{ac}\rho _{b}^{i}\frac{\partial ^{2}\breve{h}}{\partial
p_{c}\partial x^{i}}+g_{bc}\rho _{a}^{i}\frac{\partial ^{2}\breve{h}}{%
\partial p_{c}\partial x^{i}}\right)  \label{cnclca}
\end{equation}%
\ defines a canonical N--connection structure $~^{h}\mathbf{\breve{N}=}%
\left( ~^{h}N_{ab}\right) \}$ on $\mathcal{L}^{\ast \pi }E$ constructed only
from $\breve{h}$ and $\ ^{h}\breve{g}^{ab}$ and its dual, $~^{h}g_{ab},$ in
a form related to the canonical N--connection $~^{l}\mathbf{N}$ (\ref{cncla}%
) defined by the corresponding Lagrangian $l(x^{i},u^{a})$ on $\mathcal{L}%
^{\pi }E.$
\end{theorem}

\begin{proof}
Let us introduce the locally adapted base for such Hamilton algebroids,
\begin{equation}
\mathbf{\breve{c}}_{A}=(\mathbf{z}_{a^{\prime }}=\mathring{z}_{a^{\prime
}}-\ N_{a^{\prime }a}\mathring{v}^{a},\ \mathbf{\breve{v}}^{a}=\mathring{v}%
^{a}=\breve{v}^{a})  \label{dder1c}
\end{equation}%
and the duals
\begin{equation}
\mathbf{\breve{c}}^{A}=(\mathbf{z}^{a^{\prime }}=\mathring{z}^{a^{\prime
}},\ \mathbf{\breve{v}}_{a}=\breve{v}_{a}+\ N_{b^{\prime }a}\mathring{z}%
^{b^{\prime }})  \label{ddif1c}
\end{equation}%
where the algebroid indices $A=(a^{\prime },a),B=(b^{\prime },b),...$
running the values $a^{\prime },...b^{\prime },a,...b=1,2,...,m.$ The
construction of such N--elongated operators is similar to that from the
Proposition \ref{resws}, but in our case we use geometrical objects defined
on prolongation coalgebroids (for instance, labelled in the form $\breve{v}$
in order to emphasize the difference from the similar ones on usual
algebroids; for simplicity, we shall omit a such label, and the left label ''%
$h",$ for $N_{b^{\prime }a}$ when denotations will not give rise to
ambiguities). By local computations we can verify that $~^{h}N_{a^{\prime
}b} $ define a local distribution which can be globalized to the Whitney sum
\begin{equation*}
\ \mathcal{L}^{\ast \pi }E=h\ \mathcal{L}^{\ast \pi }E\oplus v\ \mathcal{L}%
^{\ast \pi }E
\end{equation*}%
which is an equivalent definition of the N--connection on a Lie algebroid,
see (\ref{cncla}). We may conclude that the coefficients (\ref{cnclca})
define a N--connection on $\mathcal{L}^{\ast \pi }E.$ It is a more
cumbersome task to prove that $~^{h}N_{a^{\prime }b}=\rho _{a^{\prime
}}^{i}\partial G_{b}/\partial x^{i}$ defines a nonlinear geodesic semispray
configuration $G_{b}$ which is equivalent both to the Euler--Lagrange
equations (\ref{eleq1b}) and the equivalent (for regular Legendre
transforms) Hamilton equations (\ref{hameq}). Such computations are
equivalent to those outlined for the proof of Theorems \ref{thnge} and \ref%
{r2} but re--derived in algebroid terms on $\mathcal{L}^{\ast \pi }E.$ $%
\square $
\end{proof}

\begin{definition}
A N--connection $\breve{N}_{a^{\prime }b}$ on $\ \mathcal{L}^{\ast \pi }E$
is symmetric if its torsion d--tensor
\begin{equation*}
\tau _{ab}\doteqdot \frac{1}{2}\left( \breve{N}_{ab}-\breve{N}_{ba}\right)
=0.
\end{equation*}
\end{definition}

From (\ref{cnclca}), one follows that the canonical N--connection is
symmetric, i. e.
\begin{equation}
~^{h}\tau _{ab}=0.  \label{scnct}
\end{equation}

\begin{definition}
The curvature d--tensor $\mathbf{\breve{\Omega}}$ of a N--connection $\breve{%
N}_{ab}$ on $\ \mathcal{L}^{\ast \pi }E$ is defined by the components
\begin{equation*}
\breve{\Omega}_{ab^{\prime }e^{\prime }}\doteqdot \mathbf{z}_{e^{\prime
}}N_{b^{\prime }a}-\mathbf{z}_{b^{\prime }}N_{e^{\prime }a}.
\end{equation*}
\end{definition}

A N--connection distribution on $\ \mathcal{L}^{\ast \pi }E$ is integrable
if and only if $\mathbf{\breve{\Omega}=0.}$

There is also a d--connection (Berwald type) defined by the N--connection
coefficients:

\begin{definition}
The Lie algebroid Berwald d--connection with local coefficients
\begin{equation*}
\ \overline{\breve{N}}_{~a^{\prime }e}^{b}\doteqdot \mathbf{\breve{v}}%
^{b}(N_{a^{\prime }e})\text{\mbox{ and
}}\overline{\breve{N}}_{be}^{a}\doteqdot 0
\end{equation*}%
is associated to a N--connection $N_{ab}$ and defines a covariant derivative
$\overline{\mathcal{D}}$\ on sections \ in the vertical vector subbundle $v\
\mathcal{L}^{\ast \pi }E\mathbf{.}$
\end{definition}

One holds (the proof is similar to that for the prolongation Lie algebroids,
see formula (\ref{ncurvla}) and Proposition \ref{bcdl}) the

\begin{proposition}
The Berwald covariant derivative $\overline{\mathcal{\breve{D}}}$ \ on $%
\mathcal{L}^{\ast \pi }E$ has the local expression%
\begin{equation*}
\overline{\mathcal{\breve{D}}}_{X}\left( \ ^{\star }B\right) \doteqdot
\mathbf{\breve{X}\cdot }\overline{\mathcal{\breve{D}}}=\left[ \breve{X}%
^{b}\left( \mathbf{z}_{b}B^{a}-\mathbf{\breve{v}}^{c}\ (\breve{N}_{\ cb}))\
B^{a}\right) +\ ^{\star }\breve{X}^{e}\mathbf{\breve{v}}_{e}B^{a}\right]
\mathbf{\breve{v}}_{a.}.
\end{equation*}
\end{proposition}

\paragraph{Canonical almost complex structures on coalgebroids \newline
}

The N--connection splitting on prolongation coalgebroids defines a
corresponding class of almost complex structures, which for the canonical
configurations are defined by the Hamiltonians.

\begin{proposition}
A canonical N--connection $~^{h}N_{a^{\prime }b}$ (\ref{cnclca}) induces,
naturally an almost complex (coalgebroid) structure $\ \mathbf{\breve{F}}:%
\mathcal{X}(\ \mathcal{L}^{\ast \pi }E)\rightarrow \mathcal{X}(\ \mathcal{L}%
^{\ast \pi }E),$
\begin{equation*}
\ \mathbf{\breve{F}}(\mathbf{z}_{a})=\mathbf{\breve{v}}_{a}\ \mbox{
and }\ \mathbf{\breve{F}}(\ \mathbf{\breve{v}}_{a})=-\mathbf{z}_{a},
\end{equation*}%
when
\begin{equation}
\ \mathbf{\breve{F}}=~^{h}g_{ab}\ \mathbf{\breve{v}}^{a}\otimes \mathbf{z}%
^{a}-~^{h}g^{ab}~\mathbf{z}_{a}\otimes \ \mathbf{\breve{v}}_{a}
\label{acs1c}
\end{equation}%
satisfies the condition $\ \mathbf{\breve{F}\rfloor \ \ \breve{F}=-I,}$ i.
e. $\breve{F}_{\ B}^{A}\breve{F}_{\ K}^{B}=-\delta _{K}^{A},$ where $\delta
_{K}^{A}$ is the Kronecker symbol and $\mathcal{X}$ \ denotes the module of
vector fields on $\widetilde{E^{\ast }}.$
\end{proposition}

\begin{proof}
It follows from the action of the operator (\ref{acs1c}) on the N--elongated
bases.$\square $
\end{proof}

\paragraph{Canonical metric and sympletic structures\newline
}

A regular Hamiltonian induces a canonical metrics on the corresponding
Hamilton algebroid:

\begin{theorem}
There is a canonical metric structures on $\mathcal{L}^{\ast \pi }E,\ $
\begin{equation}
\ ^{h}\mathbf{\breve{g}}=\ ^{h}\mathbf{\breve{g}}_{AB}\mathbf{\breve{c}}%
^{A}\otimes \mathbf{\breve{c}}^{A}=\ ^{h}\breve{g}_{ab}\ \mathbf{z}%
^{a}\otimes \mathbf{z}^{b}+\ ^{h}\breve{g}^{ab}\ \mathbf{\breve{v}}%
_{b}\otimes \mathbf{\breve{v}}_{b}  \label{candmcal}
\end{equation}%
called distinguished metrics (d--metrics) defined by the corresponding
Lagrangians and, induced by such Lagrangians, canonical N--connecti\-ons.
\end{theorem}

\begin{proof}
The existence follows from a Sasaki type lift of $\ ^{h}\breve{g}^{ab}$ to $%
\mathcal{L}^{\ast \pi }E$ by emphasizing the $(E,E^{\ast })$ structure of
fibers.$\square $
\end{proof}

In modern gravity, one considers models with cofiber metrics resulting in
more general d--metrics then (\ref{candmcal}), $\ $
\begin{equation}
\ \mathbf{\breve{g}}=\ \mathbf{\breve{g}}_{AB}\mathbf{\breve{c}}^{A}\otimes
\mathbf{\breve{c}}^{A}=\ \breve{g}_{ab}\ \mathbf{z}^{a}\otimes \mathbf{z}%
^{b}+\ \breve{h}^{ab}\ \mathbf{\breve{v}}_{b}\otimes \mathbf{\breve{v}}_{b}
\label{candmcalg}
\end{equation}%
where $\ \breve{g}_{ab}$ and $\ \breve{h}^{ab}$ are respectively some
independent z- and v-- components.

\begin{definition}
A Lie algebroid (coalgebroid) with associated vector bundle $\mathcal{E}=(%
\mathbf{E},\pi ,M)$ (covector bundle $\mathcal{E}=(\mathbf{E}^{\ast },\pi
^{\ast },M)$) is said to be sympletic if it admits a sympletic structure $%
\omega $ on the sections of the bundle $\wedge ^{2}E^{\ast }\rightarrow M$
such that 1) the map $\omega (x):E_{x}\times E_{x}\rightarrow \R$ is
nondegenerate and $d^{E}\omega =0.$\footnote{%
In a more general context, the Hamilton algebroids may be described in terms
of the Lichnerowicz--Poison and/or H--Chevalley--Eilengerg cohomologies \cite%
{dlmp1,dlmp2} where the Lie algebroids of Jacobi manifolds were considered.
For simplicity, in this work we do not concern topological properties of the
Lie N--algebroids.}
\end{definition}

A regular Hamiltonian defines a canonical almost sympletic structure derived
from the canonical N--connection (\ref{cnclca}), almost complex structure (%
\ref{acs1c}) and the canonical Poisson structure (\ref{linpostr}) which in
terms of the N--elongated partial derivatives (\ref{dder1c}) defines the
N--adapted Poisson bracket
\begin{equation*}
\{f_{1},f_{2}\}_{\mathbf{E}^{\ast }}=\mathbf{\breve{v}}^{a}\left(
f_{1}\right) ~\mathbf{z}_{a}\left( f_{2}\right) -\mathbf{\breve{v}}%
^{a}\left( f_{2}\right) ~\mathbf{z}_{a}\left( f_{1}\right)
\end{equation*}%
for any $f_{1}$ and $f_{2}$ on $\ \mathcal{L}^{\ast \pi }E.$

\begin{theorem}
A Hamilton algebroid is a sympletic algebroid with the cano\-ni\-cal sympletic
structure defined by the 2--form
\begin{equation}
\ ^{h}\mathbf{\omega }=\ ^{h}\mathbf{\omega }_{AB}\mathbf{\breve{c}}%
^{A}\wedge \mathbf{\breve{c}}^{B}\doteqdot \ \mathbf{\breve{v}}_{a}\otimes
\mathbf{z}^{b}  \label{s2fc}
\end{equation}%
and canonical Poisson structure $\{f_{1},f_{2}\}_{\mathbf{E}^{\ast }}$ on $%
\mathcal{L}^{\ast \pi }E.$
\end{theorem}

\begin{proof}
We compute, using formulas$\ $(\ref{dder1c}) and (\ref{cnclca}),
\begin{equation*}
^{h}\mathbf{\omega }=\mathbf{\ \breve{v}}_{a}\otimes \mathbf{z}^{b}=\ \breve{%
v}_{a}\otimes \mathring{z}^{b}+~^{h}\tau _{ab}~\mathring{z}^{a}\wedge
\mathring{z}^{b}
\end{equation*}%
and
\begin{equation*}
\{f_{1},f_{2}\}_{\mathbf{E}^{\ast }}=\breve{v}^{a}\left( f_{1}\right) ~%
\mathring{z}_{a}\left( f_{2}\right) -\breve{v}^{a}\left( f_{2}\right) ~%
\mathring{z}_{a}\left( f_{1}\right) -2~^{h}\tau _{ab}~\breve{v}^{a}\left(
f_{1}\right) \mathring{z}_{a}\left( f_{2}\right)
\end{equation*}%
where $~^{h}\tau _{ab}=0$ (\ref{scnct}).

The next step is to show that $d\ ^{h}\mathbf{\omega }=0$\textbf{\ }which
implies $d\breve{v}_{a}\wedge \mathring{z}^{a}=0$ for the N--connection $%
~^{h}N_{ab}.$ We have that the exterior differential
\begin{equation*}
d\breve{v}_{a}=-d\left( ~^{h}N_{ab}\right) \wedge \mathring{z}^{b},
\end{equation*}%
where $d\left( ~^{h}N_{ab}\right) $ is symmetric on indices $a$ and $b.$ So,
the antisymmetric product with $\mathring{z}^{a}$ vanishes, which proofs $d%
\breve{v}_{a}\wedge \mathring{z}^{a}=0$ and $d\ ^{h}\mathbf{\omega }=0.$%
\textbf{\ }We can conclude that $\ ^{h}\mathbf{\omega }$ is a closed 2--form
defining a sympletic structure.$\square $
\end{proof}

\subsubsection{Canonical anchors for the Lagrange--Hamilton algebroids}

It is possible to model the Hamilton algebroids as almost Kahlerian
structures:

\begin{theorem}
There is canonical almost Kahlerian model of the Hamilton algebroids defined
by the respective unique almost Kahlerian d--connection $~^{\ast }\widehat{%
\mathcal{D}}$ on $\ \mathcal{L}^{\ast \pi }E$ \ which preserves by
parallelism the vertical distribution and satisfies the conditions:

\begin{enumerate}
\item there is compatibility with the almost Kahlerian structure
\begin{equation*}
~^{\ast }\widehat{\mathcal{D}}_{\breve{X}}\ ^{h}\mathbf{g}=0,~^{\ast }%
\widehat{\mathcal{D}}_{X}\ ^{h}\mathbf{\omega }=0\mbox{ and }\widehat{%
\mathcal{D}}_{\breve{X}}\ \mathbf{\breve{F}}=0\mathbf{;}
\end{equation*}

\item the complete ''horizontal'', i.e. z-component, and ''vertical''
torsions vanish, i. e.
\begin{equation*}
z~^{\ast }\widehat{\mathcal{T}}(\mathbf{z}_{a},\mathbf{z}_{b})=0\mbox{ and }%
v~^{\ast }\widehat{\mathcal{T}}(\mathbf{\breve{v}}^{a},\mathbf{\breve{v}}%
^{a})=0;
\end{equation*}%
for any vector field $\ \mathbf{\breve{X}}$ on $TE^{\ast }.$
\end{enumerate}
\end{theorem}

\begin{proof}
The proof is similar to that of Theorem \ref{alhmla}\ but re--written for $%
~^{\ast }\widehat{\mathcal{D}}$ on $\ \mathcal{L}^{\ast \pi }E$ in terms of
dual objects indices. The almost Kahlerian d--connection $\ ^{h}\widehat{%
\mathbf{\Gamma }}_{~\beta \gamma }^{\alpha }=\left( \ ^{h}\ \widehat{L}%
_{ab}^{e},\ ^{h}\widehat{K}_{a}^{~bc}\right) $, equivalent to $~^{\ast }%
\widehat{\mathcal{D}},$ is defined by the coefficients
\begin{eqnarray}
\ ^{h}\ \widehat{L}_{ab}^{e} &=&\frac{1}{2}\ ^{h}g^{ac}\left\{ \mathbf{z}%
_{b}(\ ^{h}g_{ca})+\mathbf{z}_{a}(\ ^{h}g_{cb})-\mathbf{z}_{c}(\
^{h}g_{ab})\right\} ,  \label{dccanlah} \\
\ ^{h}\widehat{K}_{a}^{~bc} &=&\frac{1}{2}\ ^{h}g_{ac}\left\{ \ \mathbf{v}%
^{b}(\ ^{h}g^{ca})+\ \mathbf{v}^{a}(\ ^{h}g^{cb})-\mathbf{\ v}^{c}(\
^{h}g^{ab})\right\} .  \notag
\end{eqnarray}%
By straightforward calculations with covariant derivatives defined by the
the coefficients (\ref{dccanlah}) we can verify that one holds true all
conditions of the theorem.$\square $
\end{proof}

Any sympletic structure on a sympletic manifold $(M,\omega )$ induces a
corresponding isomporphism related to the bracket operator $\left\{
.,.\right\} ,$%
\begin{equation*}
T^{\ast }M\overset{\widetilde{\left\{ ,\right\} }=\widetilde{\omega }^{-1}}{%
\longrightarrow }TM
\end{equation*}%
where $\widetilde{\omega }(v)\doteqdot \omega (v,.)$ and $\widetilde{\left\{
.,.\right\} }=\widetilde{\omega }^{-1}$ denotes the pulling back of the
standard bracket on $\mathcal{X}(M)$ to define the bracket operation for the
differential 1--forms $\Omega ^{1}(M)\doteqdot Sec(T^{\ast }M).$ This
transforms $T^{\ast }M$ into a Lie algebroid with anchor $\rho =-\widetilde{%
\omega }^{-1}$ defined by the sympletic structure (as it was observed in %
\cite{weinst2} and investigated in details for the case of Lie algebroids of
the Poisson manifolds in \cite{vais2} and \cite{acsw})

The algebroids considered in this work (Lagrange--Finsler and
Hamil\-ton--Cartan ones) posses canonical sympletic structures defined by
the corresponding canonical N--connections induced by the fundamental
Lagrange or Hamilton functions. The sympletic structure induces canonical
anchor maps on such N--anholonomic manifolds.

Finally, we note that the torsions and curvatures on Hamilton algebroids may
be globally defined to be compatible to the N--connection structure and
computed in z-- and v--component form following the geometric formalism
presented in section \ref{dcons}\ \ (on d--connections, in the general
canonical case related to d--metrics of type )\ and in section \ref{torscurv}
(on d--torsions and d--curvatures, adapted to the N--elongated bases on
algebroids, see (\ref{candmcalg})).

\section{Einstein--Cartan Algebroid Structures}

In the previous sections, we demonstrated that the theory of Lie algebroids
provided with nontrivial N--connections has a natural background from
geometric mechanics and Finsler geometry. The aim of this section is to
demonstrate that certain nontrivial Lie algebroid and N--connection
structures can be defined by generic off--diagonal metrics and nonholonomic
frames in gravity theories and that such gravity configurations may be also
modelled by analogous optic--mechanical geometries. Further developments on
exact solutions in gravity possessing Lie algebroid symmetry can be found in
Refs.\cite{v501,v502,v503,v504}, see also some recent results on gerbe
extensions \cite{v505}.

\subsection{N--connections and algebroid structures in gravity}

For the geometric models of gravity and string theories with nonholonomic
frame (vielbein) structure, one does not work on the tangent bundle $TM$ but
on a general manifold $\mathbf{V},~$\ $\ dim\mathbf{V}=n+m,$ which is a
(pseudo) Riemannian space or a certain generalization with possible torsion
and nonmetricity fields \cite{vsjmp,v0408121}.

\subsubsection{N--anholonomic manifolds}

Let us consider a metric tensor $\mathbf{g}$ on the manifold $\mathbf{V}$
with the coefficients defined with respect to a local coordinate basis $%
du^{\alpha }=\left( dx^{i},du^{a}\right) ,$ \footnote{%
the indices run correspondingly the values $i,j,k,...=1,2,...,n$ and $%
a,b,c,...=1,2,...,m.$}
\begin{equation*}
\mathbf{g}=\underline{g}_{\alpha \beta }(\mathbf{u})du^{\alpha }\otimes
du^{\beta }
\end{equation*}%
where
\begin{equation}
\underline{g}_{\alpha \beta }=\left[
\begin{array}{cc}
g_{ij}+N_{i}^{a}N_{j}^{b}h_{ab} & N_{j}^{e}h_{ae} \\
N_{i}^{e}h_{be} & h_{ab}%
\end{array}%
\right] .  \label{ansatzc}
\end{equation}

\begin{definition}
A manifold $\mathbf{V}$ $\ $is N--anholonomic if it is provided with a
N--connection structure $\mathbf{N}=\{N_{j}^{a}\}$ defining a global
splitting
\begin{equation}
T\mathbf{V=}h\mathbf{V\oplus }v\mathbf{V,}  \label{withnsc}
\end{equation}%
which, in general, is a nonholonomic distribution on $T\mathbf{V.}$
\end{definition}

Such nonintegrable distributions may be derived for any generic
off--diago\-nal metrics parametrized in the form (\ref{ansatzc}) which can
not be diagonalized by coordinate transforms.

\begin{theorem}
\ A metric $\mathbf{g}$ with the coefficients (\ref{ansatzc}) models a
N--anholo\-no\-mic manifold if and only if a nonholonomic vielbein structure
\begin{equation}
\mathbf{e}_{\nu }=(e_{i},\ v_{b})=(e_{i}=\frac{\partial }{\partial x^{i}}%
-N_{\ i}^{a}\ \frac{\partial }{\partial u^{a}},\ v_{b}=\frac{\partial }{%
\partial u^{b}})  \label{dder3}
\end{equation}%
and
\begin{equation}
\mathbf{e}^{\mu }=(e^{i},\ v^{b})=(e^{i}=dx^{i},v^{b}=du^{b}+N_{\ i}^{b}\
dx^{i}),  \label{ddif3}
\end{equation}%
is prescribed on $T\mathbf{V.}$
\end{theorem}

\begin{proof}
Performing a frame transform
\begin{equation*}
\mathbf{e}_{\alpha }=\mathbf{e}_{\alpha }^{\ \underline{\alpha }}\partial _{%
\underline{\alpha }}\mbox{ and }\mathbf{e}_{\ }^{\beta }=\mathbf{e}_{\
\underline{\beta }}^{\beta }du^{\underline{\beta }}
\end{equation*}%
with the coefficients%
\begin{equation}
\mathbf{e}_{\alpha }^{\ \underline{\alpha }}(u)=\left[
\begin{array}{cc}
e_{i}^{\ \underline{i}}(\mathbf{u}) & N_{i}^{b}(\mathbf{u})e_{b}^{\
\underline{a}}(\mathbf{u}) \\
0 & e_{a}^{\ \underline{a}}(\mathbf{u})%
\end{array}%
\right]  \label{vt1c}
\end{equation}%
and
\begin{equation}
\mathbf{e}_{\ \underline{\beta }}^{\beta }(u)=\left[
\begin{array}{cc}
e_{\ \underline{i}}^{i\ }(\mathbf{u}) & -N_{k}^{b}(\mathbf{u})e_{\
\underline{i}}^{k\ }(\mathbf{u}) \\
0 & e_{\ \underline{a}}^{a\ }(\mathbf{u})%
\end{array}%
\right] ,  \label{vt2c}
\end{equation}%
we write equivalently the metric $\mathbf{g}$ in the form
\begin{equation}
\mathbf{g}=\mathbf{g}_{\alpha \beta }\left( \mathbf{u}\right) \mathbf{e}%
^{\alpha }\otimes \mathbf{e}^{\beta }=g_{ij}\left( \mathbf{u}\right)
e^{i}\otimes e^{j}+h_{cb}\left( \mathbf{u}\right) \ v^{c}\otimes \ v^{b},
\label{dmetrgr}
\end{equation}%
where $g_{ij}\doteqdot \mathbf{g}\left( e_{i},e_{j}\right) $ and $%
h_{cb}\doteqdot \mathbf{g}\left( v_{c},v_{b}\right) $ and $\mathbf{e}_{\nu
}=(e_{i},\ v_{b})$ and $\mathbf{e}^{\mu }=(e^{i},\ v^{b})$\ are,
respectively, just the values (\ref{dder3}) and (\ref{ddif3}), i. e.
vielbeins of type (\ref{dder1}) and (\ref{ddif1}), but in our case
considered for arbitrary $N_{i}^{b}(\mathbf{u})$ of dimension $n\times m$
stating a splitting of the manifold into submanifolds of dimensions $n$ and $%
\ m.$ This defines a special class of nonholonomic manifolds provided with a
global splitting into conventional ''horizontal'' and ''vertical'' subspaces
(\ref{withnsc}) (similar to (\ref{withns}) but not for the vector bundles)
induced by the ''off--diagonal'' terms $N_{i}^{b}(\mathbf{u})$ and the
prescribed type of nonholonomic frame structure. The d--metric (\ref{dmetrgr}%
) is a generalization of the (\ref{slm}) to the case of arbitrary
(non--Lagrange) metric and N--connection coefficients.
\end{proof}

If the manifold $\mathbf{V}$ is (pseudo) Riemannian, there is a unique
linear connection (the Levi--Civita connection) $\nabla ,$ which is metric, $%
\nabla \mathbf{g=0,}$ and torsionless, $\ ^{\nabla }T=0$ but this connection
is not adapted to the nonintegrable distribution induced by $N_{i}^{b}(u).$
In order to construct exact solutions parametrized by generic off--diagonal
metrics, or for investigating nonholonomic frame structures in gravity
models with nontrivial torsion, it is more convenient to work with more
general classes of linear connections which are N--adapted but contain
nontrivial torsion coefficients because of nontrivial nonholonomy
coefficients $W_{\alpha \beta }^{\gamma }.$ Such geometric constructions can
be adapted to the prescribed N--connection structure on Riemmann--Cartan
spaces provided with corresponding classes of nonholonomic (N--adapted)
frames. \footnote{%
certain models of Finsler geometry posses nontrivial nonmetricity filelds
which motivates further generalizations to metric--affine manifolds with
prescribed N--anholonomic structure, called generalized Finsler--affine
spaces \cite{vmaf}}

For splitting into subspaces of dimensions $n$ and $m$ of a
Riemann--Ca\-rtan space of dimension $(n+m),$ (the (pseudo) Riemannian
configurations can be treated as particular cases), the Lagrange and Finsler
type geometries were modelled by N--anholonomic structures as exact
solutions of gravitational field equations \cite{vsjmp}. It was concluded
that the geometry of any Riemann space of dimension $n+m$ (where $n,m\geq 2;$
we can consider $n,m=1$ as special degenerated cases), provided with
off--diagonal metric structure of type (\ref{ansatzc}) can be equivalently
modelled, by vielbein transforms of type (\ref{vt1c}) and (\ref{vt2c}) as a
geometry of nonholonomic manifolds enabled with N--connection structure $%
N_{i}^{b}(\mathbf{u})$ and d--metric (\ref{dmetrgr}), see details in \cite%
{vmaf}. For certain special conditions when $n=m,$ $N_{i}^{b}=\
^{L}N_{i}^{b} $ (\ref{cncl}) and the metric (\ref{dmetrgr}) is of type (\ref%
{slm}), a such Riemann--Cartan space of even dimension is 'nonholonomically'
equivalent to a Lagrange space (for the corresponding homogeneity conditions
one obtains the equivalence to a Finsler space).

Roughly speaking, by prescribing a corresponding vielbein structure, we can
model a Lagrange, or Finsler, geometry on a Riemannian manifold and,
inversely, a Riemannian geometry is 'not only a Riemannian one' but could be
also a generalized Finsler space.

It is known the fact that the first example of Finsler geometry was
considered in 1854 in the famous B. Riemann's hability thesis (see
historical details and discussion in Refs. \cite{bcs,ma2,vmaf}) who, for
simplicity, restricted his considerations only to the curvatures defined by
quadratic forms on hypersurfaces. Sure, for B. Riemann, it was unknown the
fact that if we consider general (nonholonomic) frames with associated
nonlinear connections (the E. Cartan's moving frame geometry, see Refs. in %
\cite{car1}) and off--diagonal metrics, the Lagrange and Finsler geometry
may be derived naturally even from quadratic metric forms adapted to the
N--connection structure.

\subsubsection{Lie N--aglebroids modelled on N--anholonomic manifolds}

For some additional parametrizations, a subclass of d--metrics of type (\ref%
{dmetrgr}) (equivalently, a subclass of metrics of type (\ref{ansatzc}))
models a prolongation Lie algebroid provided with N--connection structure\
and Sasaki type d--metric:

\begin{theorem}
\label{th9}A d--metric (\ref{dmetrgr}) on a N--anholonomic manifold $\mathbf{%
V,}$ for\\ $\dim \mathbf{V}=n+m,$ defines a d--metric $\ ^{\circ }\mathbf{g}$ of
type (\ref{block2}) on $(v\mathbf{V,}v\mathbf{V})$ of dimension $2m,$
modelling a Lie N--algebroid with structure functions $\rho _{b^{\prime
}}^{j}(x)$ and $C_{ab}^{d}(x)$, if and only if one holds the
parametrizations:%
\begin{eqnarray}
g_{ij}(x,u) &=&g_{a^{\prime }b^{\prime }}(x,u)\rho _{i}^{a^{\prime
}}(x,u)\rho _{j}^{b^{\prime }}(x,u),  \notag \\
~h_{ab}(x,u) &=&\ ^{\star }h_{a^{\prime }b^{\prime }}(x,u)\ e_{\
a}^{a^{\prime }}(x)\ e_{\ b}^{b^{\prime }}(x)  \notag \\
N_{\ i}^{a}(x,u) &=&\rho _{i}^{a^{\prime }}(x,u)N_{\ a^{\prime }}^{a}(x,u)
\label{aux7}
\end{eqnarray}%
for any values $\rho _{i}^{a^{\prime }}(x,u)$ and $e_{\ a}^{a^{\prime }}(x)$
for which the inverse $e_{a^{\prime }\ }^{\ a}(x)$ define a v--subspace
nonholonomic frame $e_{a^{\prime }}\doteqdot $ $e_{a^{\prime }\ }^{\ a}(x)\
\partial /\partial u^{a}$ satisfying the conditions
\begin{equation}
e_{a^{\prime }}e_{b^{\prime }}-e_{b^{\prime }}e_{a^{\prime }}=C_{a^{\prime
}b^{\prime }}^{d^{\prime }}(x)e_{d^{\prime }}  \label{aux8}
\end{equation}%
and
\begin{equation}
\rho _{b^{\prime }}^{j}(x)=g^{ji}(x,u)\ ^{\star }h_{a^{\prime }b^{\prime
}}(x,u)\rho _{i}^{a^{\prime }}(x,u).  \label{aux9}
\end{equation}
\end{theorem}

\begin{proof}
Let us introduce the values (\ref{aux7}) into (\ref{dmetrgr}) and define
\begin{equation*}
c^{a^{\prime }}\doteqdot \rho _{i}^{a^{\prime }}(x,u)dx^{i}\mbox{
and }v^{a}=e_{\ a^{\prime \prime }}^{a}(x)du^{a^{\prime \prime }}
\end{equation*}%
which together with theirs duals prescribe a class of N--adapted (to $N_{\
a^{\prime }}^{a})$ frames of type $\mathbf{c}_{A}=(\mathbf{z}_{a^{\prime }},%
\mathbf{v}_{a})$ (\ref{dder1a}) and $\mathbf{v}^{A}=(\mathbf{z}^{a^{\prime
}},\mathbf{v}^{a})$ (\ref{ddif1a}) derived by using vielbeins of type $%
(z_{a^{\prime }},v_{a})$ (\ref{bas2a}) and $(\widetilde{z}_{a},\widetilde{v}%
_{a})$ (\ref{basis1a}). Now we can consider the d--metric%
\begin{eqnarray}
\ ^{\circ }\mathbf{g} &\mathbf{\doteqdot }&g_{a^{\prime }b^{\prime }}\mathbf{%
z}^{a^{\prime }}\otimes \mathbf{z}^{b^{\prime }}+\ ^{\star }h_{ab}\mathbf{v}%
^{a}\otimes \mathbf{v}^{b},  \label{dmaux8} \\
\mathbf{z}^{a^{\prime }} &=&z^{a^{\prime }}\mbox{ and }\mathbf{v}%
^{a}=v^{a}+N_{\ a^{\prime }}^{a}z^{a^{\prime }},  \notag
\end{eqnarray}%
of type (\ref{block2}) on $(v\mathbf{V,}v\mathbf{V}).$ The values $\rho
_{b^{\prime }}^{j}=g^{ji}\ ^{\star }h_{a^{\prime }b^{\prime }}\rho
_{i}^{a^{\prime }},$ see formula (\ref{aux9}), the metric coefficients and $%
\rho _{i}^{a^{\prime }}$ depending on $x$-- and $u$--variables must be
related in a form to generate $\rho _{b^{\prime }}^{j}(x)$ depending only on
the $x$--variables, and $C_{ab}^{d}(x)$ from (\ref{aux8}): this defines the
structure constants of a Lie N--algebroid. We conclude that the data
\begin{equation*}
\left( g_{ij},h_{ab},N_{i}^{a};\rho _{i}^{a^{\prime }},C_{ab}^{d}\right)
\end{equation*}%
for a N--anholonomic manifold $\mathbf{V}$ can be transformed into the data
\begin{equation*}
\left( g_{ij},\ ^{\star }h_{ab},N_{\ b^{\prime }}^{a};\rho _{b^{\prime
}}^{j}=g^{ji}\ ^{\star }h_{a^{\prime }b^{\prime }}\rho _{i}^{a^{\prime
}},C_{ab}^{d}\right)
\end{equation*}%
modelling a Lie N--algebroid on $(v\mathbf{V,}v\mathbf{V})$ (the inverse
statement is also true) if the conditions of the Theorem are satisfied. $%
\square $
\end{proof}

The Theorem \ref{th9} motivates the concept:

\begin{definition}
A Lie N--algebroid is called Riemann--Cartan if it is provided with a
d--metric and d--connection structure induced from a Riemann--Cartan
manifold.
\end{definition}

The next two sections are devoted to general properties and explicit
examples of Riemann--Cartan and related Einstein--Cartan algebroids.

\subsection{Mechanical and optical modelling of gravitational processes}

During the last decade, an intensive research is devoted to analogous models
of gravity when some important gravitational effects like black hole
radiation, gravitational--electromagnetic interactions, nearly horizon
effects ... are modelled by optic and continuous media models, see outlines
of results and references in \cite{nov,sch}. The unified geometric approach
both to mechanics and gravity theories elaborated as some examples of
N--anholonomic manifolds and/or algebroids give rise to new principles in
ideas on such analogous modelling of field interactions.

Any gravitational theory defined by a generic off--diagonal metric on a
N--manifold $\mathbf{V,}$ $\dim \mathbf{V}=n+m,$ induces a canonical
Lagrange N--algebroid provided with a d--metric $\ ^{\circ }\mathbf{g}$ of
type (\ref{block2}) on $(v\mathbf{V,}v\mathbf{V})$ of dimension $2m.$ By a
corresponding class of nonholonomic transforms of the v--subspace, we can
'map' a such gravity into a Lagrange algebroid:

\begin{theorem}
Any d--metric (\ref{dmetrgr}) with the coefficients satisfying the
conditions (\ref{aux7})--(\ref{aux9}), transforms into a d--metric of type (%
\ref{dmgl}) for a generalized Lagrange algebroid,
\begin{equation}
\ ^{gl}\mathbf{g}=\ \mathbf{g}_{AB}\mathbf{\tilde{c}}^{A}\otimes \mathbf{%
\tilde{c}}^{A}=\ g_{a^{\prime }b^{\prime }}(x,u)\ \mathbf{z}^{a^{\prime
}}\otimes \mathbf{z}^{b^{\prime }}+\ g_{ab}(x,u)\ \mathbf{\check{v}}%
^{a}\otimes \mathbf{\check{v}}^{b},  \label{dmetraux9}
\end{equation}%
where $g_{ab}(x,u)=A_{a}^{\ a^{\prime }}A_{b}^{\ b^{\prime }}\ ^{\star
}h_{a^{\prime }b^{\prime }},$ see the Lie N--algebroid d--metric (\ref%
{dmaux8}), and the v--vielbeins $\mathbf{\check{v}}^{a}\doteqdot A_{\
a^{\prime }}^{a}\ v^{a^{\prime }},$ $A_{\ a^{\prime }}^{a}$ being inverse to
$A_{a}^{\ a^{\prime }},$ are subjected to a nonholonomic relation of type
\begin{equation*}
\mathbf{\check{v}}_{a}\mathbf{\check{v}}_{b}-\mathbf{\check{v}}_{a}\mathbf{%
\check{v}}_{b}=\ ^{\star }W_{ab}^{c}\mathbf{\check{v}}_{c}
\end{equation*}%
for $\mathbf{\check{v}}_{a}=$ $A_{a}^{\ a^{\prime }}v_{a^{\prime }}.$
\end{theorem}

\begin{proof}
The Theorem \ref{th9} states that (\ref{dmetrgr}) transforms into (\ref%
{dmaux8}) which is not a generalized Lagrange algebroid d--metric because,
in \ general, $g_{ab}\neq $ $\ ^{\star }h_{ab}.$ A formal equality of the
c-- and v--components of the d--metric on the Lie N--algebroid can be
obtained by a certain nonholonomic frame transform in the v--subspaces when $%
g_{ab}(x,u)=A_{a}^{\ a^{\prime }}A_{b}^{\ b^{\prime }}\ ^{\star
}h_{a^{\prime }b^{\prime }}$ for some $\mathbf{\check{v}}^{a}\doteqdot A_{\
a^{\prime }}^{a}\ v^{a^{\prime }},$ which is just the d--metric (\ref%
{dmetraux9}). The nonholonomy coefficients $\ ^{\star }W_{ab}^{c}$ depend
both on $A_{a}^{\ a^{\prime }},C_{be}^{a}$ and $N_{\ a^{\prime }}^{a}:$ they
can be computed in explicit form by commutating the vielbeins $\mathbf{%
\check{v}}_{a}$ when $\mathbf{v}^{a}=v^{a}+N_{\ a^{\prime }}^{a}z^{a^{\prime
}},$ see (\ref{dmaux8}). $\square $
\end{proof}

For the d--metric (\ref{dmetraux9}), we can introduce the ''absolute
energy'' (\ref{dabse}), in this case of gravitational origin, and define the
nonlinear geodesic congruences, on Lagrange algebroids (derived as
Riemann--Cartan algebroids), as Euler--Lagrange equations on Lie algebroids (%
\ref{eleq1b}). We can investigate some particular cases like in section \ref%
{ssgla} when $g_{ab}$ is derived from
\begin{equation*}
g_{ij}(x,V(x))=\ ^{0}g_{ij}(x)+(1-u^{2}(x,V(x))V_{i}\mbox{ or }%
g_{ij}=e^{\sigma (x,y)}\ ^{0}g_{ij}(x),
\end{equation*}%
which can be related to various type of optic -- continuous media mechanics
processes.

\subsection{Exact solutions with algebroid structure}

We conclude this work by constructing some explicit examples of classes of
d--metrics which describe exact solutions of the Einstein equations in
string gravity and, for more particular cases, in Einstein--Cartan gravity
and general relativity. Such solutions define Lie N--alegebroid
configurations which are called Einstein--Cartan algebroids.

\subsubsection{A class four dimensional N--anolonomic manifolds}

The gravity field equations in string gravity can be written in effective
form in terms of differential forms on a four dimensional N--anholonomic
manifold (see, for instance, Refs. \cite{vmaf})
\begin{equation}
\eta _{\alpha \beta \gamma }\wedge \widehat{\mathcal{R}}_{\ }^{\beta \gamma
}=\widehat{\Upsilon }_{\alpha },  \label{eecdc2}
\end{equation}%
where $\widehat{\mathcal{R}}_{\ }^{\beta \gamma }$ is the curvature 2--form
for the canonical d--connection, $\Upsilon _{\alpha }$ denote all possible
sources defined by using the canonical d--connection and $\eta \doteqdot
\ast 1$ is the volume form with the Hodje operator ''$\ast $'', $\eta
_{\alpha }\doteqdot \mathbf{e}_{\alpha }\rfloor \eta ,$ $\eta _{\alpha \beta
}\doteqdot \mathbf{e}_{\beta }\rfloor \eta _{\alpha },$ $\eta _{\alpha \beta
\gamma }\doteqdot \mathbf{e}_{\gamma }\rfloor \eta _{\alpha \beta },...$.

Let us consider a four dimensional metric ansatz for the d--metric (\ref%
{dmetrgr}) and frame (\ref{ddif1}) when $u^{\alpha
}=(x^{1},x^{2},y^{3}=v,y^{4});i=1,2$ and $a=3,4$ and the coefficients
\begin{eqnarray}
g_{ij}
&=&diag[g_{1}(x^{1},x^{2}),g_{2}(x^{1},x^{2})],h_{ab}=diag[h_{3}(x^{k},v),h_{5}(x^{k},v)],
\notag \\
N_{i}^{3} &=&w_{i}(x^{k},v),N_{i}^{4}=n_{i}(x^{k},v)  \label{ansatz1}
\end{eqnarray}%
are some functions of necessary smooth class. The partial derivatives are
denoted $a^{\bullet }=\partial a/\partial x^{1},a^{^{\prime }}=\partial
a/\partial x^{2},a^{\ast }=\partial a/\partial v.$

\begin{theorem}
\label{ricci}The nontrivial components of the Ricci d--tensors for the
ca\-no\-nical d--connection are
\begin{eqnarray}
R_{1}^{1} &=&R_{2}^{2}=-\frac{1}{2g_{1}g_{2}}[g_{2}^{\bullet \bullet }-\frac{%
g_{1}^{\bullet }g_{2}^{\bullet }}{2g_{1}}-\frac{(g_{2}^{\bullet })^{2}}{%
2g_{2}}+g_{1}^{^{\prime \prime }}-\frac{g_{1}^{\prime }g_{2}^{\prime }}{%
2g_{2}}-\frac{(g_{1}^{^{\prime }})^{2}}{2g_{1}}],  \notag \\
R_{3}^{3} &=&R_{4}^{4}=-\frac{1}{2h_{3}h_{4}}[h_{4}^{\ast \ast }-h_{4}^{\ast
}(\ln \left| \sqrt{\left| h_{3}h_{4}\right| }\right| )^{\ast }],
\label{riccia} \\
R_{3i} &=&-w_{i}\frac{\beta }{2h_{4}}-\frac{\alpha _{i}}{2h_{4}},\ R_{4i}=-%
\frac{h_{4}}{2h_{3}}[n_{i}^{\ast \ast }+\gamma n_{i}^{\ast }],  \notag
\end{eqnarray}%
\begin{eqnarray*}
\alpha _{i} &=&\partial _{i}h_{4}^{\ast }-h_{4}^{\ast }\partial _{i}\ln
\left| \sqrt{\left| h_{3}h_{4}\right| }\right| ,\ \beta =h_{4}^{\ast \ast
}-h_{4}^{\ast }[\ln \left| \sqrt{\left| h_{3}h_{4}\right| }\right| ]^{\ast },
\\
\ \gamma &=&3h_{4}^{\ast }/2h_{4}-h_{3}^{\ast }/h_{4}
\end{eqnarray*}%
for $h_{3}^{\ast }\neq 0$ and $h_{4}^{\ast }\neq 0.$
\end{theorem}

\begin{proof}
Details for a such computation are provided in Ref. \cite{v2}.
\end{proof}

\begin{corollary}
The Einstein equations (\ref{eecdc2}) for the ansatz (\ref{ansatz1}) are
compatible for vanishing sources and if and only if the nontrivial
components of the source, with respect to the frames (\ref{dder1}) and (\ref%
{ddif1}), are any functions of type
\begin{equation*}
\widehat{\Upsilon }_{1}^{1}=\widehat{\Upsilon }_{2}^{2}=\Upsilon
_{1}(x^{1},x^{2},v),\ \widehat{\Upsilon }_{3}^{3}=\widehat{\Upsilon }%
_{4}^{4}=\Upsilon _{3}(x^{1},x^{2}).
\end{equation*}

\begin{proof}
The proof, see details in \cite{v2}, follows from the Theorem \ref{ricci}
with the nontrivial components of the Einstein d-tensor, $\widehat{\mathbf{G}%
}_{\ \beta }^{\alpha }=\widehat{\mathbf{R}}_{\ \beta }^{\alpha }-\frac{1}{2}%
\delta _{\ \beta }^{\alpha }\overleftarrow{\mathbf{\hat{R}}},$ computed to
satisfy the conditions
\begin{equation*}
G_{1}^{1}=G_{2}^{2}=-R_{3}^{3}(x^{1},x^{2},v),G_{3}^{3}=G_{4}^{4}=-R_{1}^{1}(x^{1},x^{2}).
\end{equation*}
\end{proof}
\end{corollary}

Having the values (\ref{riccia}), we can prove \cite{v2} the

\begin{theorem}
\label{pent}The system of gravitational field equations (\ref{eecdc2})
defined for the ansatz (\ref{ansatz1}) can be solved in general form if
there are given certain values of functions $g_{1}(x^{1},x^{2})$ (or,
inversely, $g_{2}(x^{1},x^{2})$), $h_{3}(x^{i},v)$ (or, inversely, $%
h_{4}(x^{i},v)$) and of sources $\Upsilon _{1}(x^{1},x^{2},v)$ and $\Upsilon
_{3}(x^{1},x^{2}).$
\end{theorem}

So, we defined a class of four dimension N--anholonomic manifolds as exact
solutions of the Einstein equations for the canonical d--connection.

\subsubsection{Generation of Einstein--Cartan algebroid structures}

On $(v\mathbf{V,}v\mathbf{V}),$ which being associated to the data
satisfying the conditions of the Theorem \ref{pent} is 2+2--dimensional, we
introduce additional parametrizations modelling a Lie N--anholonomic
structure stated by the Theorem \ref{th9}.

Let us label the local bases by corresponding abstract indices, $%
z_{a^{\prime }}=(z_{1^{\prime }},z_{2^{\prime }})$ corresponding to $%
e_{i}=(e_{1,}e_{2})$ and $v_{a^{\prime \prime }}=(v_{1^{\prime \prime
}},v_{2^{\prime \prime }})$ corresponding to $v_{b}=(v_{3},v_{4}).$ For
simplicity, we shall work with d--metrics which are diagonalized with
respect to the corresponding N--anholonomic frames, i. e.
\begin{equation*}
g_{ij}=(g_{1}(x^{k}),g_{2}(x^{k})),h_{ab}=(h_{3}(x^{k},v),h_{4}(x^{k},v))
\end{equation*}%
and
\begin{equation*}
g_{a^{\prime }b^{\prime }}=(g_{1^{\prime }}(x^{k},v),g_{2^{\prime
}}(x^{k},v)),\ ^{\star }h_{a^{\prime \prime }b^{\prime \prime
}}=(h_{1^{\prime \prime }}(x^{k},v),h_{2^{\prime \prime }}(x^{k},v)),
\end{equation*}%
and consider nontrivial components of projections%
\begin{equation*}
\rho _{i}^{a^{\prime }}=\left( \rho _{1}^{1^{\prime }}(x^{k},v),\rho
_{2}^{2^{\prime }}(x^{k},v)\right)
\end{equation*}%
and vielbein components $e_{1^{\prime \prime }}^{\ 3}(x)=e_{[1]}(x)$ and $%
e_{2^{\prime \prime }}^{\ 4}=e_{[2]}(x),$ which induces nontrivial $%
C_{b^{\prime \prime }e_{^{\prime \prime }}}^{a^{\prime \prime }}(x).$

The Lie algebroid structure functions are chosen $C_{b^{\prime \prime
}e_{^{\prime \prime }}}^{a^{\prime \prime }}(x)$ and
\begin{equation*}
\rho _{a^{\prime }}^{i}=\left( \rho _{1^{\prime }}^{1}(x^{k}),\rho
_{2^{\prime }}^{2}(x^{k})\right)
\end{equation*}%
when in order to satisfy the conditions (\ref{aux7})--(\ref{aux9}) we have
to satisfy the relations%
\begin{eqnarray}
g_{1}(x) &=&g_{1^{\prime }}(x^{k},v)\left[ \rho _{1}^{1^{\prime }}(x^{k},v)%
\right] ^{2},\ g_{2}(x)=g_{2^{\prime }}(x^{k},v)\left[ \rho _{2}^{2^{\prime
}}(x^{k},v)\right] ^{2},  \label{aux11} \\
h_{3}(x,u) &=&\ h_{1^{\prime \prime }}(x,u)\ \left[ e_{\ 3}^{1^{\prime
\prime }}(x)\right] ^{2},\ h_{4}(x,u)=\ h_{2^{\prime \prime }}(x,u)\ \left[
e_{\ 4}^{2^{\prime \prime }}(x)\right] ^{2}\   \notag
\end{eqnarray}%
and
\begin{eqnarray}
\rho _{1^{\prime }}^{1}(x^{k}) &=&\frac{1}{g_{1}(x^{k})}h_{1^{\prime \prime
}}(x^{k},v)\rho _{1}^{1^{\prime }}(x^{k},v),  \label{aux12} \\
\rho _{2^{\prime }}^{2}(x^{k}) &=&\frac{1}{g_{2}(x^{k})}h_{2^{\prime \prime
}}(x^{k},v)\rho _{2}^{2^{\prime }}(x^{k},v).  \notag
\end{eqnarray}%
Having chosen the data $\rho _{i}^{a^{\prime }}=\left( \rho _{1}^{1^{\prime
}},\rho _{2}^{2^{\prime }}\right) ,$ we can compute $N_{\ i}^{a}=\rho
_{i}^{a^{\prime }}N_{\ a^{\prime }}^{a}.$

The data $\left\{ g_{1},g_{2},h_{3},h_{b}\right\} $ are given from an ansatz
(\ref{ansatz1}) solving the Einstein equations (\ref{eecdc2}). We may
consider in infinite set of Lie algebroid anchors $\left\{ \rho _{1^{\prime
}}^{1},\rho _{2^{\prime }}^{2}\right\} $ and algebroid d--metrics $\left\{
g_{1^{\prime }},g_{2^{\prime }},h_{1^{\prime \prime }},h_{2^{\prime \prime
}}\right\} $ related in a compatible way, via (\ref{aux11}) and (\ref{aux12}%
), to some functions $\left\{ \rho _{1}^{1^{\prime }},\rho _{2}^{2^{\prime
}},e_{[1]},e_{[2]}\right\} $ with $e_{[1]}$ and $e_{[2]}$ related to the
structure coefficients $C_{b^{\prime \prime }e^{\prime \prime }}^{a^{\prime
\prime }}(x).$ This points to the fact that an exact solution of the
Einstein equations (vacuum or nonvacuum type) may parametrize an infinite
set of Lie algebroid structures which was emphasized in Refs. \cite%
{strobl2,strobl1}. This is not surprising because the Lie algebroids are
specific spaces defined by singular maps. We can induce a more explicit Lie
algebroid configuration by fixing compatible frames of reference, boundary
conditions and some classes of symmetries describing two (interrelated)
theoretical models on the N--anholonomic manifold and on a corresponding Lie
N--algebroids.

\end{document}